\documentclass[useAMS,usenatbib]{mn2e}

\usepackage{graphicx}
\usepackage{amssymb}
\usepackage{amsmath}
\usepackage{times} 
\usepackage{multirow}

\usepackage[usenames,dvipsnames]{xcolor}
\usepackage[normalem]{ulem}

\title[Tomographic local analyses of the WISE{\sf x}SuperCOSMOS catalogue]
      {Tomographic local 2D analyses of the WISE{\sl x}SuperCOSMOS all-sky galaxy catalogue}

\author[C. P. Novaes, A.  Bernui, H. S. Xavier, and G. A. Marques]{
C. P. Novaes,$^{1}$\thanks{e-mail: camilapnovaes@gmail.com}
A. Bernui,$^{1}$ 
H. S. Xavier$^{2}$ 
and G. A. Marques$^{1}$ \\ 
$^{1}$Observat\'orio Nacional, Rua General Jos\'e Cristino 77, 
          S\~ao Crist\'ov\~ao, 20921-400 Rio de Janeiro, RJ, Brazil \\
$^{2}$Instituto de Astronomia, Geof\'isica e Ci\^encias Atmosf\'ericas, Universidade de S\~ao Paulo, Rua do Mat\~ao, 1226, 05508-090, \\ 
S\~ao Paulo - SP, Brazil 
}

\begin{document}

\date{Accepted xxxx. Received xxxx; in original form xxxx}

\pagerange{\pageref{firstpage}--\pageref{lastpage}} \pubyear{2017}

\maketitle

\label{firstpage}

\begin{abstract}

The recent progress in obtaining larger and deeper galaxy catalogues is of fundamental importance for cosmological studies, especially to robustly measure the large scale density fluctuations in the Universe.
The present work uses the Minkowski Functionals (MF) to probe the galaxy density field from the WISExSuperCOSMOS (WSC) all-sky catalogue by performing tomographic local analyses in five redshift shells (of thickness $\delta z = 0.05$) in the total range of $0.10 < z < 0.35$. 
Here, for the first time, the MF are applied to 2D projections of the galaxy number count (GNC) fields with the purpose of looking for regions in the WSC catalogue 
with unexpected features compared to $\Lambda$CDM mock realisations. 
Our methodology reveals 1 -- 3 regions of the GNC maps in each redshift shell with an uncommon behaviour (extreme regions), i.e., $p$-value $<$ 1.4\%. 
Indeed, the resulting MF curves show signatures that suggest the uncommon behaviour to be associated 
with the presence of over- or under-densities there, but contamination due to residual foregrounds is not discarded. 
Additionally, even though our analyses indicate a good agreement among data and simulations, 
we identify 1 highly extreme region, seemingly associated to a large clustered distribution of 
galaxies. 
Our results confirm the usefulness of the MF to analyse GNC maps from photometric galaxy datasets. 

\end{abstract}


\section{Introduction}\label{Introduction}

The complementarity among the cosmic microwave background (CMB) and tracers of the large-scale structure (LSS) of the Universe has been essential to the expressive progress of the cosmology.
In fact, the former probes the early Universe ($z \sim 1100$), estimating cosmological parameters 
and structure formation, only possible with precise measurements by the Wilkinson Microwave Anisotropy 
Probe~\citep[WMAP;][]{2013/wmapCollab,2013/wmapCollab2} and Planck~\citep{2016/plaI,2016/plaXIII}  satellites.
The LSS tracers, on the other hand, provide information about the evolution of the Universe today ($z \sim 0 - 2$) by mapping the distribution of the luminous matter as done by, 
e.g., the 2MASS \citep{2004/2mass} and SDSS \citep{2015/sdss} projects, and other future 
surveys \citep[see, e.g.,][]{2014/jpas,2016/euclid}. 

The CMB temperature field has imprinted information about the primordial density perturbations, which, according to the current inflationary paradigm is represented by a nearly Gaussian field, well described 
by linear physics \citep{2016/plaXVII}. 
In contrast, the non-linear process of gravitational instability occurring during the structure formation leads to the cosmic density field which today appears highly non-Gaussian \citep{2002/takada,2010/berge,2010/kratochvil,2011/yang}.
For this reason, the two-point correlation function, although important and natural descriptors of such fields, is no longer enough to capture all the cosmological information.

The Minkowski Functionals \citep[MF;][]{1903/minkowski,1999/novikov,2001/sato, 2003/komatsu} are widely used to investigate the statistical properties, in particular the departure from Gaussianity, of the 2 dimensional (2D) cosmic microwave background temperature field and the 3D distribution of galaxies in the Universe.  
In fact, their success in constraining non-Gaussianity of CMB data is not restricted to the primordial type \citep[see, for example,][and references therein]{2003/komatsu, 2012/hikage-matsubara, 2013/munshi, 2016/plaXVII, 2014/novaes, 2015/novaes}, but have also been shown their efficiency in discriminating it from the secondary ones, i.e., signal originated after the last scattering surface, or even instrumental noise \citep{2016/novaes}. 
The non-Gaussianity from an observable carries a substantial amount of information beyond the one in the power spectrum. 
The capability of the MF in probing it has also been applied to differentiate between cosmological models, constraining cosmological parameters \citep{2014/shirasaki,2012/pratten,2015/petri,2017/matilla} and probing modifications of the gravity \citep{2017/fang,2017/shirasaki}. 
In this sense, several authors have also used these morphological tools to analyse volume distributed samples, as, for example, in \cite{2000/schmalzing,2003/hikage,2007/saar,2010/kerscher,2013/choi}, in general comparing the results to the expected in the $\Lambda$CDM cosmology.

Our aim in this paper is to perform a detailed analyses of the LSS density fields as traced by the distribution of galaxies.
Currently, efforts are being done to reproduce, through large and sophisticated N-body hydrodynamical simulations, the formation and evolution of the structures today observed in the Universe, where the foam-like structures displayed in the cosmic web include large voids, walls, filaments, and superclusters. 
For this, one expects that such structures be present in the sky, although to disclose their features in the datasets (i.e., space location, size, morphology, etc.) is not an easy task. 
This motivates our search for the plausible signatures that such structures might leave in the LSS density fields from deep galaxy surveys. 
Accordingly, our analyses intend to reveal regions with unexpected excess-of or lack-of luminous matter, suggestive of the presence of superclusters of galaxies or giant voids, respectively. 
For these analyses we shall use the MF, successfully employed in recent morphological analyses of the large-scale galaxy distribution, where they capture the imprints left by the clustering of galaxies and the presence of voids [\cite{2010/kerscher}; see, e.g.,~\cite{2016/novaes} for the application of MF to the inspection of the CMB temperature fluctuations field].

In the present paper, for the first time, the analyses of the galaxy distribution by the MF are not performed upon a 3D volume distributed sample, but a 2D projection of the data sample. 
We use these tools to do the morphological analyses of the recently released galaxy catalogue WISExSuperCOSMOS \citep[WSC;][]{2016/bilicki}.
In fact, for the study of the LSS density fields, the WSC catalogue present some 
advantages with respect to other datasets, namely: 
(i) its sky coverage is much larger than in other surveys, giving us a more complete picture of the local Universe's structure diversity; after the application of a severe cut-sky mask one still has left a sky fraction $f_{\mathrm{sky}} \simeq 0.55$, almost twice the area surveyed by the SDSS. 
(ii) The number density of galaxies and the angular resolution of the WSC data are good enough for our planned local analyses using MF, in particular to reveal peculiar sky patches with an uncommon behaviour as compared to simulations.
(iii) There are two versions of the WSC catalogue, each one corresponding to different foreground cleaning process, which allows for a comparison of the results derived from each version, minimizing the influence of systematics in our outcomes.

Our methodology consists in comparing the MF calculated from the WSC catalogue to those from a set of $\Lambda$CDM lognormal realisations. 
These analyses are performed upon small sky regions of tomographic redshift bins of the sample, that is, after splitting the galaxy distribution in five shells defined at disjoint photo-$z$ ranges in the total interval of 0.10-0.35. 
This allowed us to identify a total of 10 regions (each redshift bin containing from 1 to 3 of them) appearing in disagreement with the simulations, 
with $p$-value $<$ 1.4\%. 
We choose this threshold $p$-value as a compromise in order to have at least 1 patch selected in each photo-$z$ bin. 

The outline of this paper is as follows. 
Section \ref{sec:section2} presents a description of the data and simulations used in our analyses.
The basic concepts of the MF and their calculation are introduced in Section \ref{sec:section3}.
Details of how the local tomographical analyses were employed and the results obtained are described in Sections \ref{sec:section4} and \ref{sec:section5}, while Section \ref{sec:section6} compiles our 
conclusions and final remarks.

\section{Data description}
\label{sec:section2}

The analyses presented in this paper were performed upon two different samples, both derived from the same  galaxy catalogue, the WISExSuperCOSMOS, through different cleaning processes as will be discussed. 
The present section compile the description of these two data samples and details about the generation of the $\Lambda$CDM mock realisations used for comparison analyses. 

\subsection{WISExSuperCOSMOS galaxy catalogue}

The WSC catalogue was constructed by cross-matching the currently largest all-sky photometric 
samples, namely, the WISE \citep{2010/wright}, in the mid-infrared, and the SuperCOSMOS \citep{2016/peacock}, in the optical. 
The cleaning procedure used by \cite{2016/bilicki} to separate galaxies from stars and quasars is based in color cuts, aiming to obtain a robust sample for the purpose of estimating their photometric redshift (photo-$z$). 
The released WSC sample has extinction-corrected magnitudes within $B < 21$, $R < 19.5$ (both AB-like), and $13.8 < W1 < 17$ (in Vega system), resulting in a total of $\sim$ 20 million galaxies with a median redshift of $z_0 \simeq 0.2$ ($z < 0.4$). 
Since the contamination is smaller when removing objects at the lowest and highest photo-$z$ ranges, we applied an extra cut to select only the sources in the range $0.10 < z < 0.35$. 
Moreover, because the stellar contamination is highly sensitive to a $W1 - W2$ color cut, we also apply the fixed cut of $W1 - W2 > 0.2$ upon the whole sky. 
Hereafter, this sample will be referred as WSC-$clean$.
We then split it in five photo-$z$ bins of the same thickness, $\delta z = 0.05$, as summarized in Table \ref{tab:redshif_bins}, and build the first set of galaxy number-count (GNC) maps for analysis. 
Such maps were constructed in the {\sc healpix} (Hierarchical EqualArea iso-Latitude Pixelization) pixelization scheme \citep{2005/gorski} with N$_{\rm side}$ = 128, such that each pixel contains the objects encompassed by its area (pixel area of $\sim 0.05 ~{\rm deg}^2$).

The second set of GNC maps was produced from the catalogue constructed by \cite{2016/krakowski} from the same WSC photometric redshift catalogue but using the alternative approach of support vector machines (SVM) learning for an automatized identification of the galaxies. 
The released catalogue comprises the whole WISE and SuperCOSMOS samples, in a total of $\sim$ 40 million sources, each of them flagged by the SVM according to the probability of being a galaxy, quasar, or star ($p_{\mathrm{gal}}$, $p_{\mathrm{star}}$, and $p_{\mathrm{QSO}}$, respectively). 
In order to use a sample as pure as possible, in our analyses we consider the conservative cut of $p_{\mathrm{gal}} > 0.67$, resulting in a catalogue (hereafter called WSC-$svm$) with a median redshift of $z_0 \simeq 0.15$.
Again, some details about the samples originating the GNC maps investigated here are summarized in table \ref{tab:redshif_bins}.

Additionally, in order to remove areas of contamination not accounted by the above mentioned cuts, we also use the cut-sky mask released by \cite{2016/bilicki} jointly to the WSC catalogue.
This mask was constructed with N$_{\rm side}$ = 256 taking into account the Galaxy extinction, including regions at 
$E(B-V) > 0.25$ and regions of the sky with abnormal source density when compared to a lognormal distribution. 
It was downgraded to N$_{\rm side}$ = 128, in such a way that a pixel in the new resolution is unmasked if at least 50\% of the high resolution pixels are unmasked, and, in order to be even more rigorous, we also include in the masked region all the pixels with at least one source obeying $E(B-V) > 0.10$. 
The resulting mask leave for analyses a sky fraction of $f_{\mathrm{sky}} \simeq 0.55$. 
Note that we apply this mask upon all the GNC maps from the previously described samples and each photo-$z$ bin.

\begin{table}
\centering
\begin{tabular}{ c | c c c c c }
\hline
\multirow{ 2}{*}{Bin \#} & \hspace{-0.1cm} \multirow{ 2}{*}{photo-$z$ range} & \multicolumn{2}{|c|}{WSC-$clean$} & \multicolumn{2}{|c|}{WSC-$svm$} \\
                         &                                                   &  counts        &           $z_0$  &  counts     &            $z_0$   \\
\hline
1 & \hspace{-0.1cm} 0.10 - 0.15 & \hspace{-0.2cm} 1435113 & \hspace{-0.1cm} 0.130 & \hspace{-0.2cm} 2108545 & \hspace{-0.1cm} 0.127 \\
2 & \hspace{-0.1cm} 0.15 - 0.20 & \hspace{-0.2cm} 2307948 & \hspace{-0.1cm} 0.176 & \hspace{-0.2cm} 2190296 & \hspace{-0.1cm} 0.174 \\ 
3 & \hspace{-0.1cm} 0.20 - 0.25 & \hspace{-0.2cm} 2682398 & \hspace{-0.1cm} 0.226 & \hspace{-0.2cm} 1754763 & \hspace{-0.1cm} 0.223 \\
4 & \hspace{-0.1cm} 0.25 - 0.30 & \hspace{-0.2cm} 2336109 & \hspace{-0.1cm} 0.272 & \hspace{-0.2cm} 1379258 & \hspace{-0.1cm} 0.274 \\
5 & \hspace{-0.1cm} 0.30 - 0.35 & \hspace{-0.2cm} 719750  & \hspace{-0.1cm} 0.316 & \hspace{-0.2cm} 859436  & \hspace{-0.1cm} 0.320 \\
\hline
\end{tabular}
\caption{Summary of the photo-$z$ range, and the total galaxy counts and median redshift ($z_0$) of each WSC sample and photo-$z$ bin, after the cut-sky masking. The median redshits of the samples in the entire range $0.10 < z < 0.35$ are $z_0^{clean}$ = 0.22 and $z_0^{svm}$ = 0.20 (see Fig. \ref{fig:nz}).} 
\label{tab:redshif_bins}
\end{table}

\begin{figure}
\includegraphics[width=1.0\columnwidth]{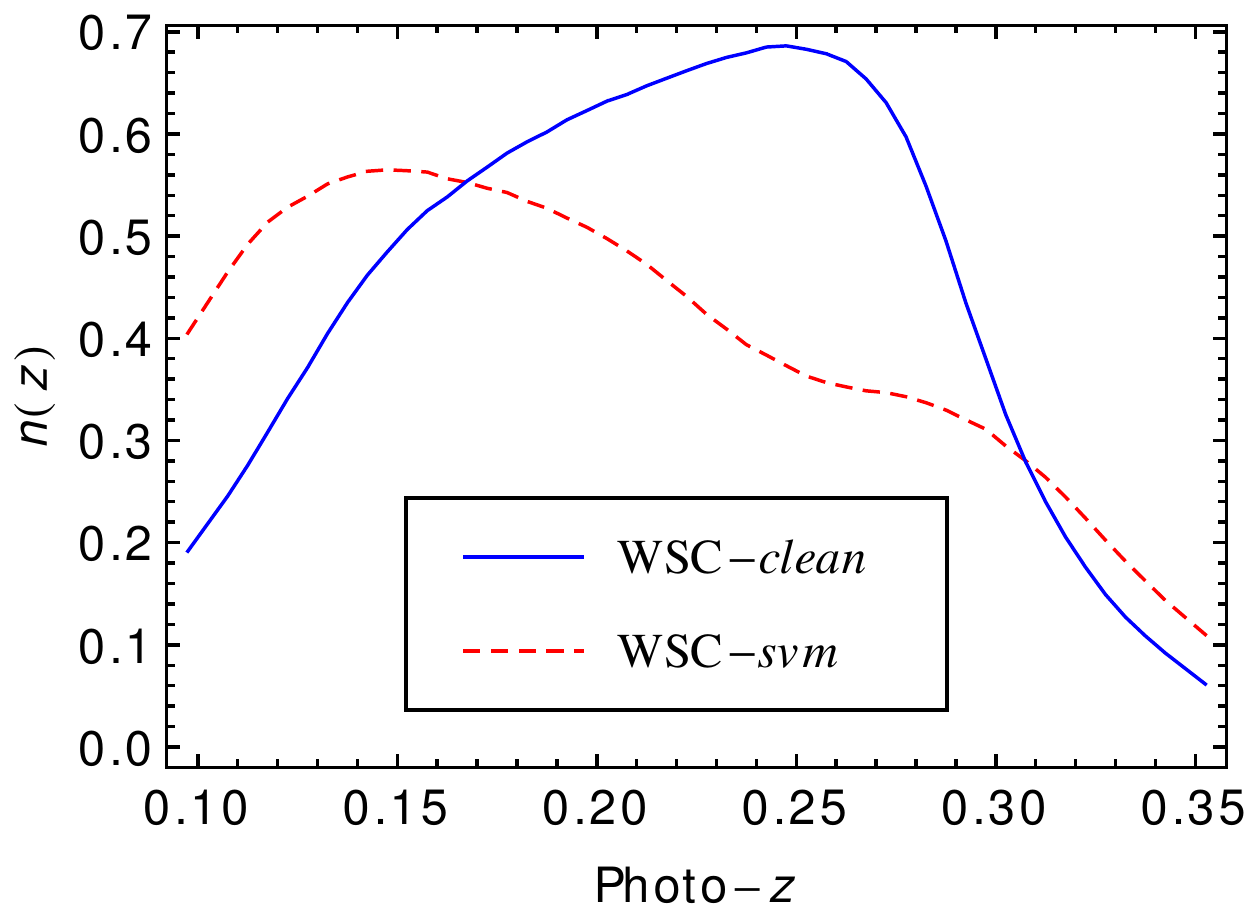}
\caption{Galaxy density $n(z)$ distribution, in counts per square arcminute and unit redshift, by the photo-$z$ for the WSC-$clean$ (blue solid line) and WSC-$svm$ (red dashed line) samples. }
\label{fig:nz}
\end{figure}

\subsection{Lognormal simulations} \label{sec:section2.2}

\begin{figure}
 \includegraphics[width=1.0\columnwidth]{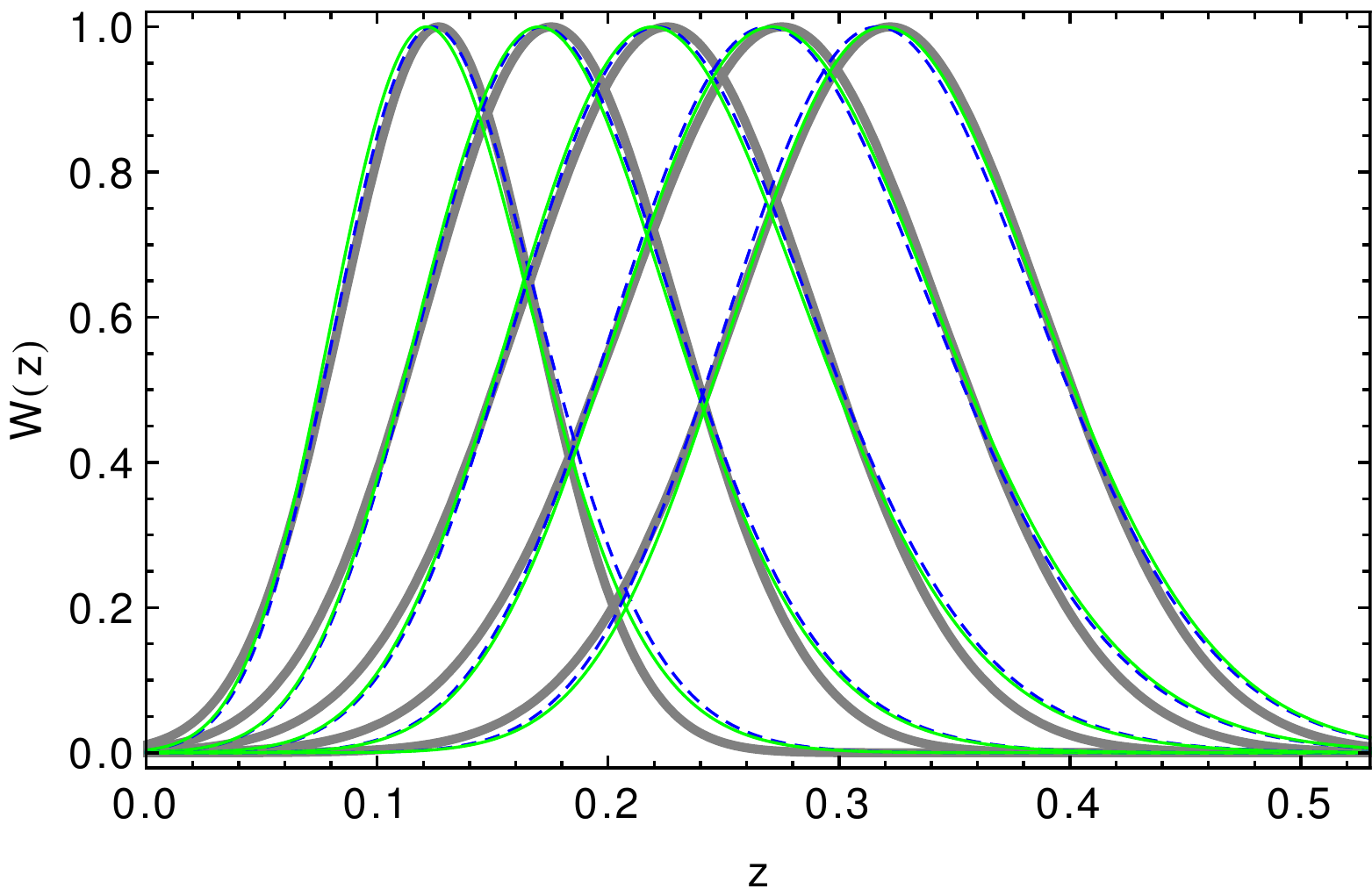}
\caption{Estimated true redshift normalized distribution of the galaxies inside 
  each one of the five photo-$z$ bins. The thin, dashed, blue curves represent the WSC-$clean$ sample; 
  the thin, solid, green lines represent the WSC-$svm$ sample; while the thick, gray curves 
  represent their Gaussian approximation (actually used to compute the $C_{\ell}^{ij}$s, see text for details). The difference 
  between the curves for the same photo-$z$ bin are much smaller than their widths, justifying 
  the approximation of both WSC-$clean$ and WSC-$svm$ by the Gaussian curves.}
 \label{fig:cl-window}
\end{figure}

The mock data used in this work were created with the {\sc flask} 
code\footnote{{\tt http://www.astro.iag.usp.br/\~{ }flask}} \citep{2016/xavier}.
{\sc Flask} takes as input a set of auto- and cross- angular power spectra $C_{\ell}^{ij}$s 
for the number counts in all photo-$z$ bins $i,j$ considered here and the radial selection function of the survey $n(z)$ (see Fig. \ref{fig:nz}), and creates correlated, Poisson-sampled lognormal realisations of the projected number counts in each bin (i.e. a set of correlated 2D {\sc healpix} maps). The input $C_{\ell}^{ij}$s were computed using {\sc camb sources}\footnote{{\tt http://camb.info/sources}} \citep{2011/challinor} assuming a flat $\mathrm{\Lambda CDM}$ cosmological model with minimal neutrino mass and cosmological parameters matching those measured by the Planck collaboration \citep{2016/plaI}. 
Since each bin of our data is selected in photo-$z$, the actual redshift window $W(z)$ it represents (for which the power spectra ought to be calculated) is not simply a rectangular function $\Pi(z)$ (representing the selected redshift range) multiplied by $n(z)$, but $\Pi(z)n(z)$ convolved with the 
probability density function (PDF) of the photo-$z$ error. 
In this work we assume this PDF is a Gaussian centered at the true redshift with standard deviation $\sigma_z$ given the equation
\begin{equation}
\sigma_z = (1+z)(0.011 + 0.185 z + 0.382 z^2 - 1.803 z^3),
\end{equation}
a phenomenological fit to the data of \citet{2016/bilicki}. The distribution resulting from 
the convolution was also approximated by a Gaussian. Fig. \ref{fig:cl-window} compares the 
window functions for the WSC-$clean$ and WSC-$svm$ samples justifying their Gaussian approximations.

The computed $C_{\ell}^{ij}$s include the effects of lensing, redshift space distortions and 
other linear terms \citep[see equation 30 of][for details]{2011/challinor}, as well as non-linear 
corrections given by {\sc halofit} \citep{2003/smith, 2012/takahashi}, and were re-scaled 
to match the observed variance in the counts in pixels (using $N_{\mathrm{side}}=128$) -- an 
effect similar to galaxy bias. The $n(z)$ for each sample was measured directly from the 
data's photo-$z$ histogram, assuming the selection function is separable in radial and 
angular parts (the latter taken to be isotropic outside the mask), and thus our simulations 
also reproduce the observations' mean number counts in pixels.

\begin{figure}
 \includegraphics[width=1.0\columnwidth]{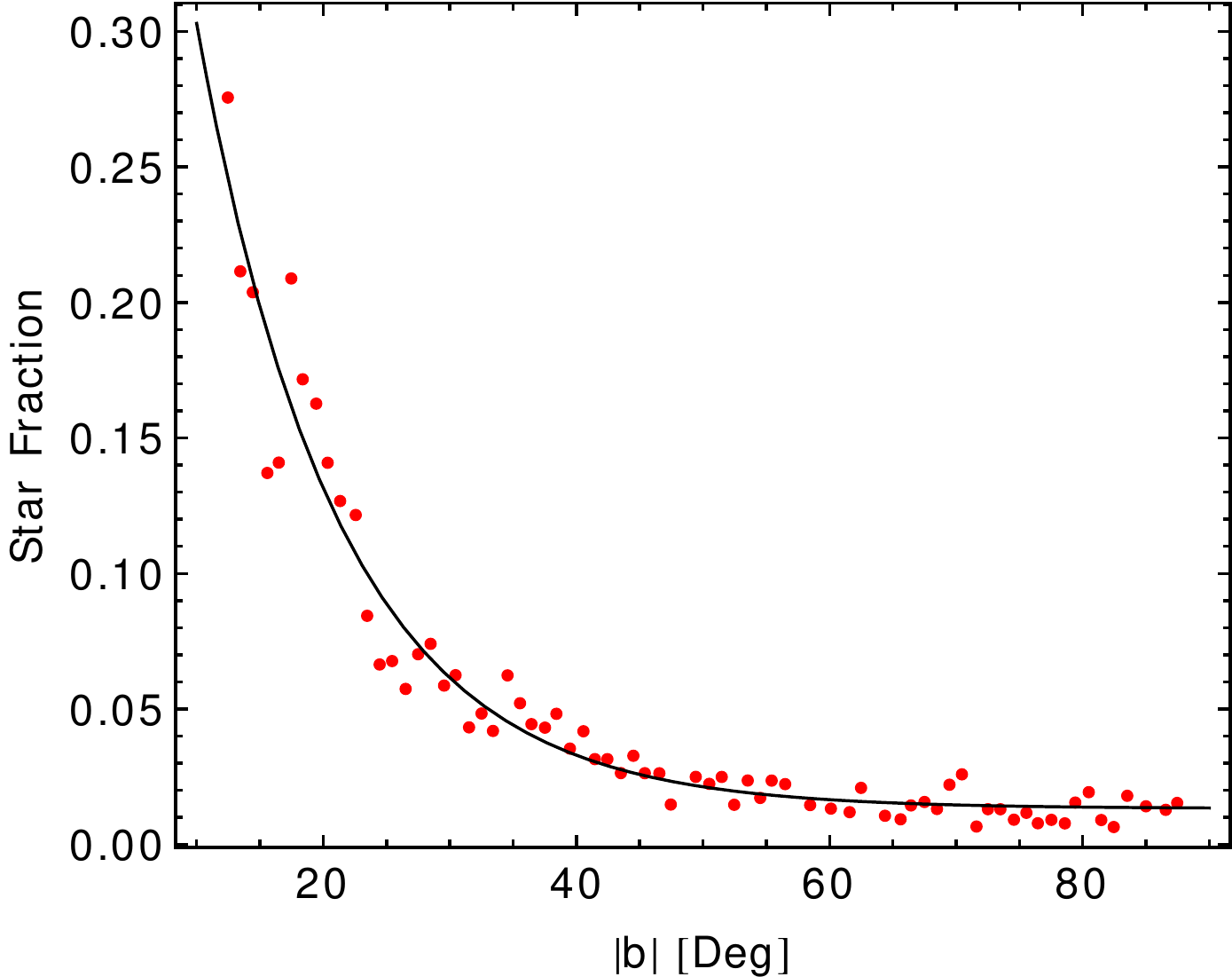}
\caption{Expected stellar contamination in the WSC-$clean$ sample. The red dots are the fraction of sources in the 
 $\delta=30^{\mathrm{\circ}}$ strip classified as stars by SDSS, and the black curve is our exponential fit.}
 \label{fig:star-contam}
\end{figure}

\begin{figure*}
\includegraphics[width=0.49\textwidth]{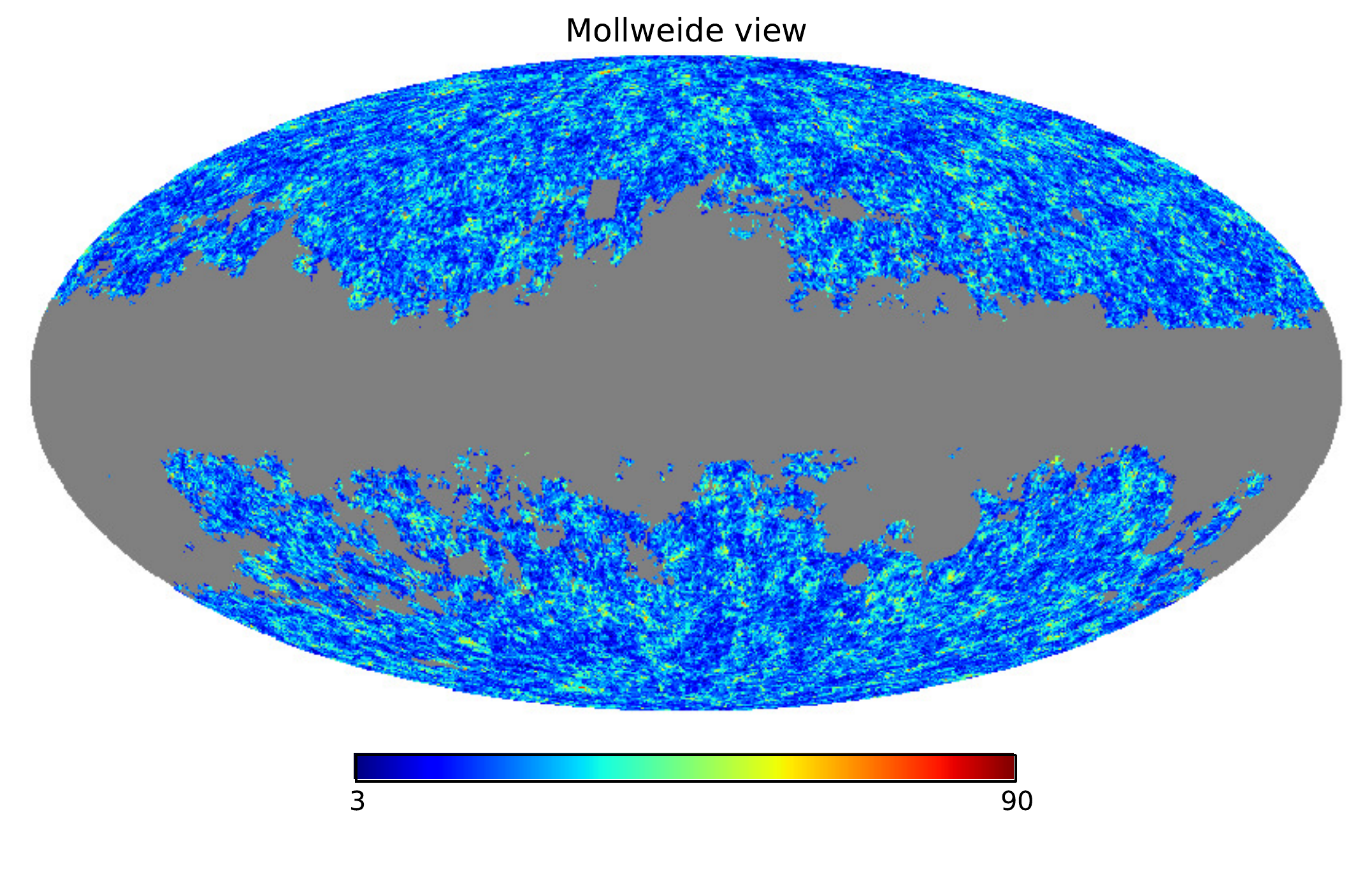}
\includegraphics[width=0.49\textwidth]{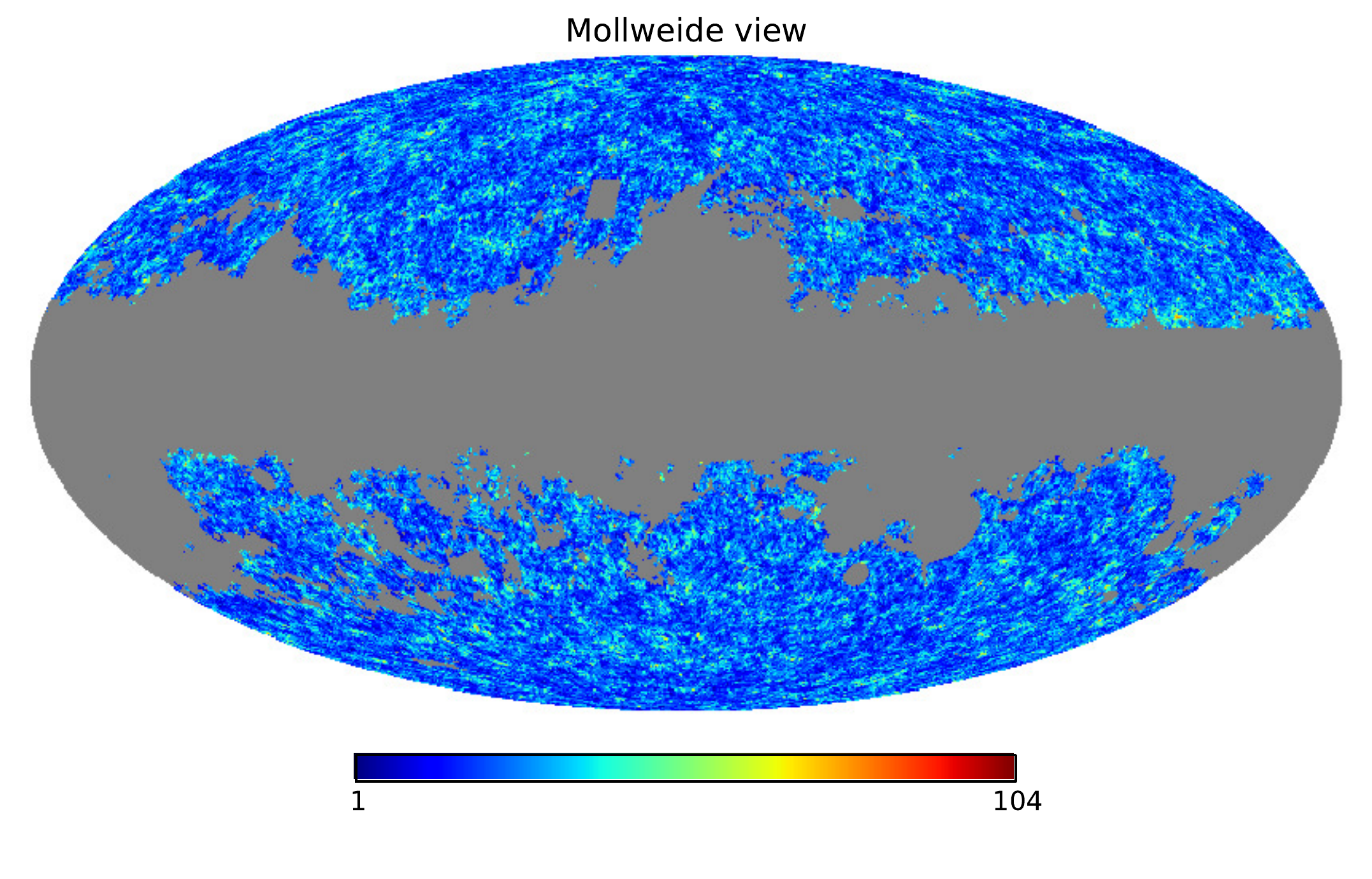}
\caption{Simulated GNC maps in the photo-$z$ range 0.20--0.25 of the WSC-$clean$ sample, without (left panel) and with (right panel) stellar contamination. Their means and variances are the same, however it is possible to see a subtle increase in counts on the region closest to the Galactic plane in the contaminated map.}
\label{fig:sim-maps}
\end{figure*}

We have also created one set of simulated maps that includes contamination by stars. By comparing 
the WSC sources selected as `galaxies' with the SDSS photometric classification (expected to 
be more accurate due to a larger colour space and better resolution) on a $1^{\mathrm{\circ}}$-wide 
strip centered on declination $\delta=30^{\mathrm{\circ}}$, \citet{2016/bilicki} estimated the 
remaining contamination in the WSC-$clean$ sample as a function of Galactic latitude $b$. As 
Fig. \ref{fig:star-contam} shows, the fraction $f_{\mathrm{star}}$ of WSC-$clean$ sources classified as stars by SDSS is well fitted by $f_{\mathrm{star}} = 0.013 + 0.71\exp{(-0.09|b|)}$. 
We assume the contamination fraction to be the same in all redshift bins [a hypothesis that \citet{2016/bilicki} verified to be reasonable, with a possible exception for the 0.30--0.35 range, likely more contaminated] and used this fit to compute the expected star counts in each map, such that the mean number counts in the simulations (galaxies plus stars in the unmasked region) equals the mean number counts in the observations. 
For each realisation, we Poisson-sampled the stars according to their expected number; the 
$C_{\ell}^{ij}$s were re-scaled so the simulations with extra noise from the stars match the 
variance observed in the data. An example of mock maps with and without contamination is presented in 
Fig. \ref{fig:sim-maps}.

\section{The Minkowski Functionals} \label{sec:section3}

Unlike the power spectrum, the MF can also provide morphological information and map the shape of structures, besides informing about spatial correlation of a random field.
The morphological properties of a given random field $\mathcal{F}$ in a $d$-dimensional space can be completely described using $d+1$ MF~\citep{1903/minkowski}. 
In addition, the MF are the unique morphological descriptors obeying motion invariance, conditional continuity and additivity \citep{1957/hadwiger}. 
It means that any other functional respecting these properties can be expressed as a linear combination of the others $d+1$.
Then, for a 2D GNC field, defined as a projection on the sphere ${\cal S}^2$ of the sources in a given redshift range, 
${\cal N} = {\cal N}(\theta,\phi)$, with mean $\langle{\cal N}\rangle$ and variance $\sigma_0^2$, 
three MF would be calculated.

Given a sky patch ${\cal P}$ of the pixelized GNC sphere ${\cal S}^2$, an {\em excursion set} 
is defined as the set of pixels in ${\cal P}$ where the GNC field exceeds 
the \textit{threshold} $\nu_t$, that is, it is the set of pixels with coordinates $(\theta,\phi) 
\in {\cal P}$ such that $\delta{\cal N}(\theta,\phi) / \sigma_0 \equiv \nu > \nu_t$, with 
$\delta{\cal N} \equiv {\cal N}(\theta, \phi) - \langle{\cal N}\rangle$. 
Each excursion set $\Sigma$ and its boundary, $\partial\Sigma$, can be defined as 
\begin{eqnarray}
\Sigma & \equiv & \{(\theta, \phi) \in \mathcal{P} ~|~ \delta{\cal N} (\theta, \phi) > \nu_t	\sigma_0 \}, \\
\partial\Sigma & \equiv & \{(\theta, \phi) \in \mathcal{P} ~|~ \delta{\cal N} (\theta, \phi) = \nu_t \sigma_0 \}.
\end{eqnarray}
In a 2D case, for an excursion $\Sigma \subset {\cal S}^2$, the partial MF calculated in a 
connected\footnote{A connected region is constituted by the ensemble of pixels, with values $\nu > \nu_t$, that have at least a common vertex, i.e., they are a subset of the excursion \citep[see][for details]{2013/ducout}.} 
region $R_i$ of $\Sigma$ are: $a_i$, the Area of the connected region; $l_i$, the Perimeter or contour length of this region; and $n_i$, the number of holes inside it. 
The global MF are obtained calculating these quantities for all the connected regions in $\Sigma$, the set of pixels with $\nu > \nu_t$. 
Then, the total Area $V_0(\nu_t)$, Perimeter $V_1(\nu_t)$ and Genus $V_2(\nu_t)$ 
are \citep{1999/novikov, 2003/komatsu, 2006/naselsky_book, 2013/ducout}:

\begin{align}
\! V_0(\nu_t) &= \frac{1}{4 \pi} \int_{\Sigma} d\Omega = \sum_{i} a_i \, ,  \label{eq:v0}\\
\! V_1(\nu_t) &= \frac{1}{4 \pi} \frac{1}{4} \int_{\partial\Sigma} dl = \sum_{i} l_i \, ,  \label{eq:v1}\\
\! V_2(\nu_t) &= \frac{1}{4 \pi} \frac{1}{2 \pi} \int_{\partial\Sigma} \kappa~dl = \sum_{i} (1-n_i) =N_{\mathrm{over}} - N_{\mathrm{under}}  \label{eq:v2}\, ,
\end{align}

\noindent
where $d \Omega$ and $dl$ are, respectively, the elements of solid angle and line. 
In the Genus definition, the quantity $\kappa$ is the geodesic curvature \citep[for details see, 
e.g.,][]{2013/ducout}. 
This last MF can also be calculated as the dif\/ference between the number of regions with 
$\nu > \nu_t$ (number of connected over-dense areas, $N_{\mathrm{over}}$) and regions with $\nu < \nu_t$ (number of connected under-dense areas, or number of holes, $N_{\mathrm{under}}$) in the excursion. 
The MF used here were calculated using the algorithm developed by \cite{2013/ducout} and \cite{2012/gay}.

\begin{figure*}
\includegraphics[width=0.247\textwidth]{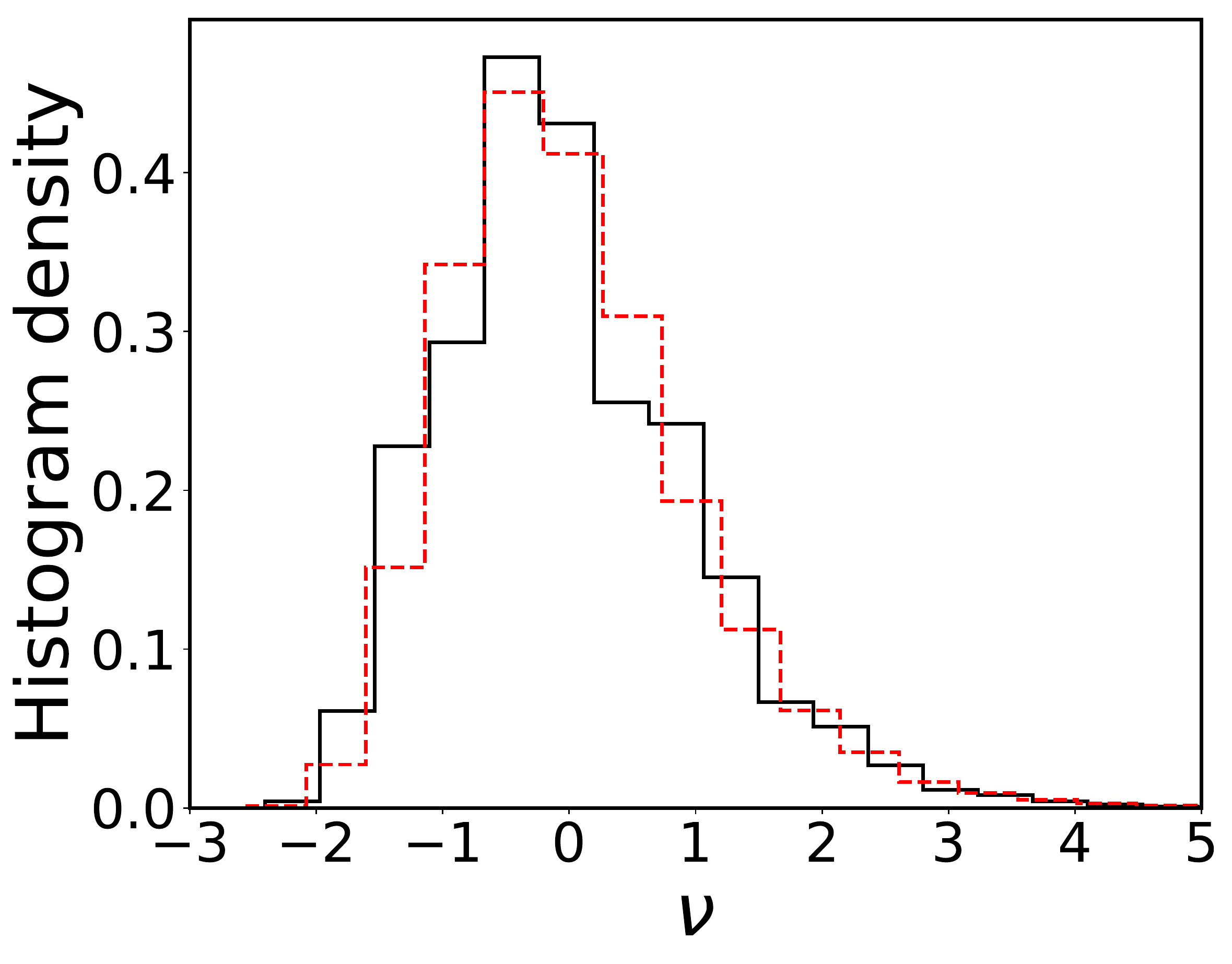}
\hspace{-0.1cm}
\includegraphics[width=0.243\textwidth]{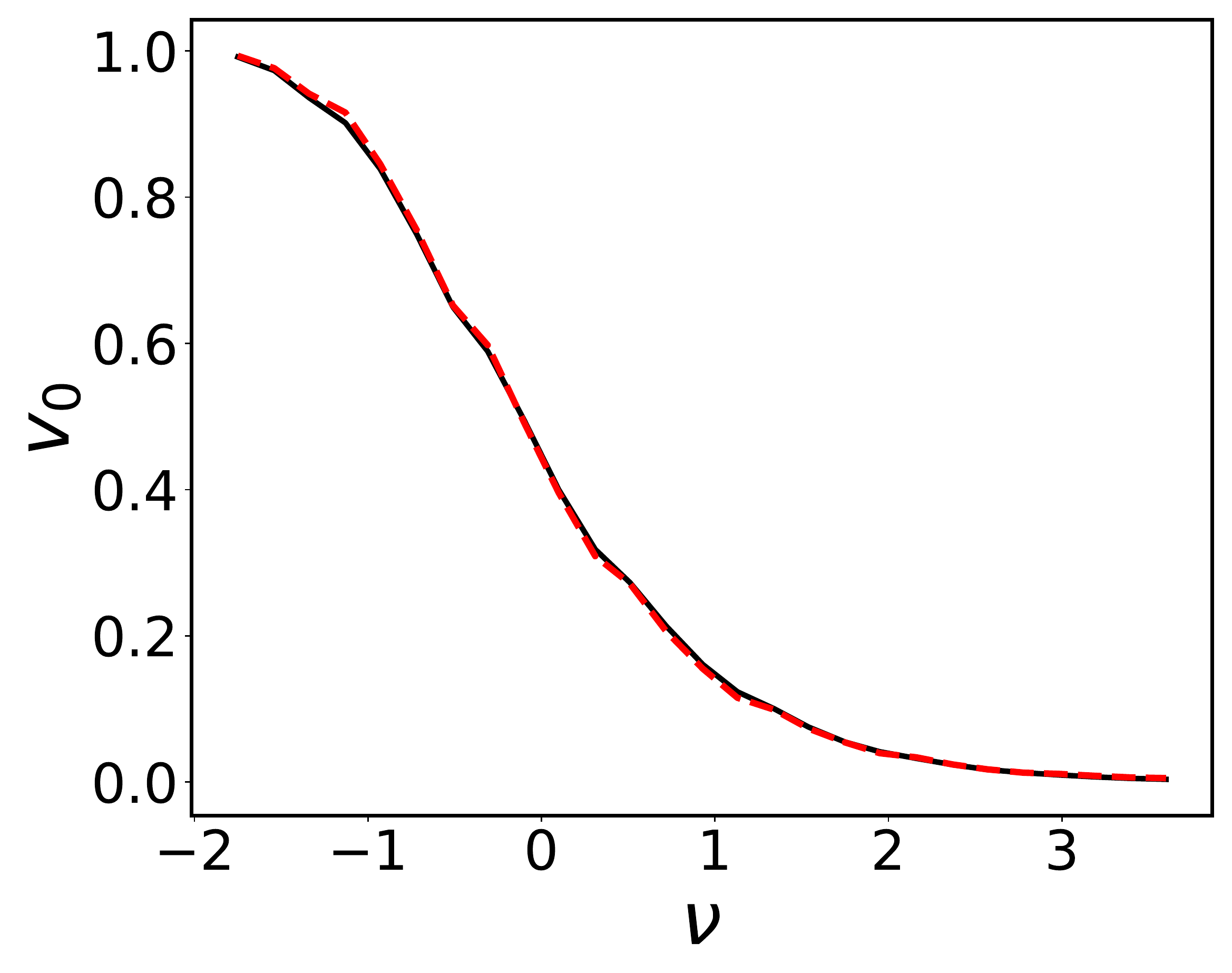}
\hspace{-0.1cm}
\includegraphics[width=0.25\textwidth]{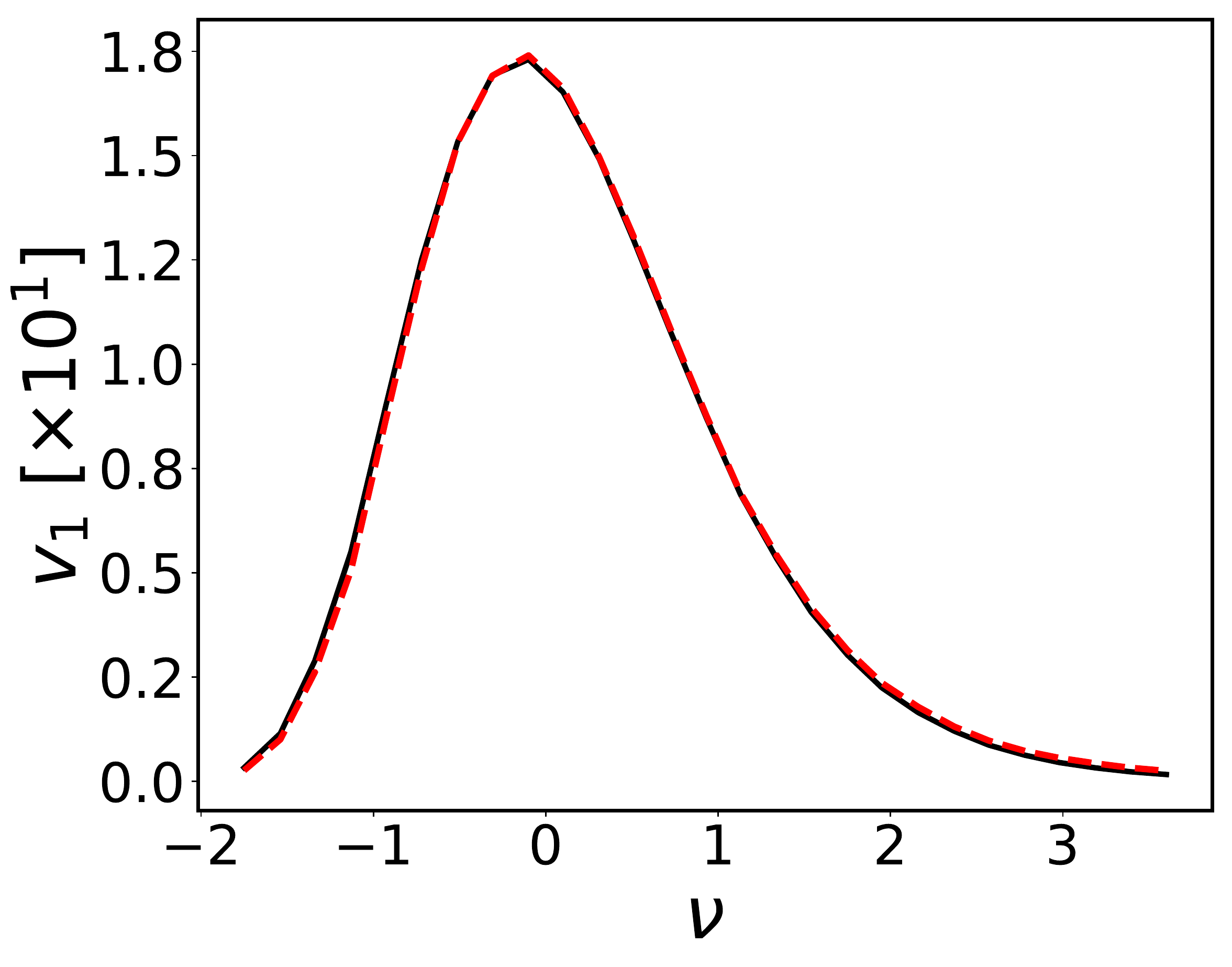}
\hspace{-0.15cm}
\includegraphics[width=0.25\textwidth]{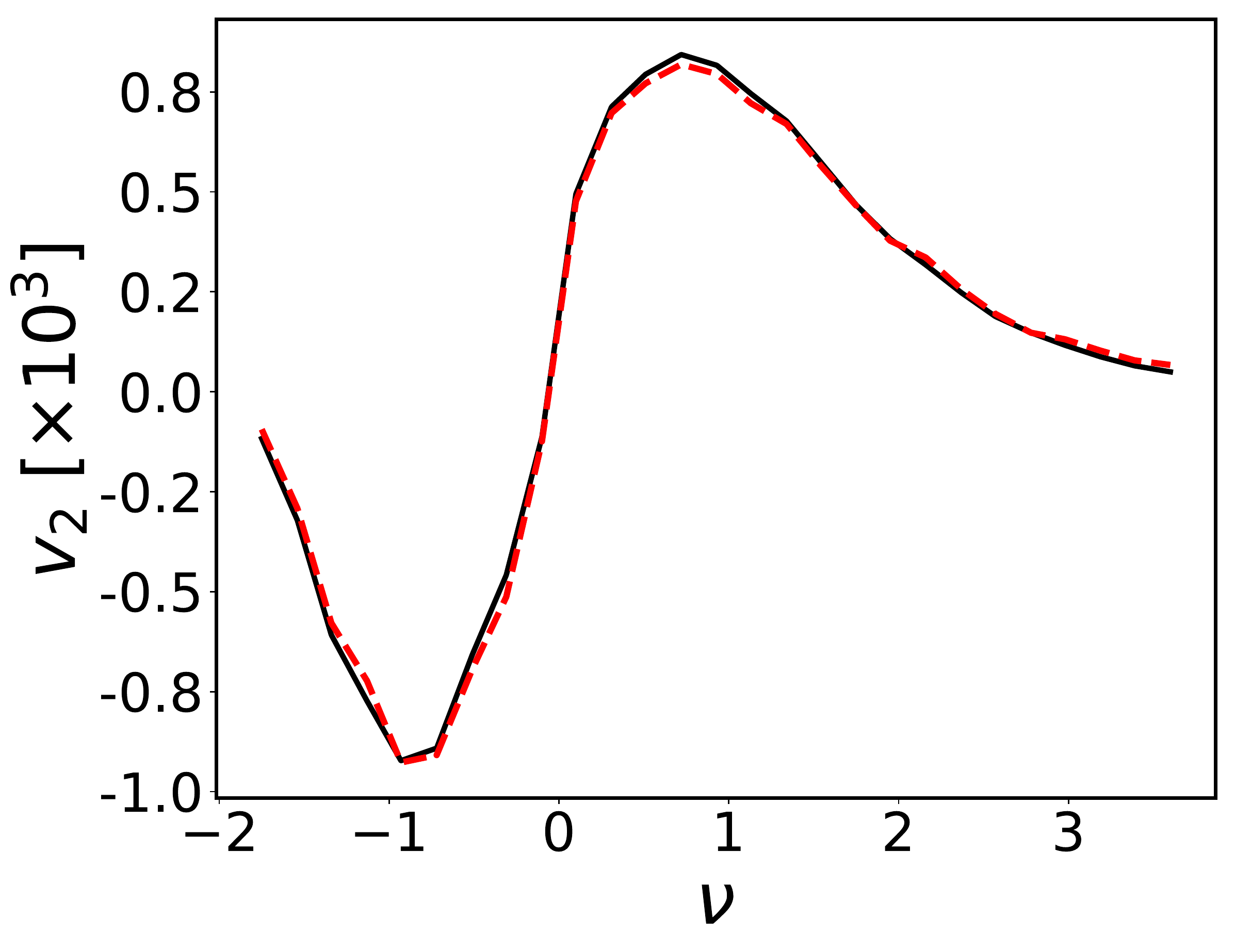}
\caption{The first panel shows the PDF distribution of the masked GNC map from the WSC-$clean$ sample at the photo-$z$ bin $0.15 < z < 0.20$ (red dashed line) and of one corresponding mock realisation (black solid line). 
The second to fourth panels present, respectively, the Area, Perimeter, and Genus curves (red dashed line) calculated from the same WSC-$clean$ GNC map. 
For comparison, it is also included, as black solid lines, the average MF over the corresponding 5000 mock realisations. 
All panels are presented in terms of the $\nu$ quantity, i.e.,  $\nu(x) = \delta{\cal N} / \sigma_0$.
Note, in all the graphs, the elongated tail for positive $\nu$ values resulting from the lognormal distribution of galaxies. }
\label{fig:ex_MFs}
\end{figure*}

\section{The local analyses}
\label{sec:section4}

All the GNC fields constructed from the two WSC galaxy samples, the corresponding lognormal realisations, and the cut-sky mask are constructed considering the resolution parameter ${\rm \,N}_{\mbox{\small side}} = 128$. 
The local analyses of these maps are performed by calculating the MF for individual sky patches of a GNC map.
These patches are defined as the pixels corresponding to a resolution parameter ${\rm \,N}_{\mbox{\small side}} = 4$, that is, the sky is divided into 192 pixels of equal area $\simeq {215\,\mathrm{deg^2}}$. 
Each of these large pixels (which we call {\em patches}) contains 1,024 small data pixels corresponding to the resolution of ${\rm \,N}_{\mbox{\small side}} = 128$. 
Notice that, since we use a mask to exclude possible contaminations, the total of 192 patches is reduced to 126 to be analysed. 
The cut-sky mask also makes the number of valid pixels in each patch to vary from one patch to another, especially near the Galactic plane. 

Considering $n$ dif\/ferent thresholds $\nu_t = \nu_{_{1}}, \nu_{_{2}}, ...\, \nu_n$, previously defined dividing the range $\nu_{min}$ to $\nu_{max}$ in $n$ equal parts, we compute the three MF $V_k, \, k=0,1,2 $, that is, the Area, Perimeter, and Genus, for every patch of a GNC map.
Then, for the $p$-th patch and for each $k$ we have the vector 
\begin{eqnarray} \label{def_v}
\mathrm{v}_k^p \equiv ( V_k(\nu_{_{1}}),V_k(\nu_{_{2}}), ... V_k(\nu_n) )|_{\mbox{\small for the $p$-th patch}} \, ,
\end{eqnarray}
for $p = 1, 2,..., 126$.
The values chosen for such variables are $[\nu_{min},\nu_{max}] = [-1.75,3.6]$ and $n = 27$. 
Note that the choice of the threshold range depends on the dispersion of the PDF distribution of the  $\delta{\cal N}$ field, which obeys a lognormal distribution, requiring a non symmetric $\nu_t$ range. 
This is shown in the example of Fig. \ref{fig:ex_MFs}, which presents the PDF distribution of the GNC map (masked sky) constructed from the WSC-$clean$ sample at the photo-$z$ bin 2, i.e., $0.15 < z < 0.20$, and the corresponding MF vectors, $\mathrm{v}_k$ for $k$ = 0, 1, and 2. 
Moreover, since the $\nu$ bin width directly influence the correlation among two consecutive thresholds, the number $n$ has to be carefully chosen, as discussed bellow \citep[for details, see][]{2013/ducout}. 
One can also see in Fig. \ref{fig:ex_MFs} the comparison of these observed MF with the mean obtained from the 5000 mock realisations of the WSC-$clean$ sample, showing the good agreement among them.

The signature contained in each sky patch, for each photo-$z$ bin and sample, is revealed by the MF when compared to the ones obtained from the corresponding mock realisation set through a $\chi^2$ analysis.
For that we define a single 81-elements vector by combining Area, Perimeter and Genus information into a joint estimator 
\begin{eqnarray}\label{eq7}
\mathcal{V}^p & \equiv & ( \mathrm{v}_0^p , \mathrm{v}_1^p , \mathrm{v}_2^p ) \nonumber \\
                       &=& (V_0^p(\nu_1),...,V_0^p(\nu_{27}),V_1^p(\nu_1),...,V_1^p(\nu_{27}), \nonumber \\
                       & & V_2^p(\nu_1),...,V_2^p(\nu_{27})),
\end{eqnarray}
calculated for each patch $p = 1, 2,..., 126$. 
To avoid excess of indices, in what follows we do not explicitly write the super-index $p$ when referring to the quantity $\mathcal{V}$, but one understands that it is being calculated for each patch $p$ of a GNC map (from data or mock sets).

Firstly, we use the joint estimator to calculate, for each patch and photo-$z$ bin, the mean vector $\langle \mathcal{V}^{\mathrm{Mock}} \rangle$ for the set of the GNC mock realisations of each sample, i.e., 5000 of WSC-$clean$ and 5000 of WSC-$svm$. 
Then, we calculate $\mathcal{V}^{\mathrm{WSC}}$ for each photo-$z$ bin and WSC sample and compare it with the corresponding mean vector $\langle \mathcal{V}^{\mathrm{Mock}} \rangle$, which contains the features expected in the simulations. 
This is done performing a $\chi^2$ analysis for each patch which takes into account the correlations among the different MF and thresholds \citep{2003/komatsu,2006/hikage,2013/ducout,2016/novaes}:
\begin{equation} \label{eq:chi2}
\chi^2 \equiv \sum_{i=1}^{81} \sum_{j=1}^{81} [ \mathcal{V}_i^{\mathrm{WSC}} - \langle \mathcal{V}_i^{\mathrm{Mock}} \rangle ] 
\,{\sf C}_{i,j}^{-1}\, [ \mathcal{V}_j^{\mathrm{WSC}} - \langle \mathcal{V}_j^{\mathrm{Mock}} \rangle ],
\end{equation}
where the $i$ and $j$ indices run over all 81 combinations of the MF and thresholds $\nu_t$ ($27$ for each of the three MF), and ${\sf C}_{i,j}^{-1}$ is the inverse of the full covariance matrix, ${\sf C}_{i,j}$, which is calculated from the mock realisations as
\begin{equation} \label{eq:Cij}
{\sf C}_{i,j} \equiv \langle \,( \mathcal{V}_i^{\mathrm{Mock}} - \langle \mathcal{V}_i^{\mathrm{Mock}} \rangle )\,( \mathcal{V}_j^{\mathrm{Mock}} - \langle \mathcal{V}_j^{\mathrm{Mock}} \rangle ) \,\rangle.
\end{equation}
Note that the number of thresholds and the quantity of GNC mocks directly influence the accuracy of ${\sf C}_{i,j}$, and, consequently, the calculation of its inverse\footnote{Notice that, although equation \ref{eq:Cij} is an unbiased estimator of the covariance matrix, its inverse is not an unbiased estimator of $C^{-1}_{i,j}$. However, in our case the multiplicative correction factor needed to make the estimator for $C^{-1}_{i,j}$ unbiased is negligible, with $(n_t - m - 2)/(n_t - 1) = 0.98$, for a total of $n_t = 81$ thresholds and $m = 5000$ mock realisations \citep[for details, see][]{2007/hartlap}.}. 
Following \cite{2013/ducout}, we verified that for $n = 27$ the amount of 5000 mocks is enough for a sufficiently accurate calculation of ${\sf C}_{i,j}^{-1}$.

\begin{figure*}
\centering
 \includegraphics[width=0.33\textwidth]{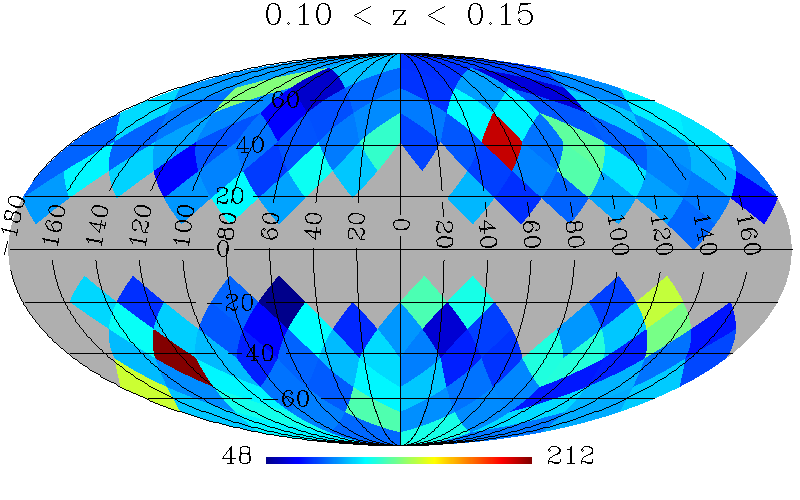}
 \includegraphics[width=0.33\textwidth]{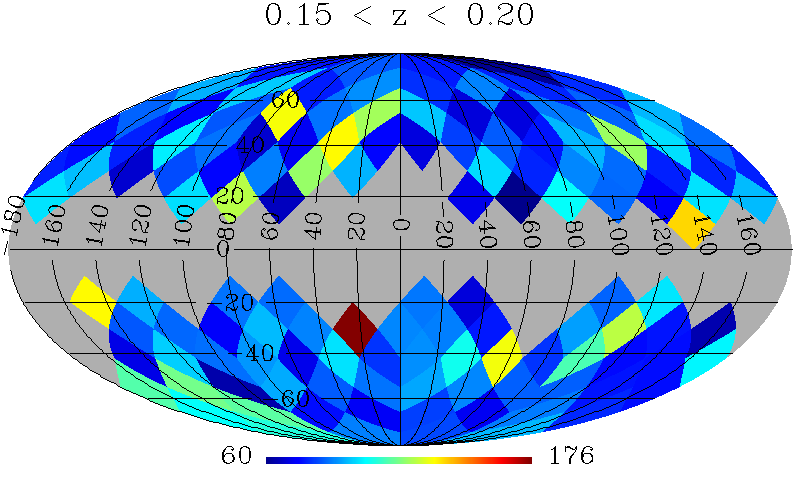}
 \includegraphics[width=0.33\textwidth]{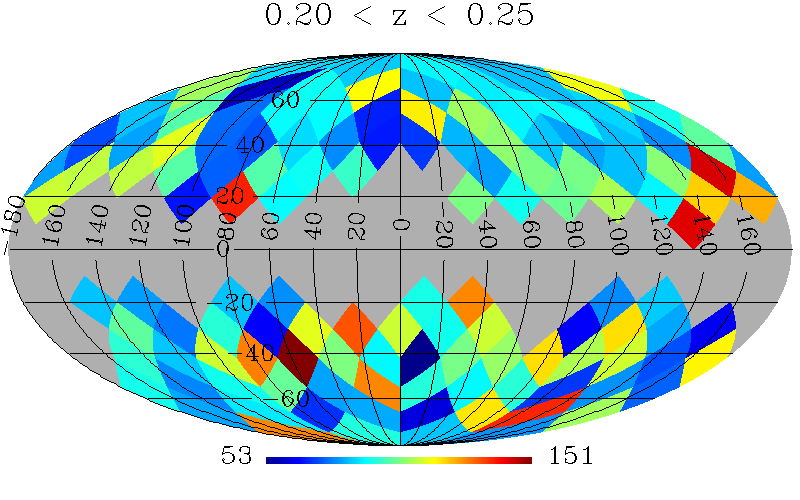}\\
 \includegraphics[width=0.33\textwidth]{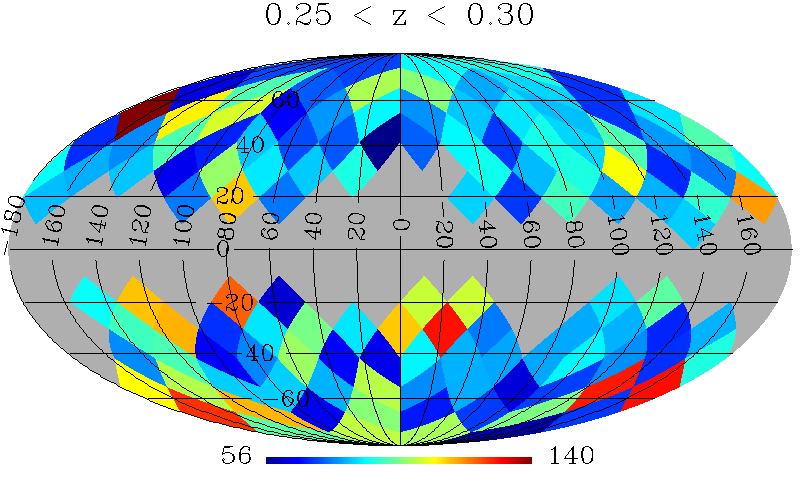}
 \includegraphics[width=0.33\textwidth]{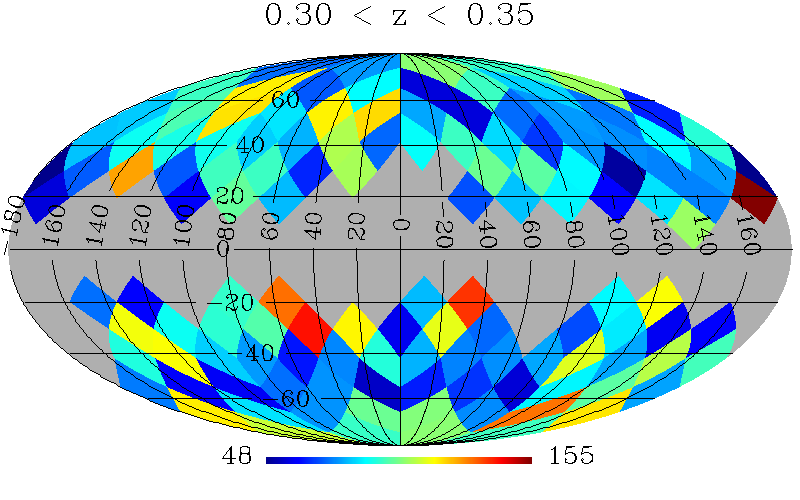}
\caption{$\chi^2$-maps resulting from the local analyses of the GNC maps constructed from the WSC-$clean$ sample. 
From left-to-right and top-to-bottom are the results for the photo-$z$ bins from 1 to 5 (see Table \ref{tab:redshif_bins}), respectively. See text for details.}
\label{fig:sigmaMapsClean}
\end{figure*}

\begin{figure*}
\centering
 \includegraphics[width=0.33\textwidth]{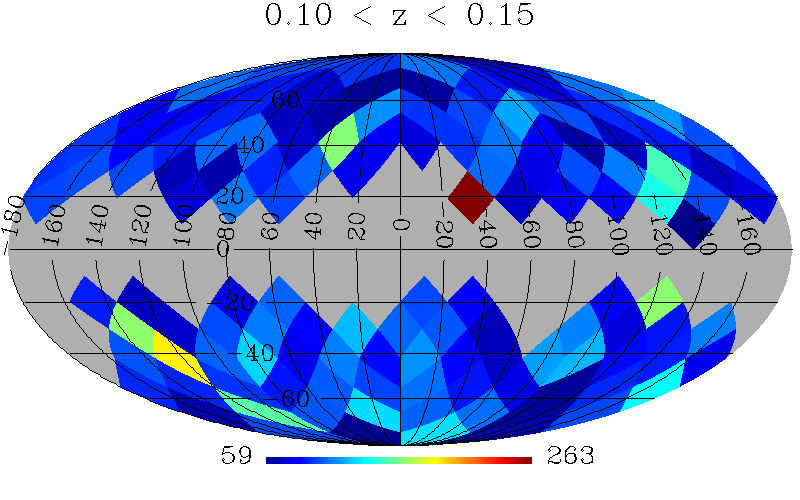}
 \includegraphics[width=0.33\textwidth]{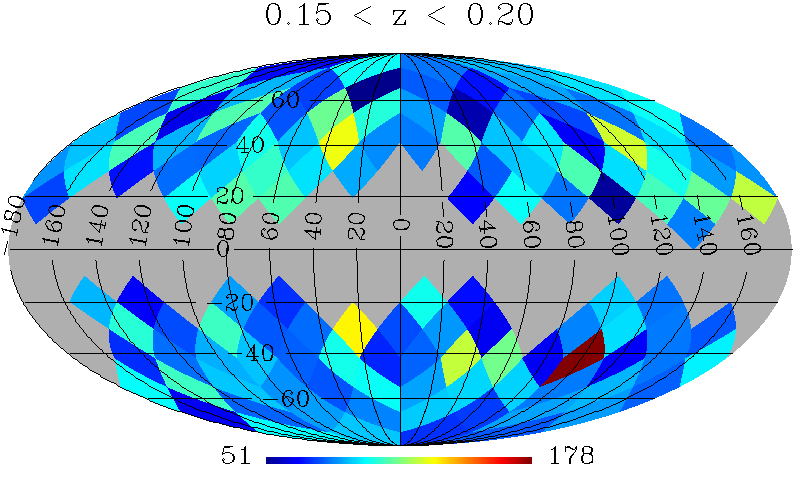}
 \includegraphics[width=0.33\textwidth]{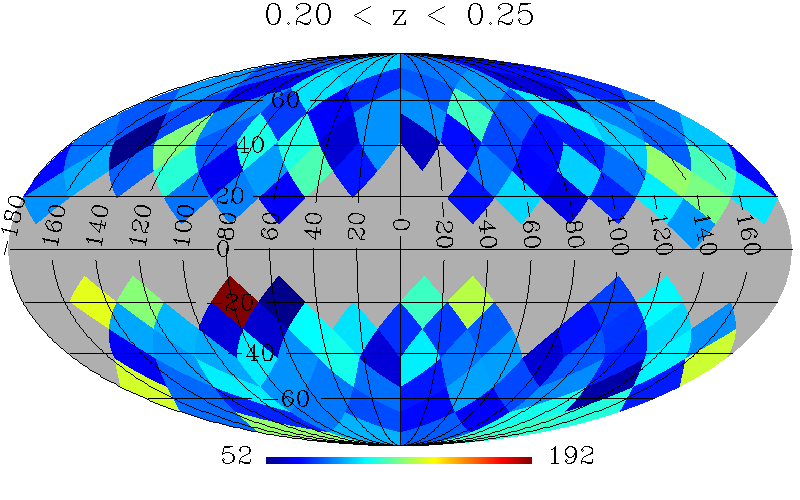}\\
 \includegraphics[width=0.33\textwidth]{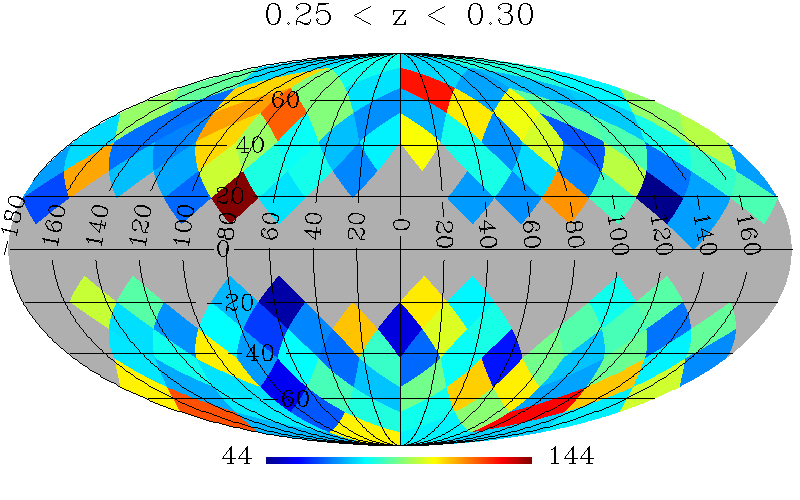}
 \includegraphics[width=0.33\textwidth]{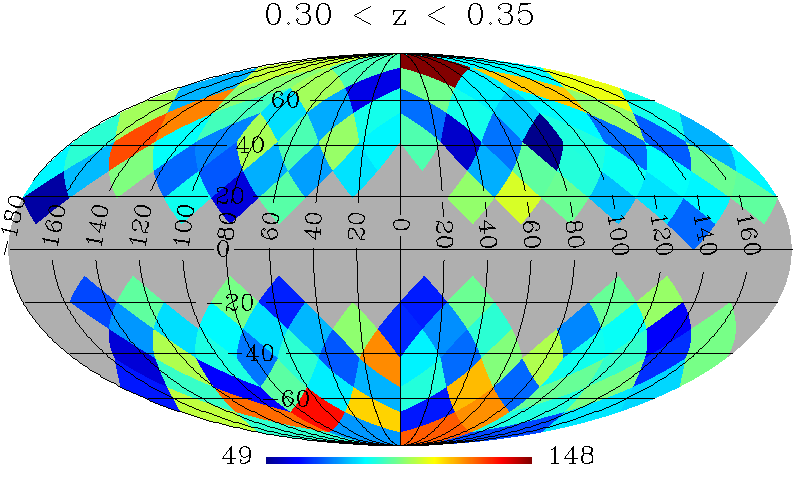}
\caption{Analogous to Fig. \ref{fig:sigmaMapsClean}, but for the WSC-$svm$ sample.}
\label{fig:sigmaMapsSVM}
\end{figure*}

Using equation \ref{eq:chi2}, we calculated the $\chi^2$ value for each patch and each photo-$z$ bin of the WSC-$clean$ and WSC-$svm$ GNC samples, obtaining the corresponding $\chi^2$-maps, presented in Figs \ref{fig:sigmaMapsClean} and \ref{fig:sigmaMapsSVM}. The $\chi^2$ values follow a colour scale and tell us how well the patches are described by the simulations; some $\chi^2$ values are very large, indicating a substantial discrepancy there. 
However, the $\mathcal{V}_i$ distribution is non-Gaussian, and this might make high $\chi^2$ values more likely than one would naively expect. 
Therefore, to properly identify true anomalous structures very unlikely to be found in $\mathrm{\Lambda CDM}$ universes with lognormal galaxy densities, we analyse in the next section how unexpected are the observed $\chi^2$ values in comparison to the $\chi^2$ distribution estimated from the simulations. 
We also examine in detail the features of the MF curves from these patches in such a way to investigate the possible reason for their disagreement with simulations.

\section{Identification and analyses of extreme regions} \label{sec:section5}

\begin{figure}
\includegraphics[width=1\columnwidth]{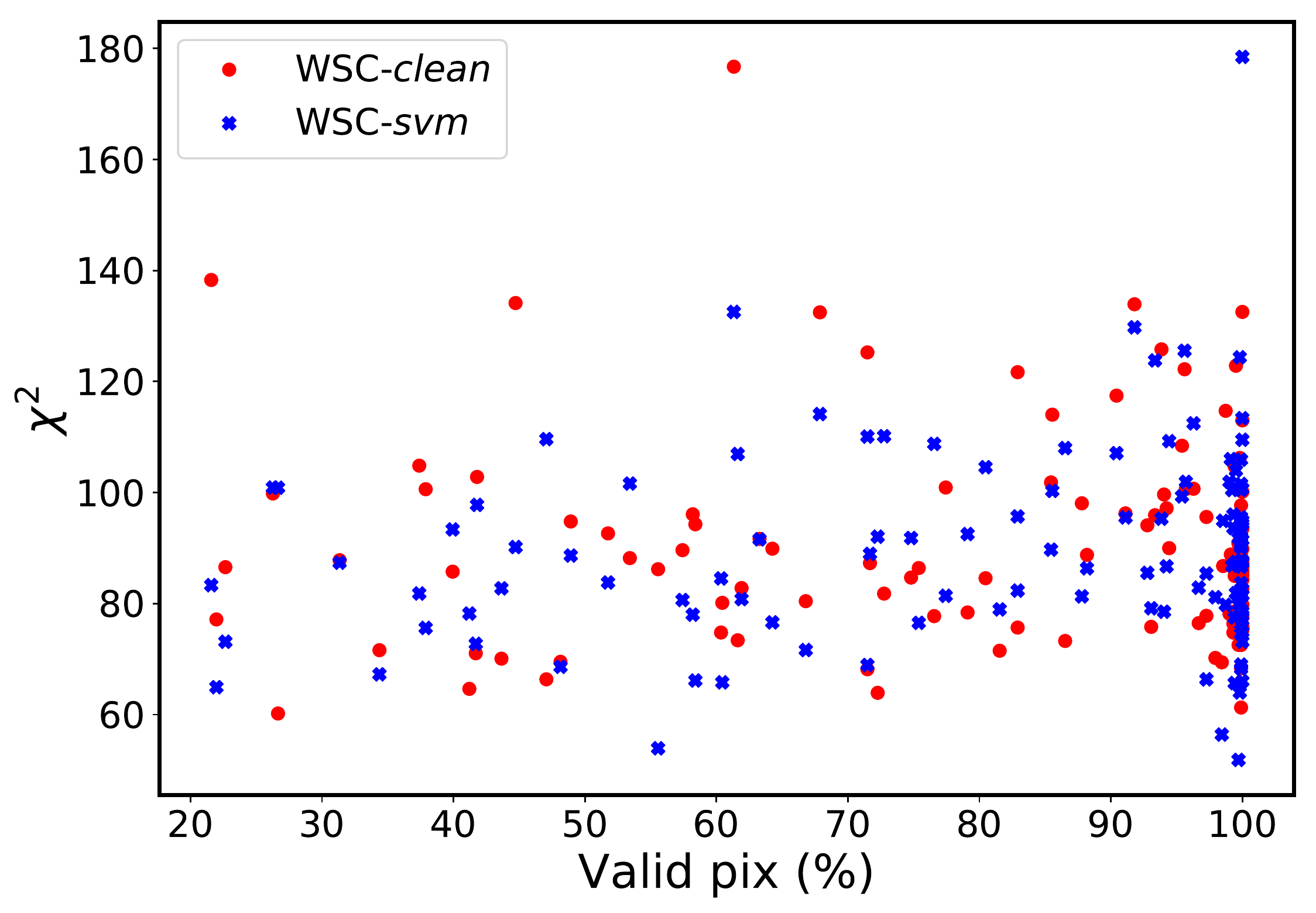}
\caption{Dependence of the $\chi^2$ estimates resulting from the analyses of the WSC-$clean$ and WSC-$svm$ samples with the fraction of valid pixels composing the patches. 
This graphic is derived by analysing photo-$z$ bin 2, that is, $0.15 < z < 0.20$, but similar results are obtained from the other bins.}
 \label{fig:sigmaXvalidpix}
\end{figure}

From the $\chi^2$-maps summarized in Figs \ref{fig:sigmaMapsClean} and \ref{fig:sigmaMapsSVM}, a first -- and necessary -- consistency check is to confirm that the different number of valid pixels composing each patch, caused by the application of a cut-sky mask upon all GNC maps, is not influencing our results concerning the $\chi^2$ values. 
For this we show in Fig. \ref{fig:sigmaXvalidpix} an example of the relation between the $\chi^2$ value and the percentage of valid pixels in each patch. 
The plot corresponds to the range $0.15 < z < 0.20$ for both WSC samples, but a very similar behaviour is observed in the other photo-$z$ bins, that is, it does not seem to exist an evident dependence between these two quantities.

Once we discarded the possibility of the mask to be influencing our statistics, we must evaluate the probability of finding such $\chi^2$ amplitudes in the mock realisations, in special those patches whose $\chi^2$ values are seemingly far above the mean (redder regions). 
For this, we estimate the frequency of occurrences of the observed $\chi^2$ amplitudes in the mock realisations by constructing $\chi^2$-maps for each of the 5000 simulations of each sample and photo-$z$ bin. 
In other words, we repeat the procedure of constructing a $\chi^2$-map using equation \ref{eq:chi2} but we replace the quantity $\mathcal{V}^{\mathrm{WSC}}$ by the one obtained from each mock realisation.
From these $\chi^2$-maps we estimate the $p$-value for each of the patches by counting the number of occurrences of $\chi^2$ values higher than those from Figs \ref{fig:sigmaMapsClean} and \ref{fig:sigmaMapsSVM}. 

\begin{figure}
\centering
\includegraphics[width=1\columnwidth]{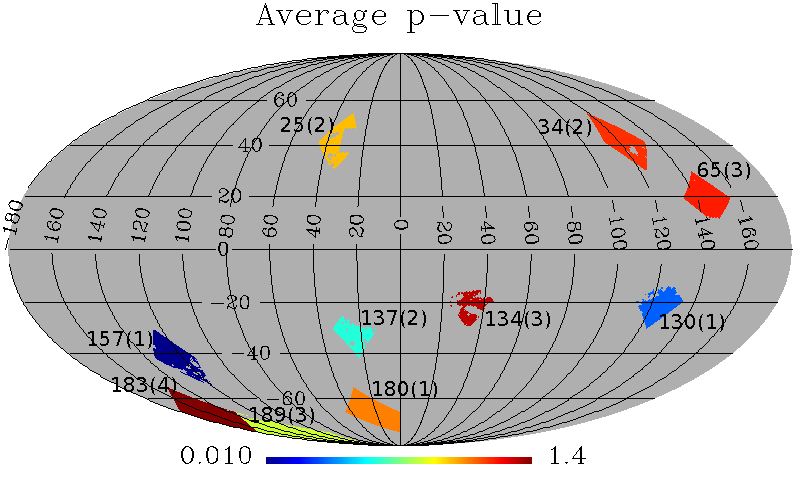}
\caption{Mollweide projection highlighting the extreme patches selected according to their $p$-value amplitudes ($< 1.4\%$ from the two samples simultaneously) estimated locally from the GNC maps constructed from the WSC samples summarized in Table \ref{tab:redshif_bins}. The color scale corresponds to the average $p$-value over the two samples.
The number $p$ of the patches and the photo-$z$ bin in which they were identified, $b$, are also indicated in the panels, referred to as $p$($b$) [for example, the patch no. 130 identified at the photo-$z$ bin 1 is indicated as 130(1)]. 
}
\label{fig:selectP}
\end{figure}

\begin{figure*}
\centering
\includegraphics[width=0.2\textwidth]{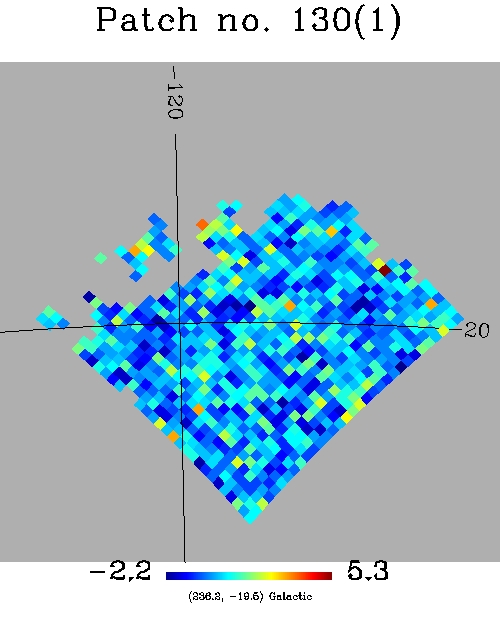}
\includegraphics[width=0.2\textwidth]{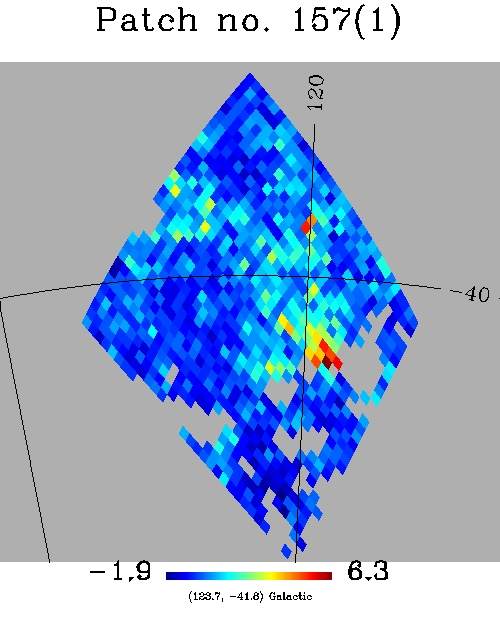}
\includegraphics[width=0.2\textwidth]{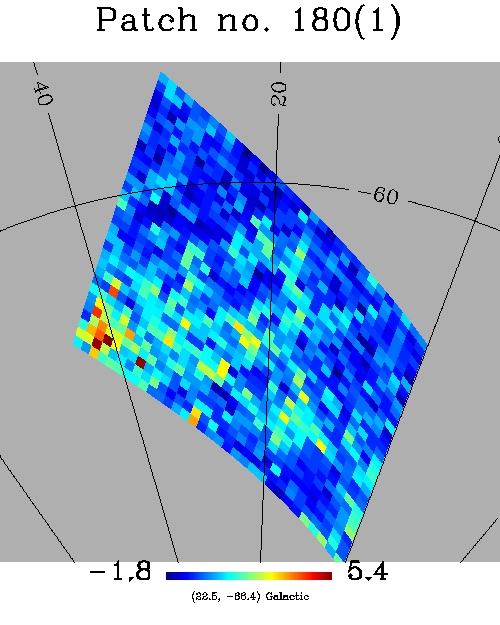}\\
\includegraphics[width=0.2\textwidth]{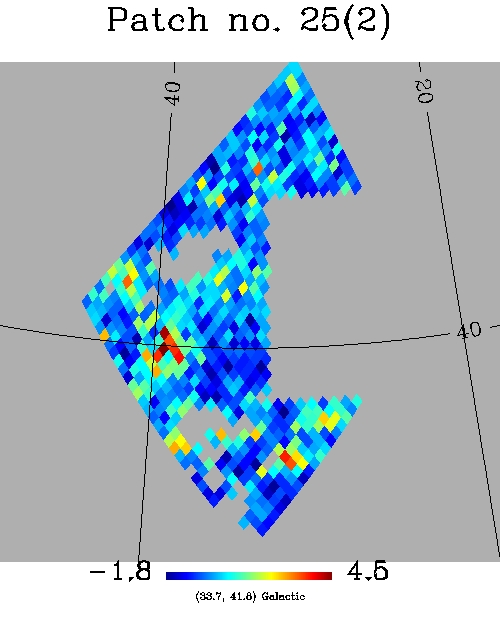}
\includegraphics[width=0.2\textwidth]{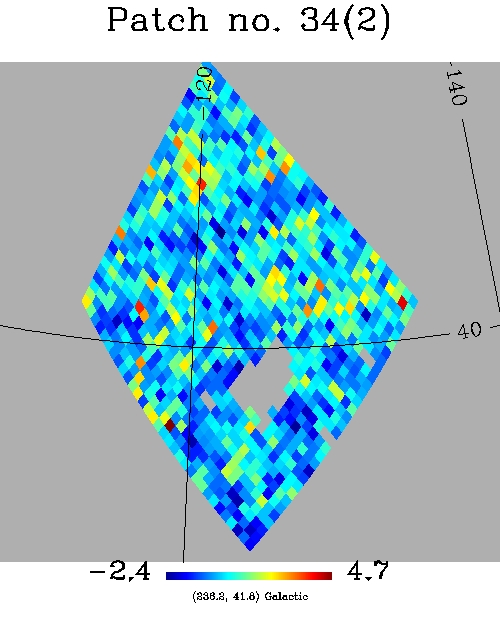}
\includegraphics[width=0.2\textwidth]{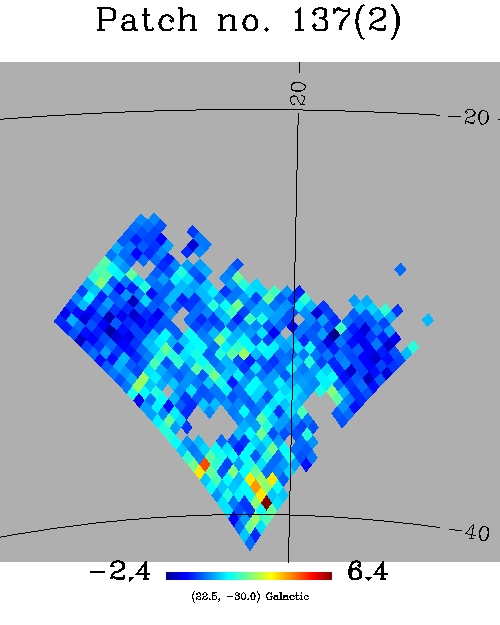}\\
\includegraphics[width=0.2\textwidth]{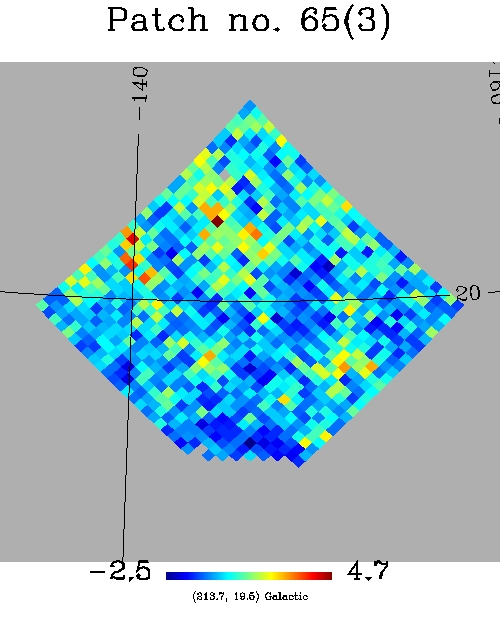}
\includegraphics[width=0.2\textwidth]{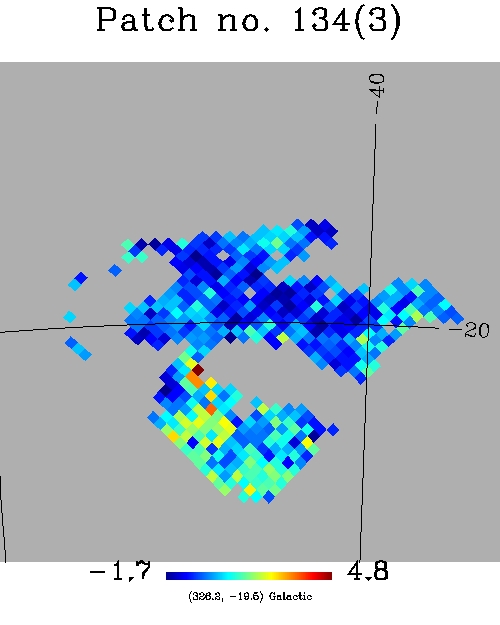}
\includegraphics[width=0.2\textwidth]{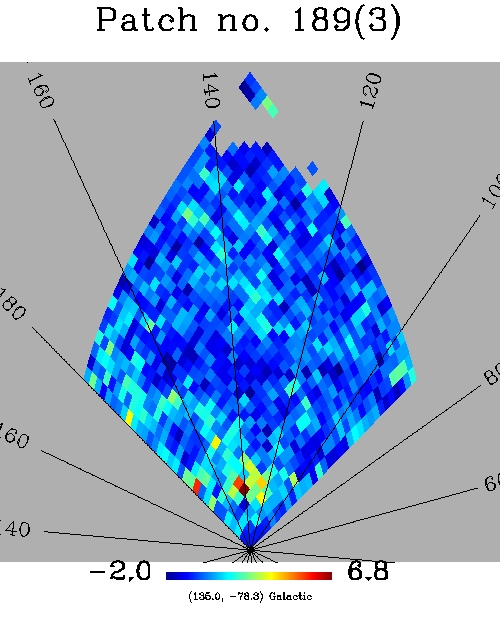}\\
\includegraphics[width=0.2\textwidth]{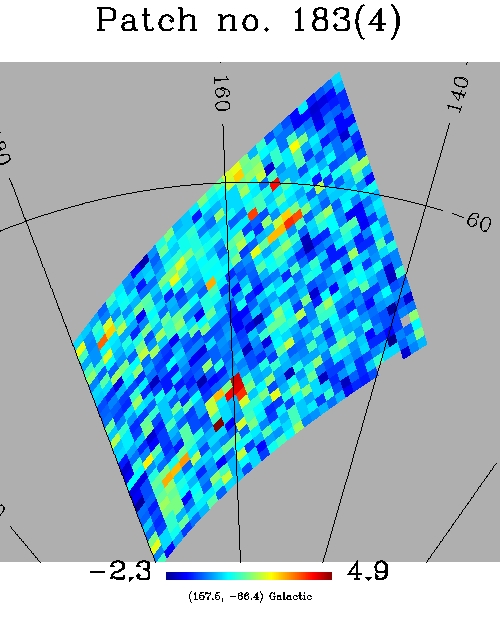}
\includegraphics[width=0.2\textwidth]{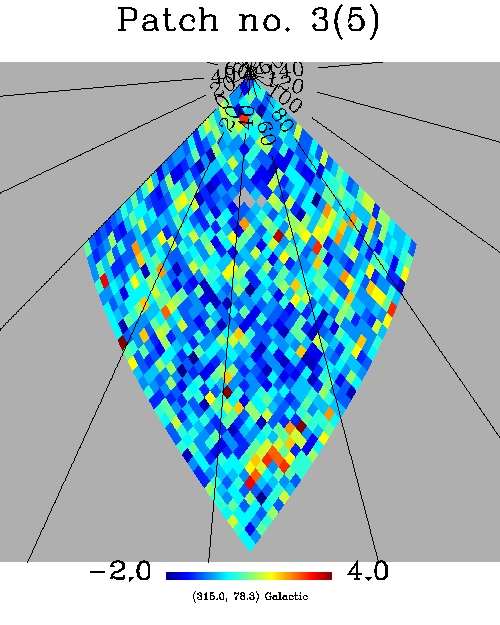}
\caption{Gnomonic projections of the GNC maps obtained from the WSC-$clean$ sample in the selected patches. 
The number $p$ of the patch and the corresponding photo-$z$ bin $b$ are indicated in the title of the panels as $p$($b$). 
The maps are presented in terms of the dimensionless $\nu$ quantity.}
\label{fig:gnomviews}
\end{figure*}

In fact, we identified some patches with significantly low $p$-values. 
However, it is worth to mention that such discrepancy can be associated to contamination, specially if a patch is found discrepant in only one of the two analysed samples. 
In this sense, our criterion to finally select an unusual patch is its statistical significance to be such that $p$-value $< 1.4\%$ in both WSC-$clean$ and WSC-$svm$ samples simultaneously: 
if a patch is highlighted in both samples, each one constructed from a different cleaning process, it suggests that its discrepant feature is not a consequence of contamination, specially if the patch is located far from the Galactic region.
A total of 10 patches were identified under this criterium, namely: 
\begin{itemize}
\item patches no. 130, 157, and 180 in the photo-$z$ bin 1 ($0.10 < z < 0.15$);
\item patches no. 25, 34, and 137 in the photo-$z$ bin 2 ($0.15 < z < 0.20$);
\item patches no. 65, 134, and 189 in the photo-$z$ bin 3 ($0.20 < z < 0.25$); and 
\item patch no. 183 in the photo-$z$ bin 4 ($0.25 < z < 0.30$).
\end{itemize}
No region obeying this criterium were found in the last photo-$z$ bin ($0.30 < z < 0.35$).
All these 10 patches are highlighted in Fig. \ref{fig:selectP}, with the color scale representing the average between the $p$-values obtained from the two samples. 
For an illustration of the distribution of sources in these patches, we show in Fig. \ref{fig:gnomviews} their Gnomonic projection for the WSC-$clean$ GNC maps.

Moreover, it is still worth to remember that we are analysing a very large number of separated regions (a total of 630 patches, 126 from each of the 5 photo-$z$ bins), implying that some patches would inevitably contain large clusters or voids by chance, which is a plausible explanation for the significantly different morphological properties of the selected regions. 
In other words, one should expect to observe some patches whose MF features appear to be in disagreement with the simulations. 
To evaluate this statement and confirm if the selected regions disprove our model or if, on the contrary, they are rare and extreme (but expected) findings, we analyse the $\chi^2$-maps obtained from the mock realisations and estimate the probability of a simulation, considering the 5 photo-$z$ bins altogether, to have at least the same number of patches with $p$-value $<$ 1.4\% as we identified in the WSC GNC maps.
We found that $\sim$38.6\% of the simulations contain 10 or more patches satisfying this condition. 
Notice that this estimative is an upper limit probability, since our criterium for selecting a patch is it to have a $p$-value $< 1.4\%$  in both samples simultaneously. 

In this context one can say that our findings show the WSC samples to be fairly well described by the simulations, that is, in agreement not only with the concordance cosmological model, but also with the selection function, the lognormal distribution of sources, among other aspects of the mocks generation. 
Aware of this, we still emphasize the importance of analysing the MF from the selected patches, hereafter called {\it extreme}, since, although not discrepant, they are very rare regions, what motivates their scrutiny in order to look for the possible reasons for their uncommon behaviour. 
Indeed, this argument is specially reinforced by the high statistical significance of patch no. 157, with an average $p$-value of $0.01$\%, identified in photo-$z$ bin 1, and possibly little contaminated given its location far from the Galactic plane.
To evaluate the probability of finding such extreme region in the GNC maps, we calculate its frequency of occurrence in any of the 5 photo-$z$ bins.
We observe that $\sim$11.8\% of the simulations have at least one patch with $p$-value $< 0.01$\%. 
Therefore, although this patch has a very low probability of occurrence, 0.01\%, when considering such specific portion of the sky, it is reasonably likely ($\sim$11.8\%) to observe at least one of such region, given the volume of Universe we are analysing (i.e., a total of 630 attempts --126 patches in each of the 5 photo-$z$ bins-- to find discrepant regions). 
In this context, it is worth to notice that the statistical problem we are dealing with here involves a large number of tests, and, for this, it can be affected by the so called {\it look elsewhere effect} or {\it multiple comparisons problem}. 
This is a statistical problem in which, an observation, apparently statistically significant, actually arise by chance due to the size of the parameter space \citep[for detailed discussions see, e.g.,][]{2010/eilam, 2008/louis}. 
Aware of this, we performed a rigorous statistical verification of our results, in the attempt to take into account such effect.

\begin{figure*}
\centering
 \includegraphics[width=0.32\textwidth]{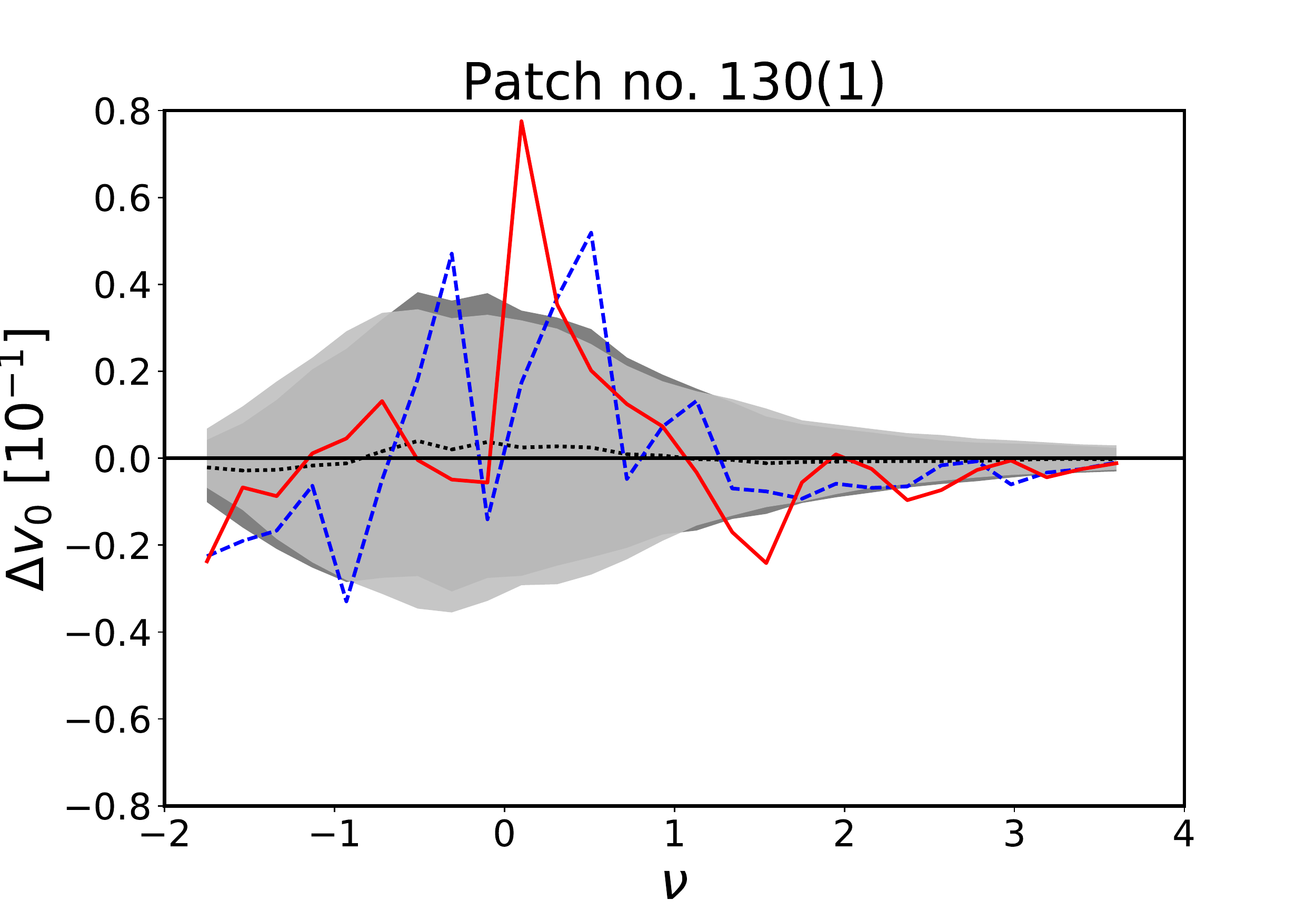}
 \includegraphics[width=0.32\textwidth]{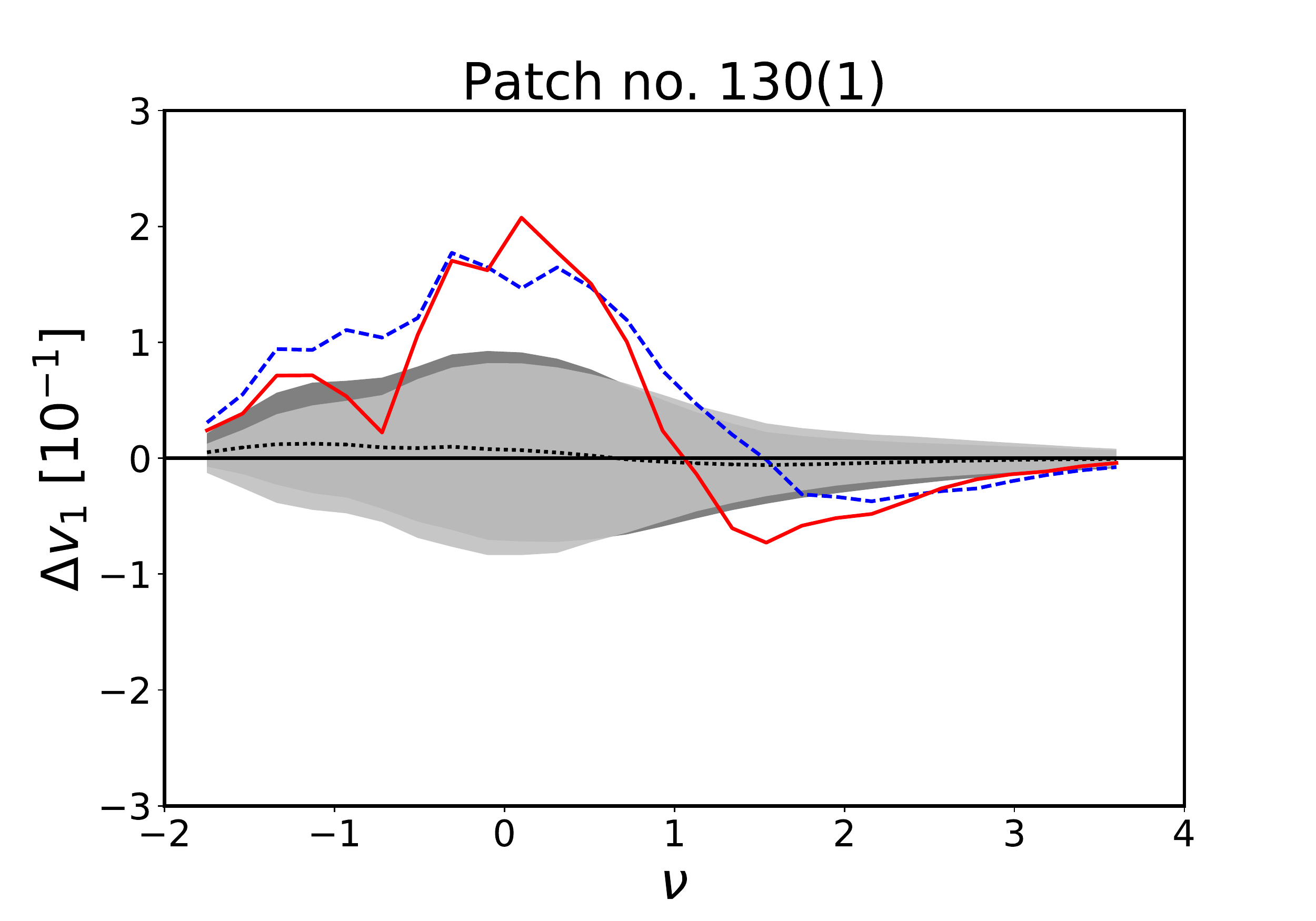}
 \includegraphics[width=0.32\textwidth]{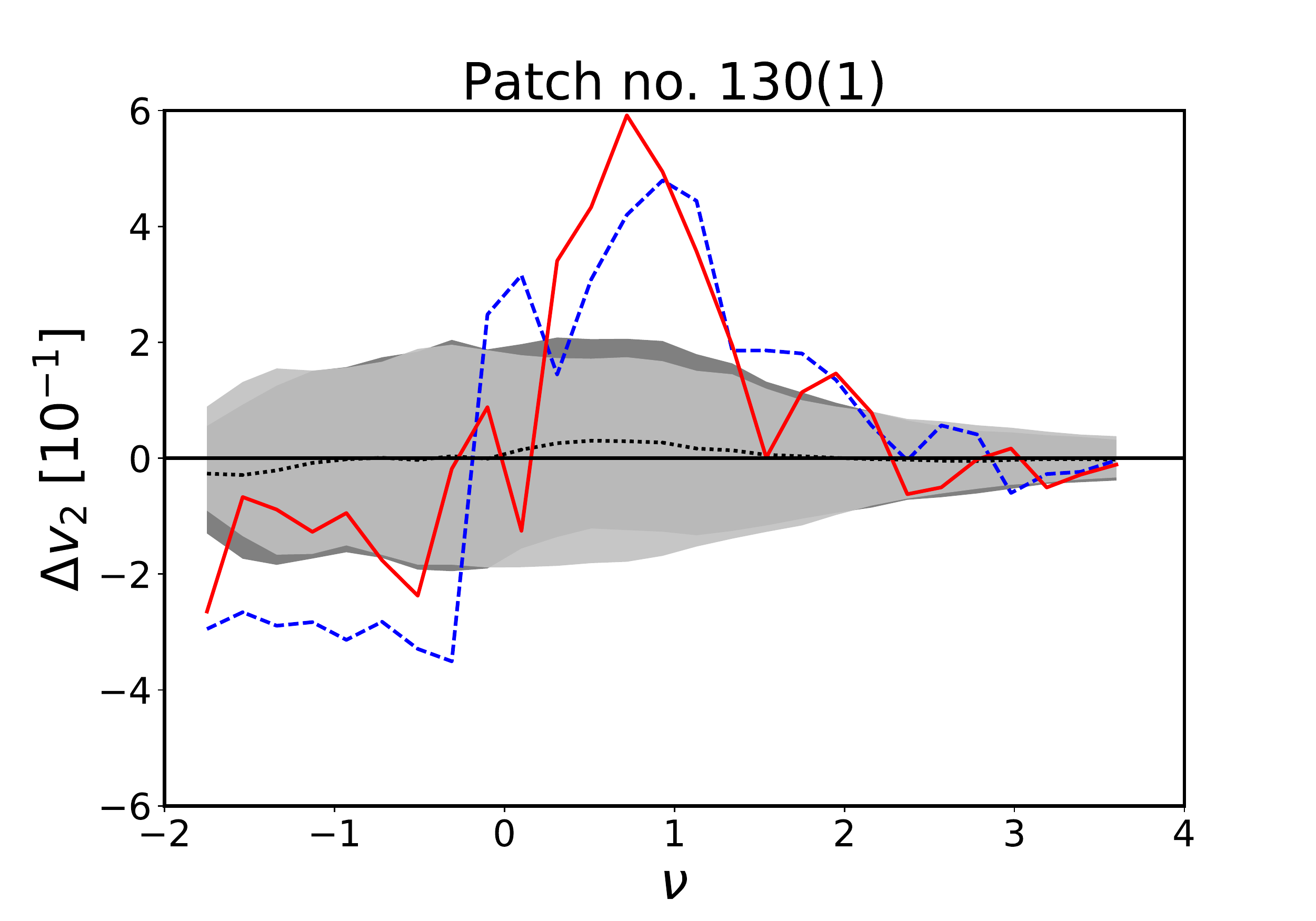}\\
 \includegraphics[width=0.32\textwidth]{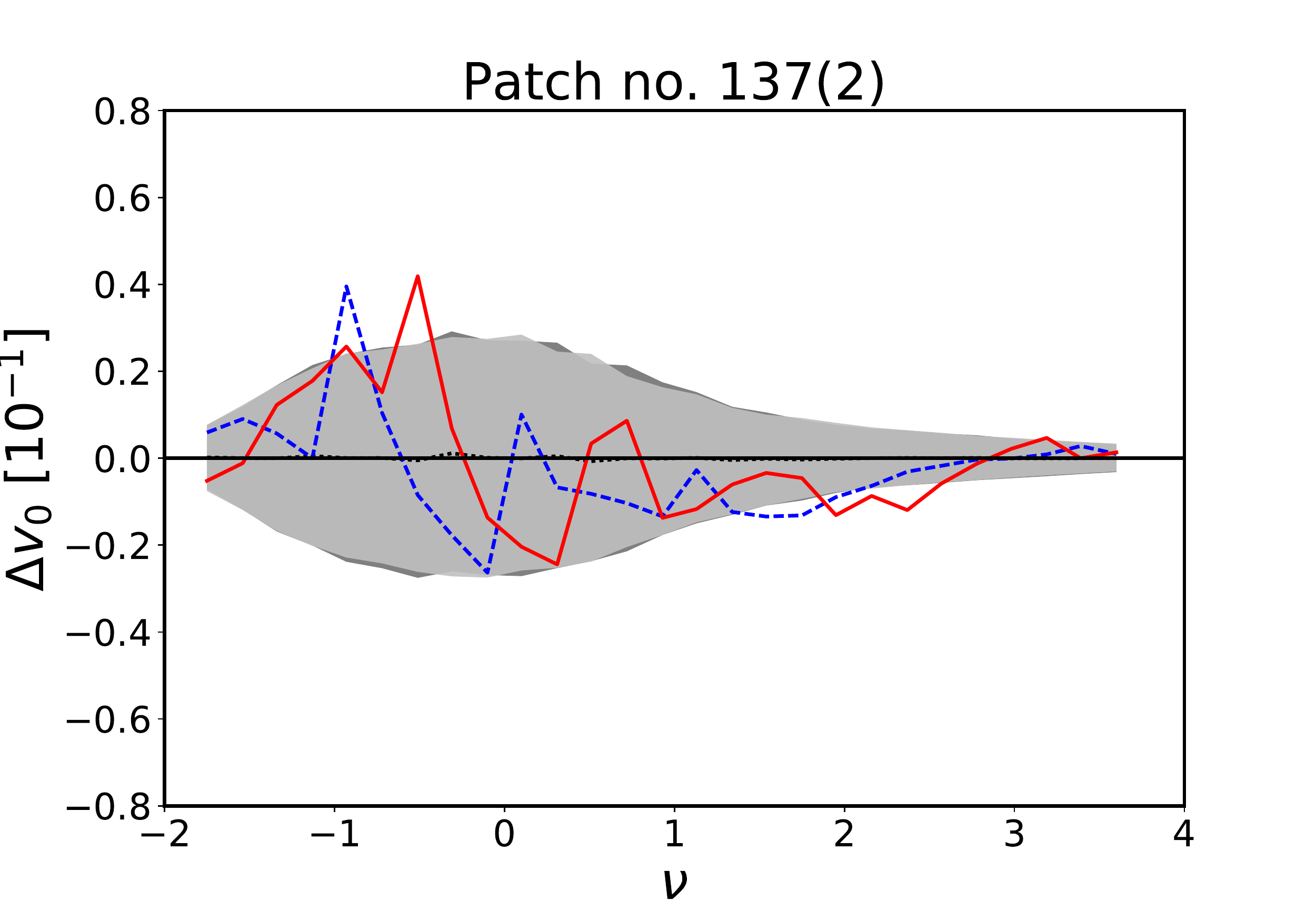}
 \includegraphics[width=0.32\textwidth]{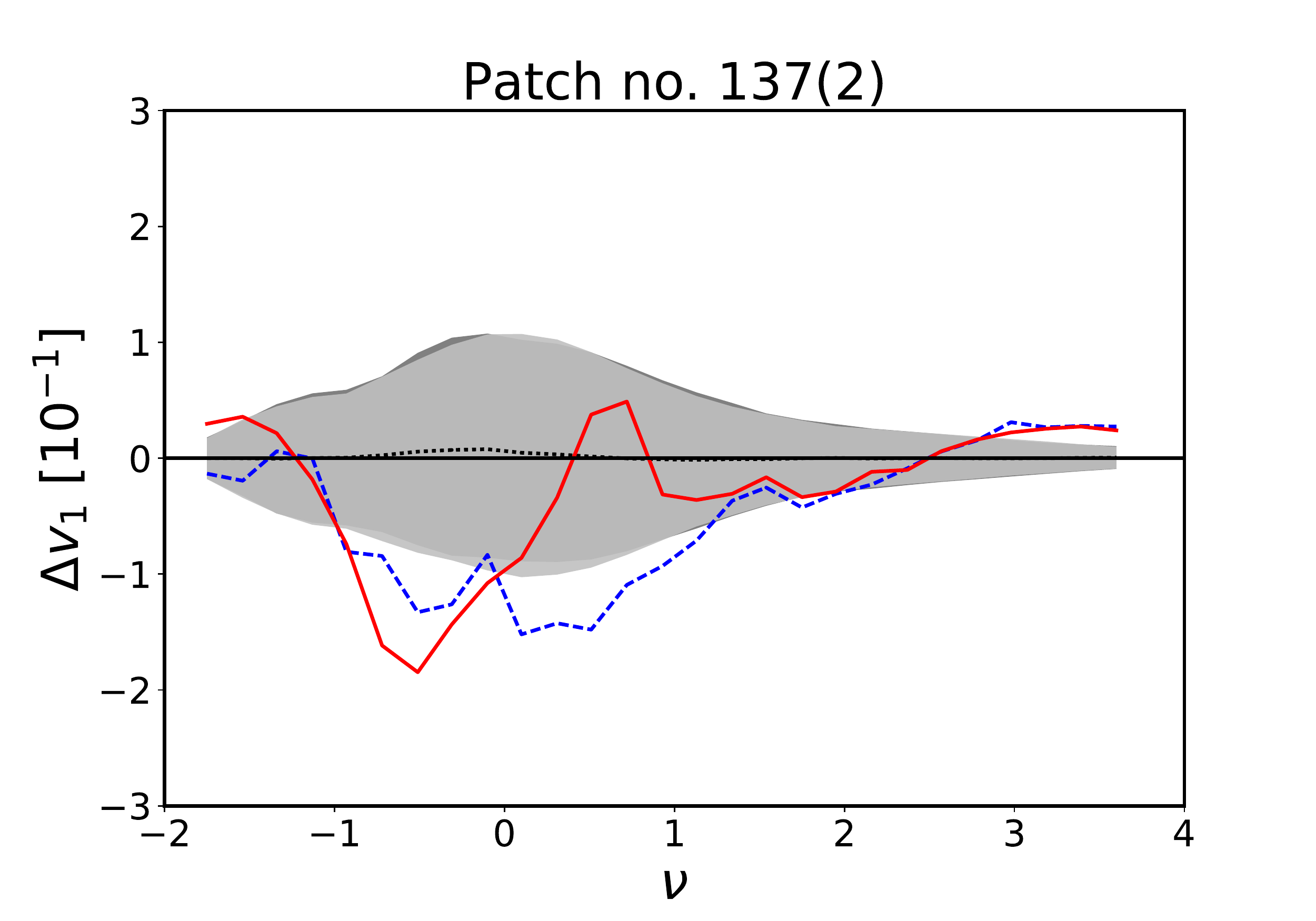}
 \includegraphics[width=0.32\textwidth]{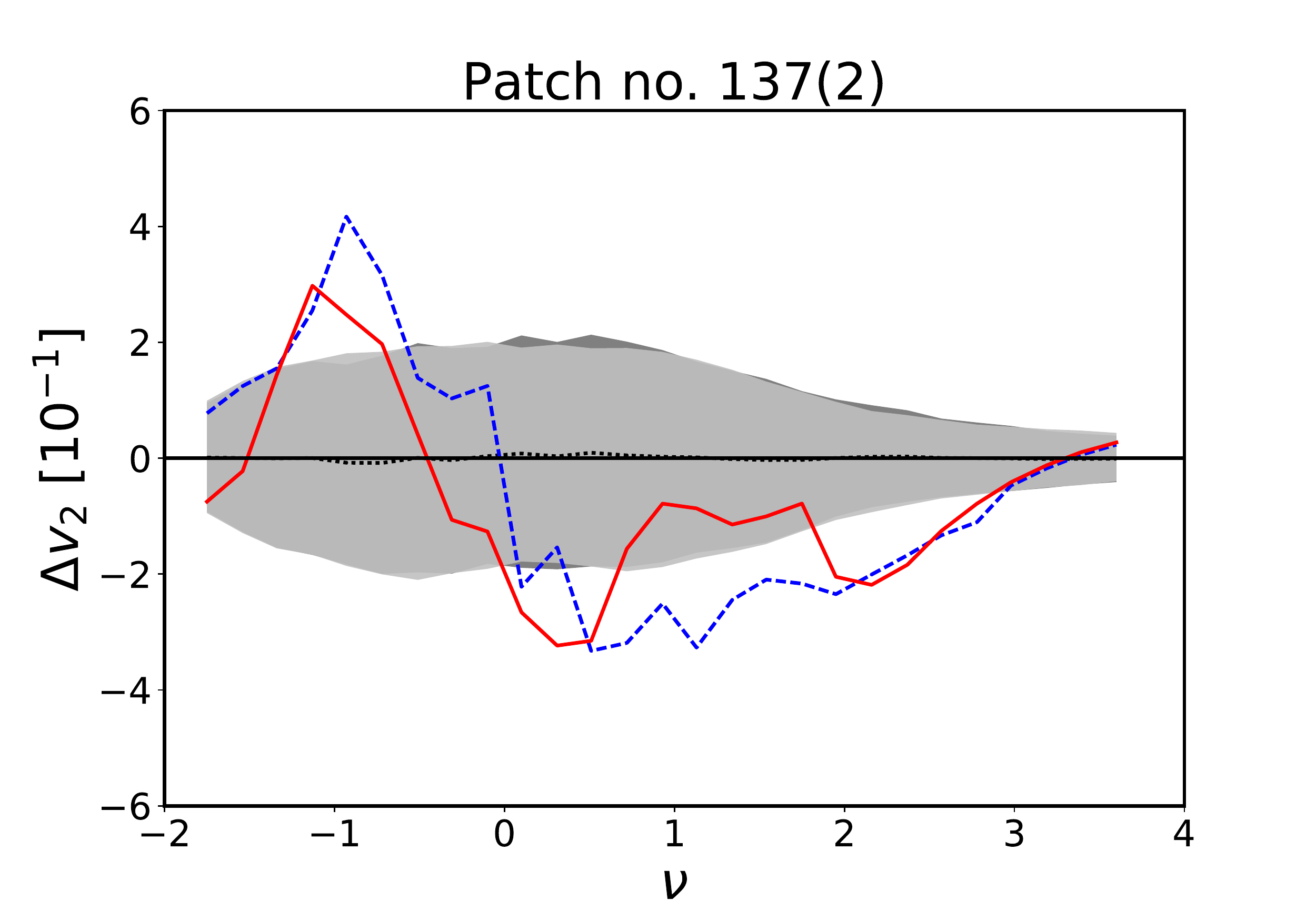}\\
 \includegraphics[width=0.32\textwidth]{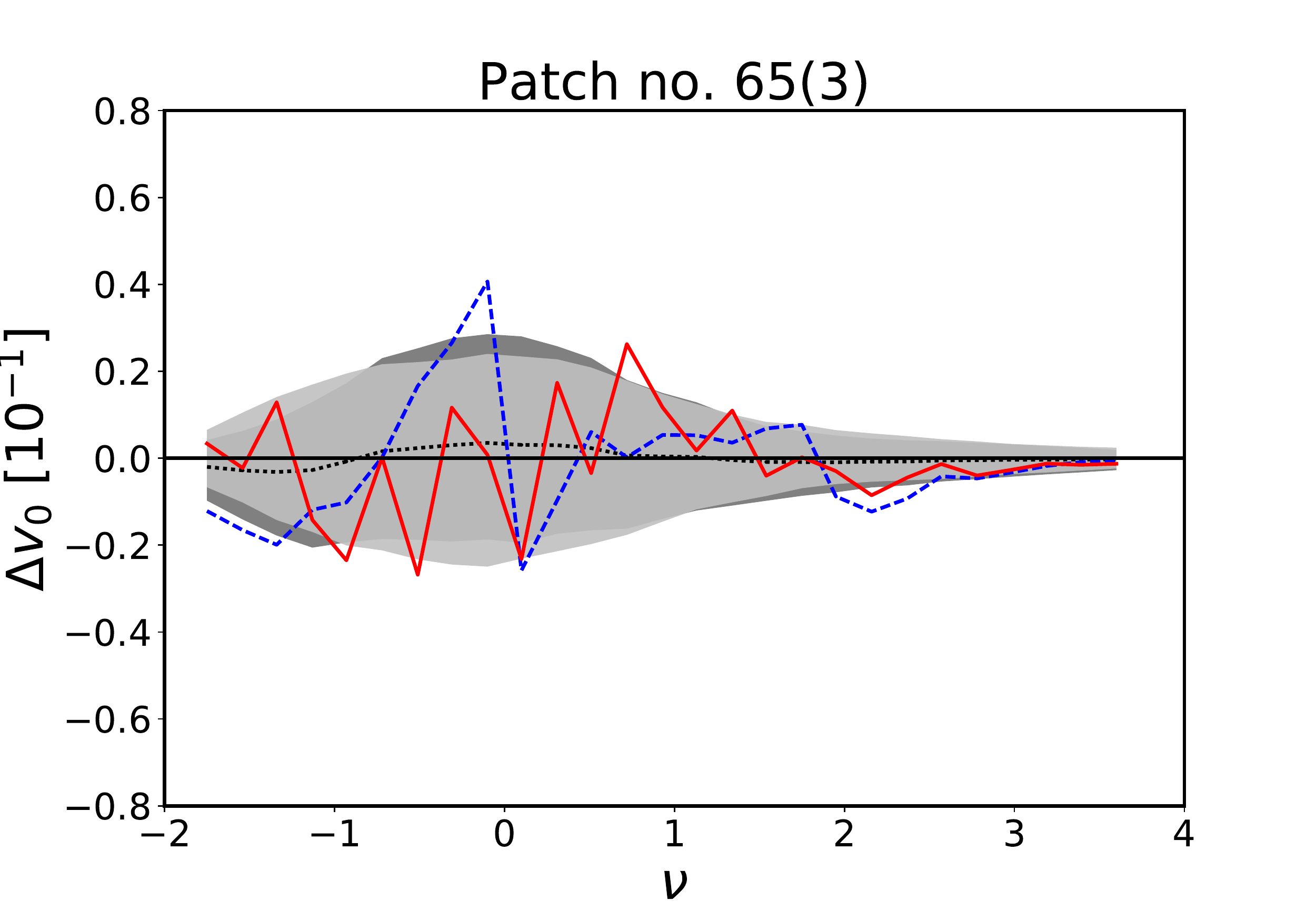}
 \includegraphics[width=0.32\textwidth]{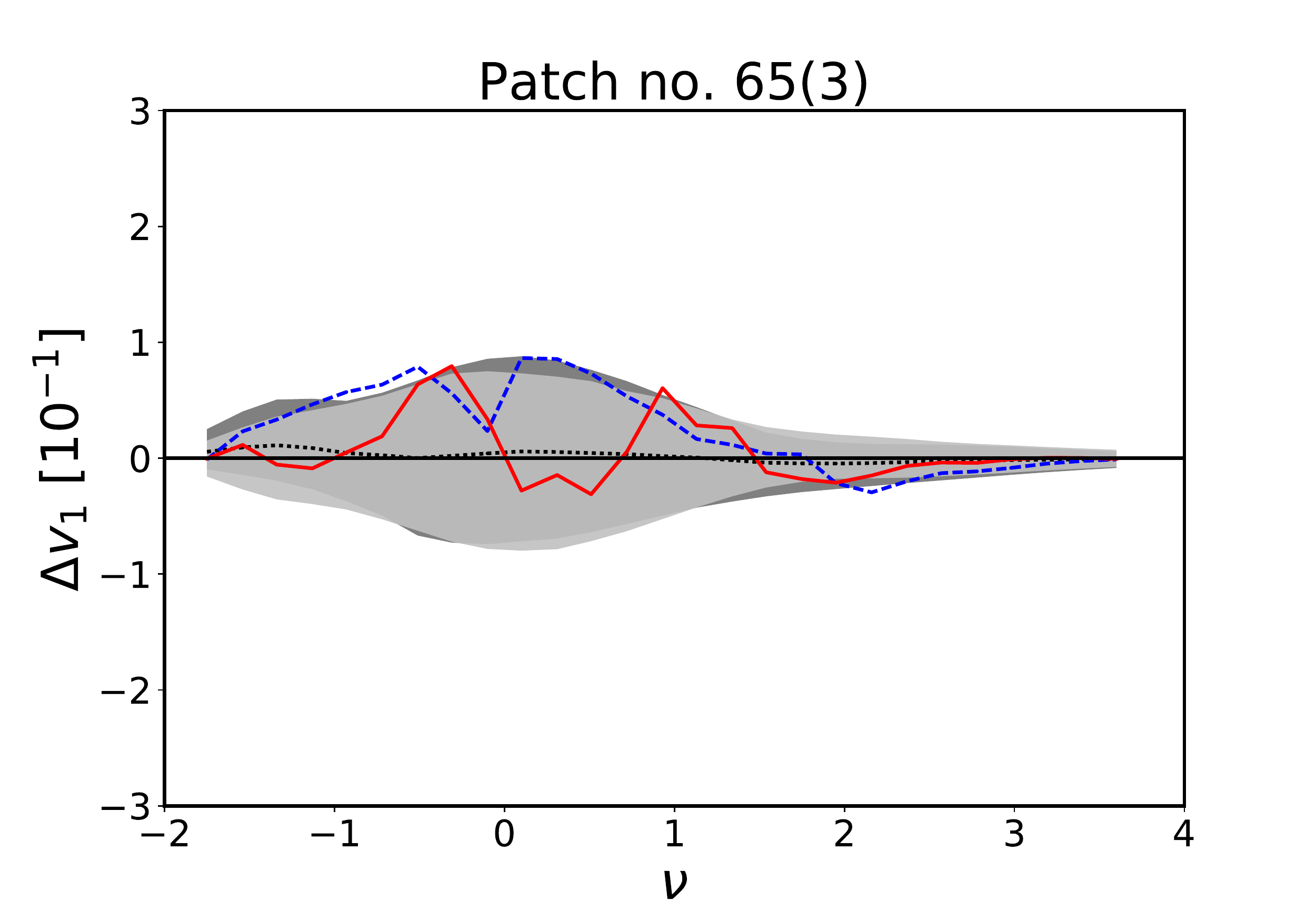}
 \includegraphics[width=0.32\textwidth]{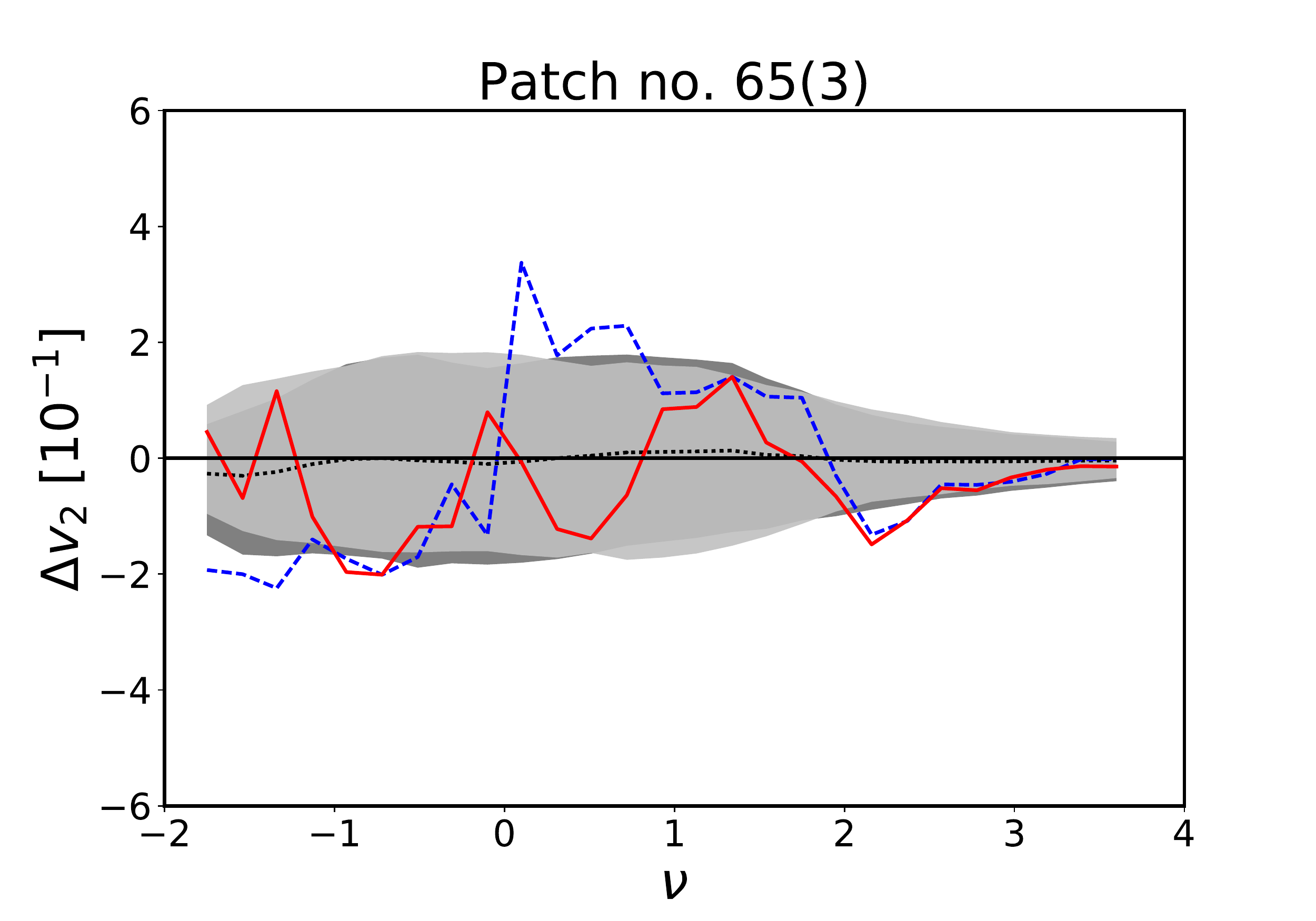}
 \includegraphics[width=0.32\textwidth]{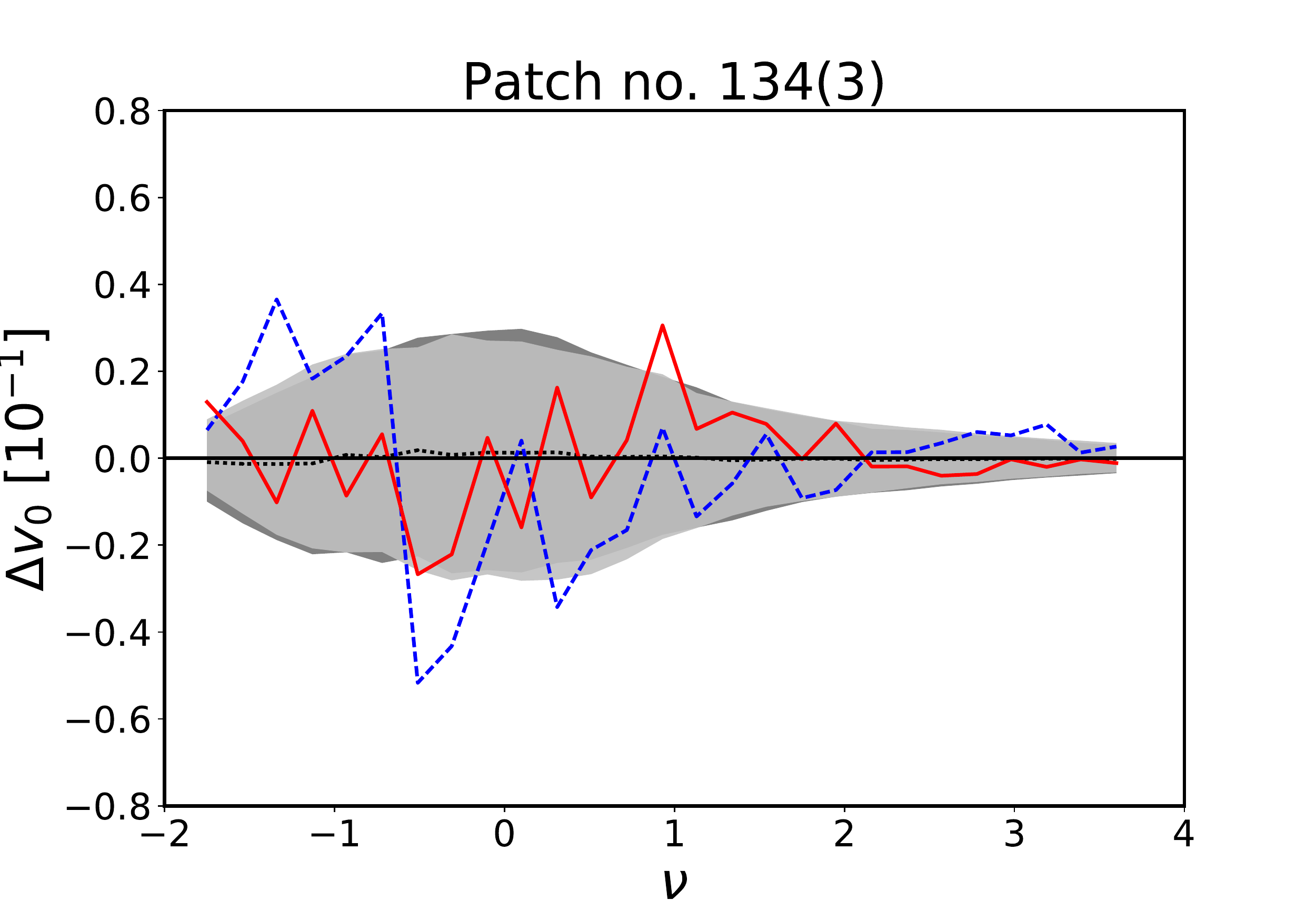}
 \includegraphics[width=0.32\textwidth]{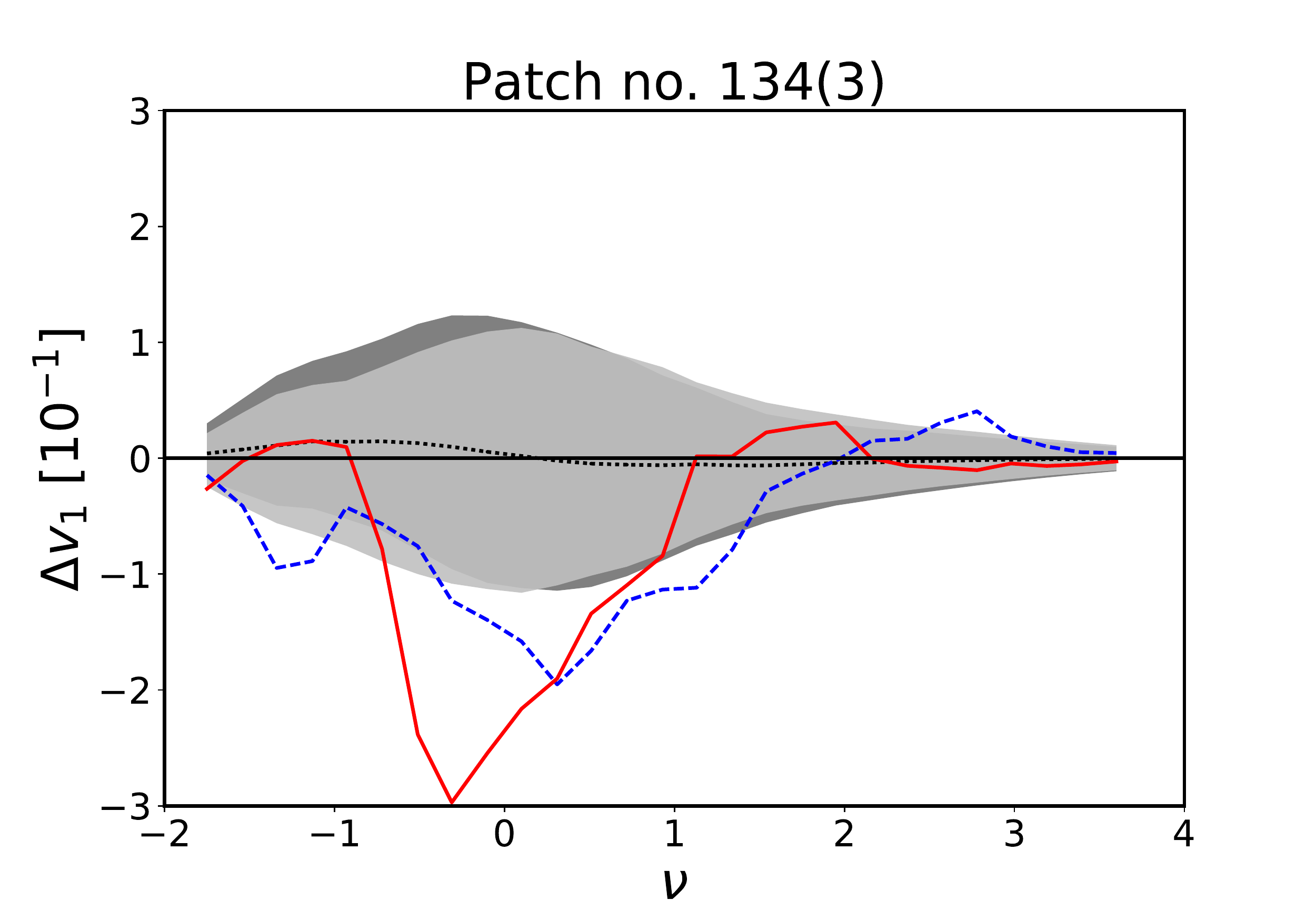}
 \includegraphics[width=0.32\textwidth]{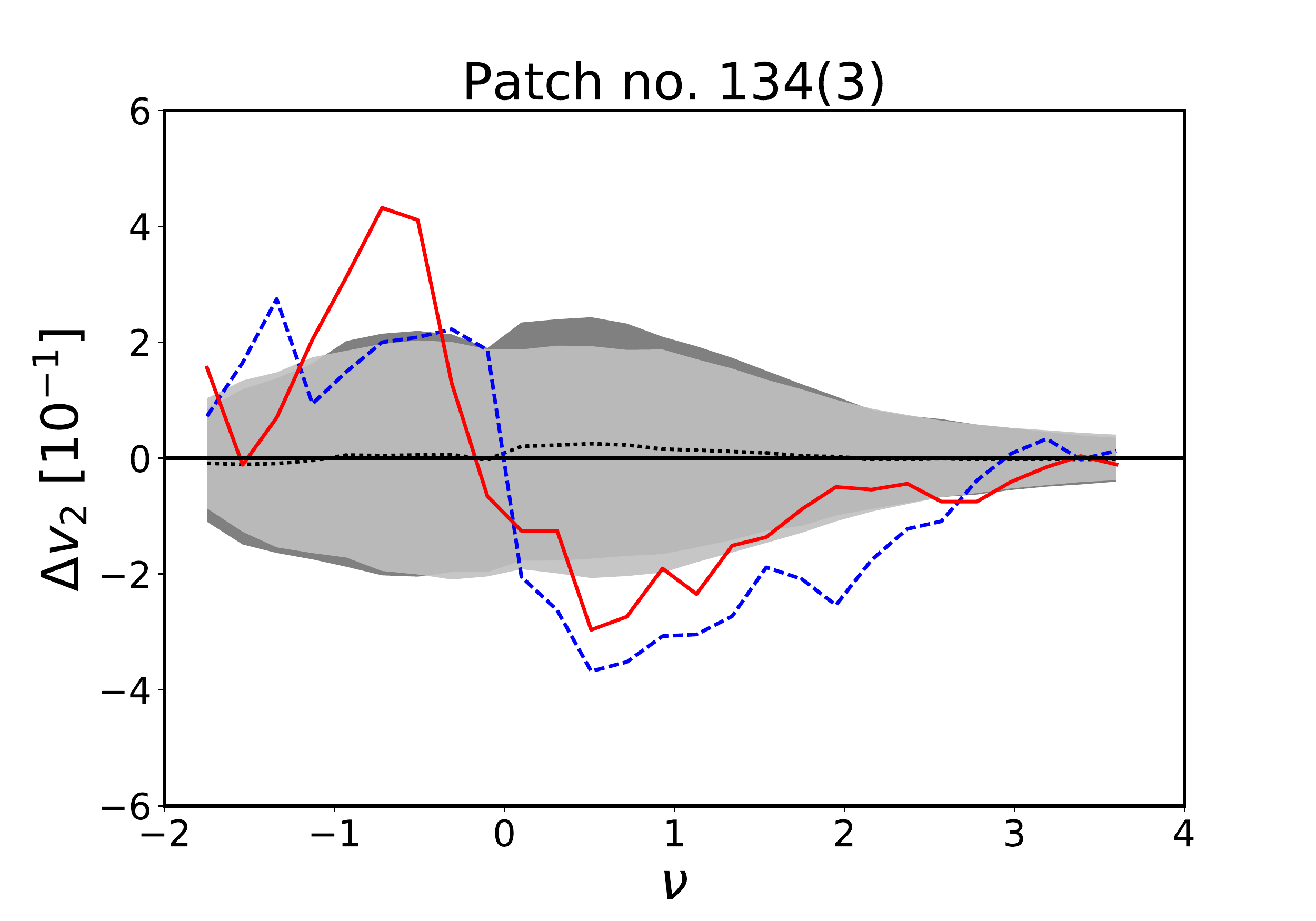}\\
\caption{Analyses of the four extreme patches located near the Galactic region. 
The plots show the relative difference curves, as given by equation \ref{eq:relat_dif}, obtained for Area, $\Delta \mathrm{v}_0$ (left panels),  Perimeter, $\Delta \mathrm{v}_1$ (middle panels), and Genus, 
$\Delta \mathrm{v}_2$ (right panels). 
The red solid and blue dashed lines correspond to the results from WSC-$clean$ and WSC-$svm$ samples, respectively. 
The black dotted line near zero represents the mean $\langle\Delta \mathrm{v}_k^i\rangle$ curve for the contaminated simulations, and the shaded regions correspond to the 2$\sigma_k$ dispersion as given by the contaminated (dark gray) and non-contaminated (light gray) mock realisations (see equation \ref{eq:relat_dif_error} and text for details).}
\label{fig:near_gal}
\end{figure*}

\begin{figure*}
\centering
 \includegraphics[width=0.32\textwidth]{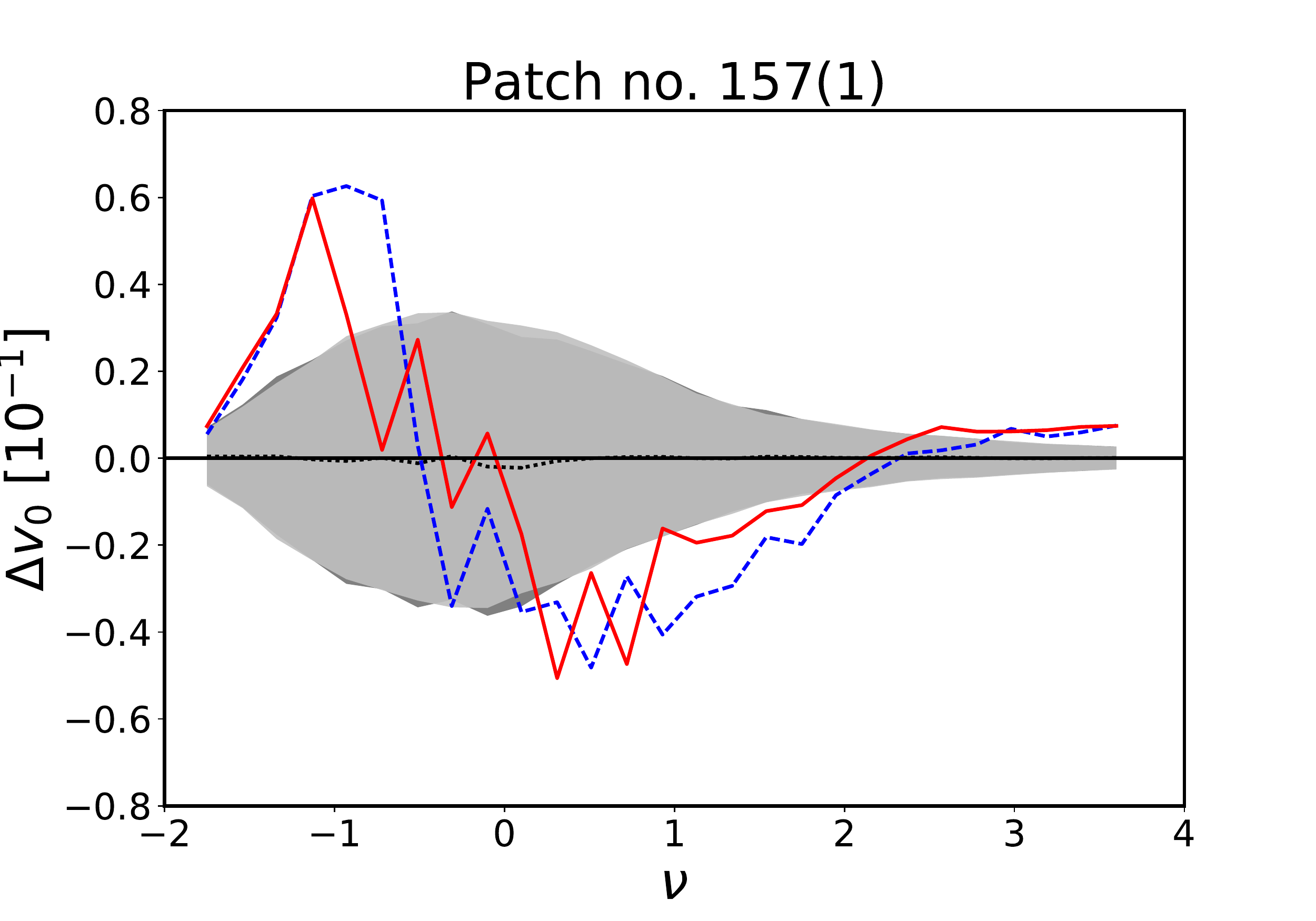}
 \includegraphics[width=0.32\textwidth]{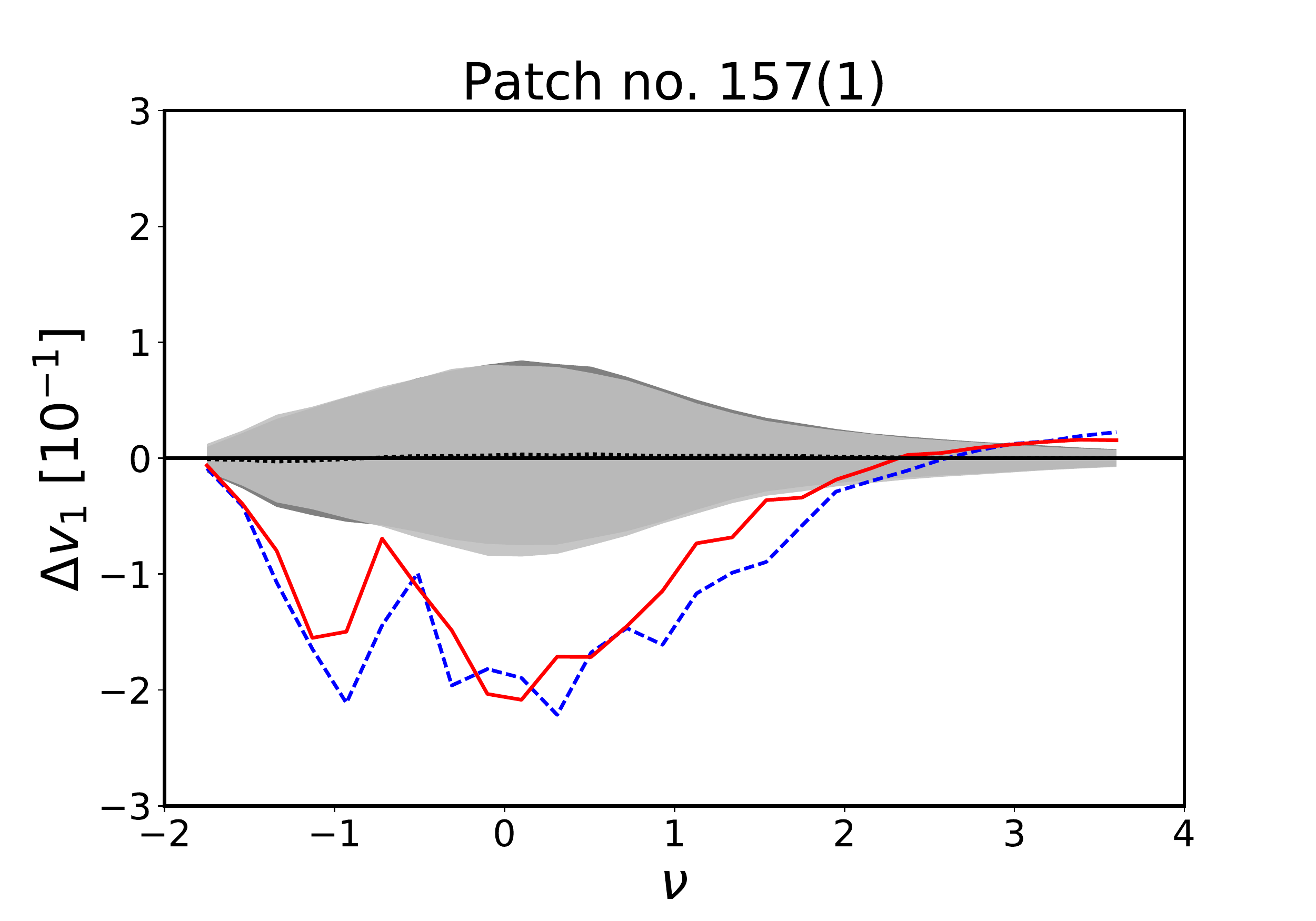}
 \includegraphics[width=0.32\textwidth]{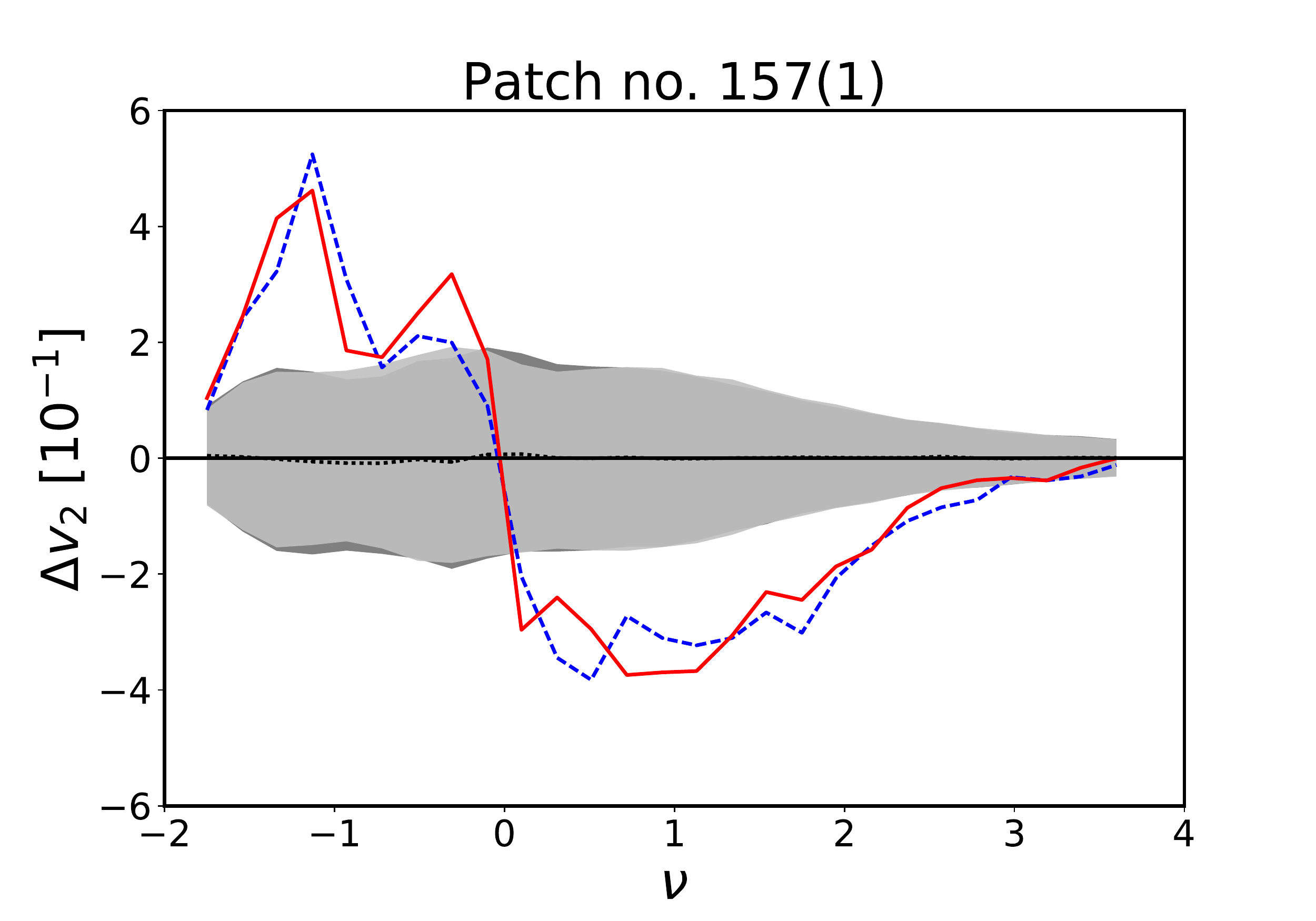}\\
  \vspace{-0.1cm}
 %
 \includegraphics[width=0.32\textwidth]{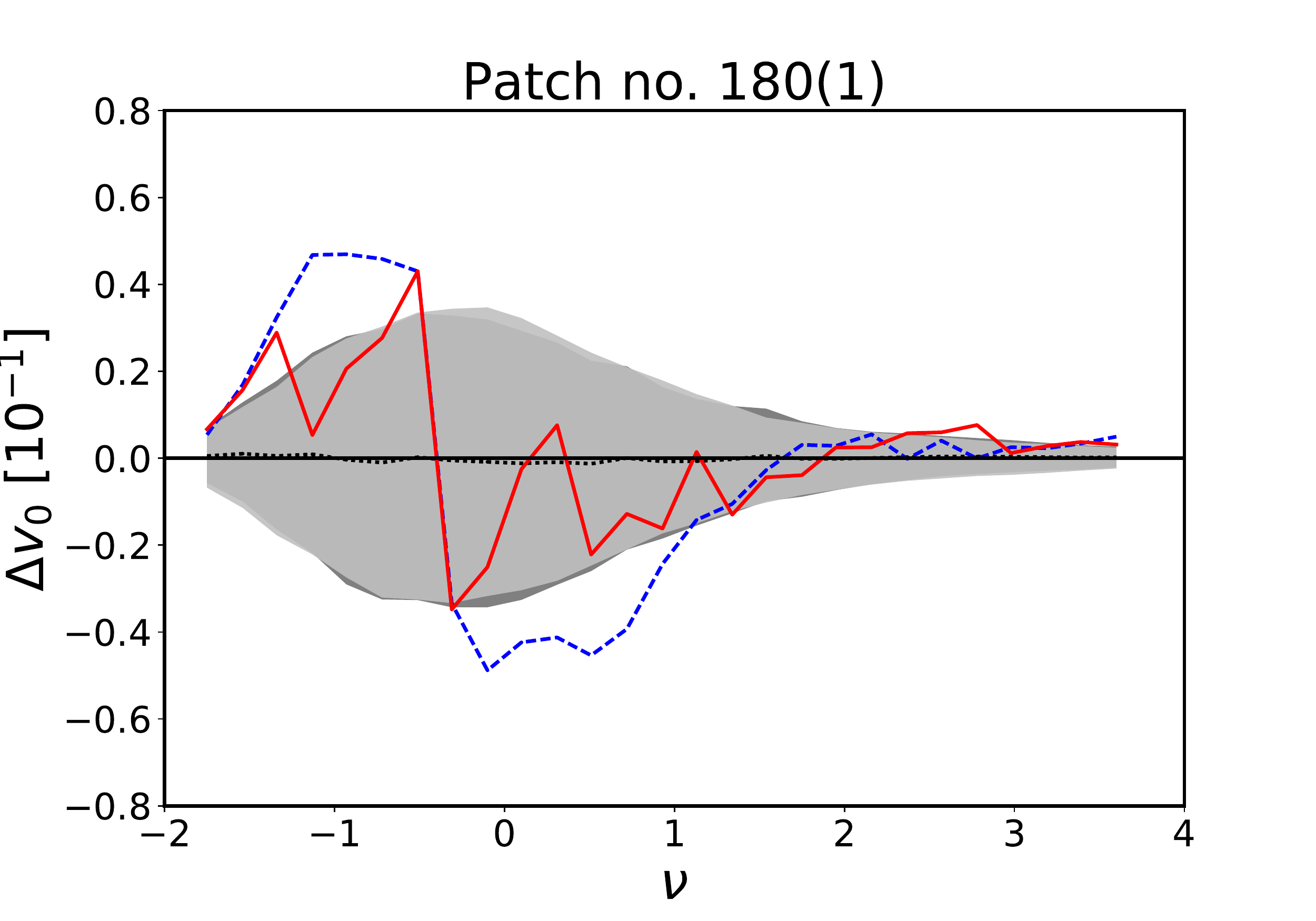}
 \includegraphics[width=0.32\textwidth]{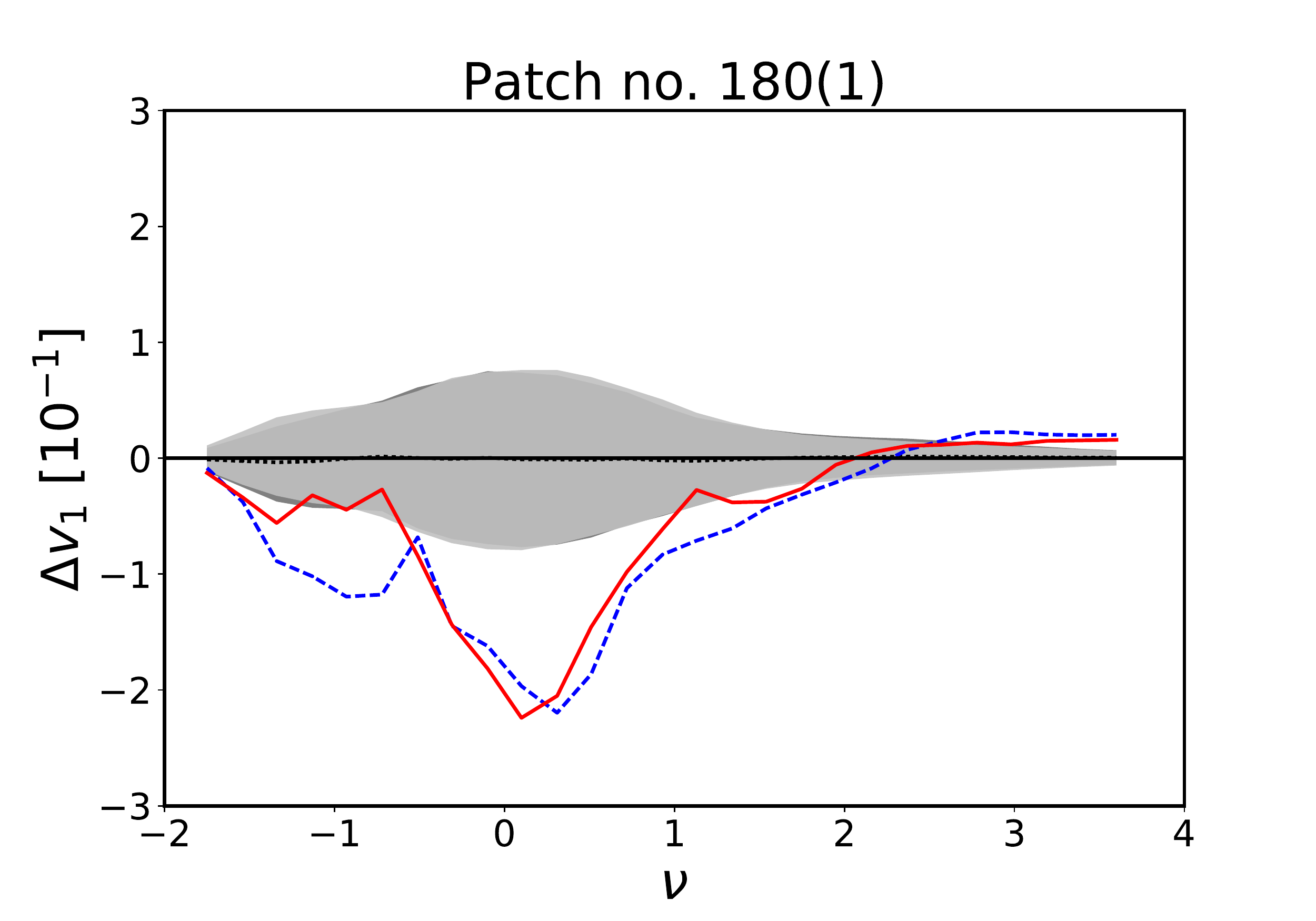}
 \includegraphics[width=0.32\textwidth]{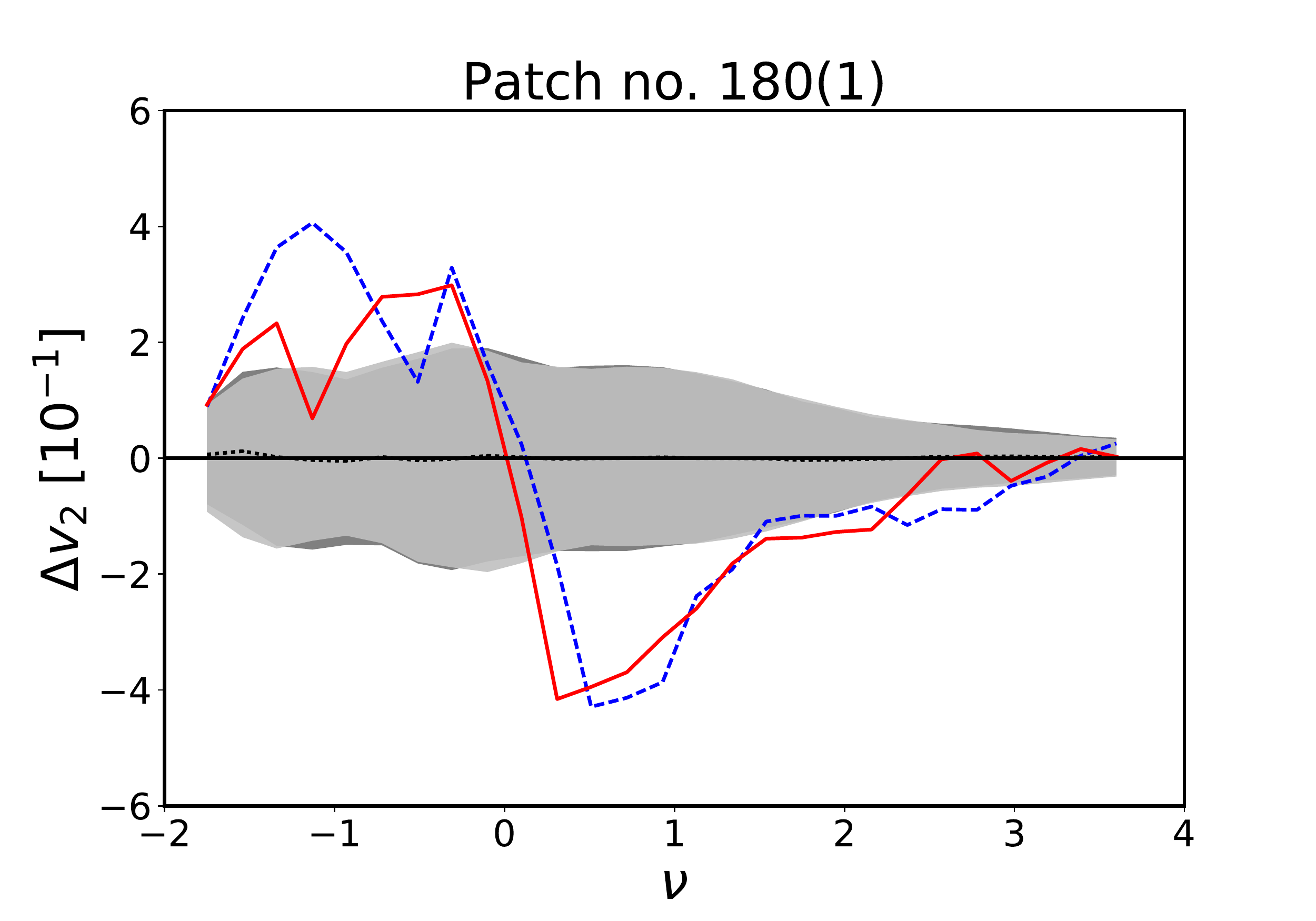}\\
 \vspace{-0.1cm}
 %
 \includegraphics[width=0.32\textwidth]{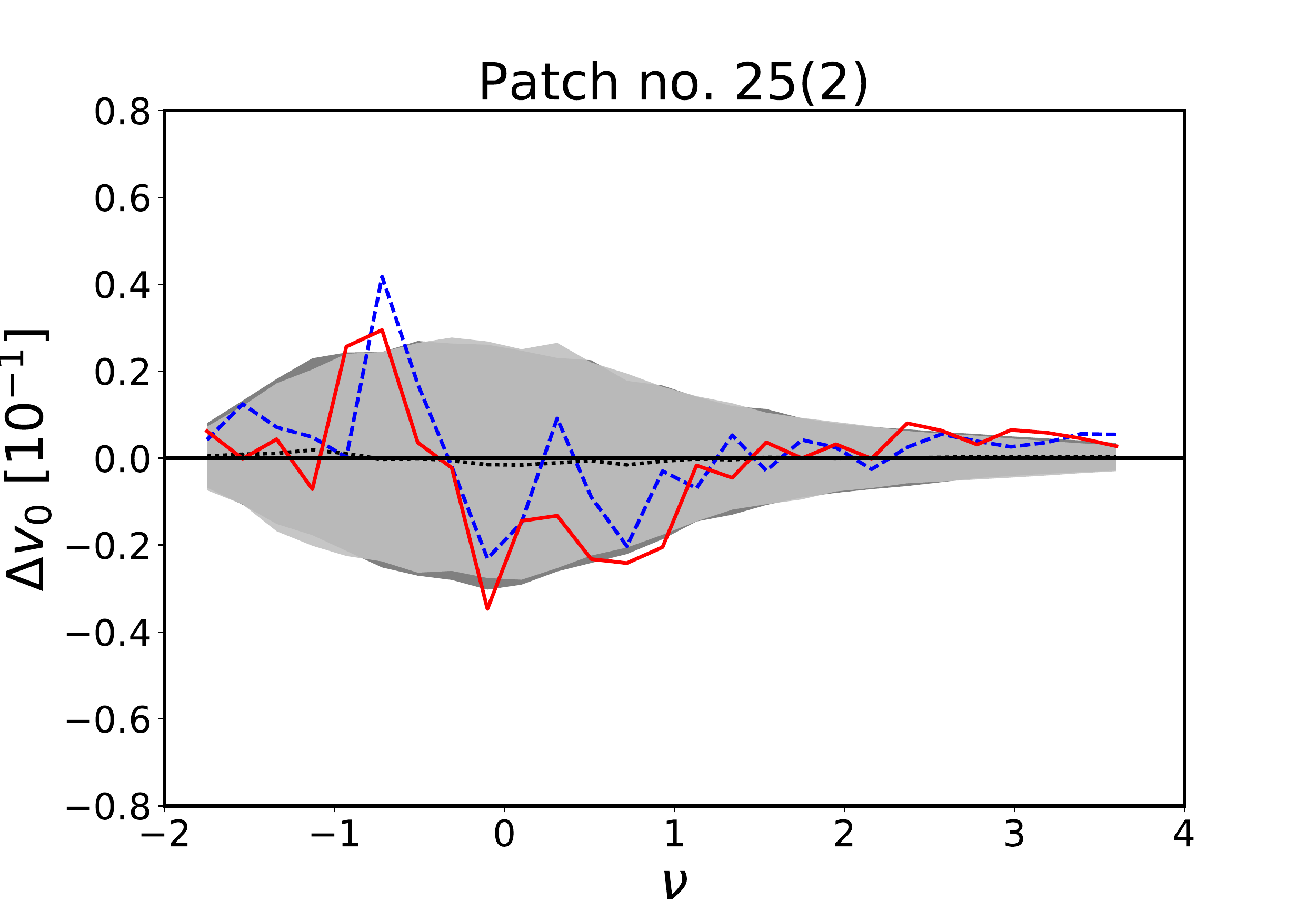}
 \includegraphics[width=0.32\textwidth]{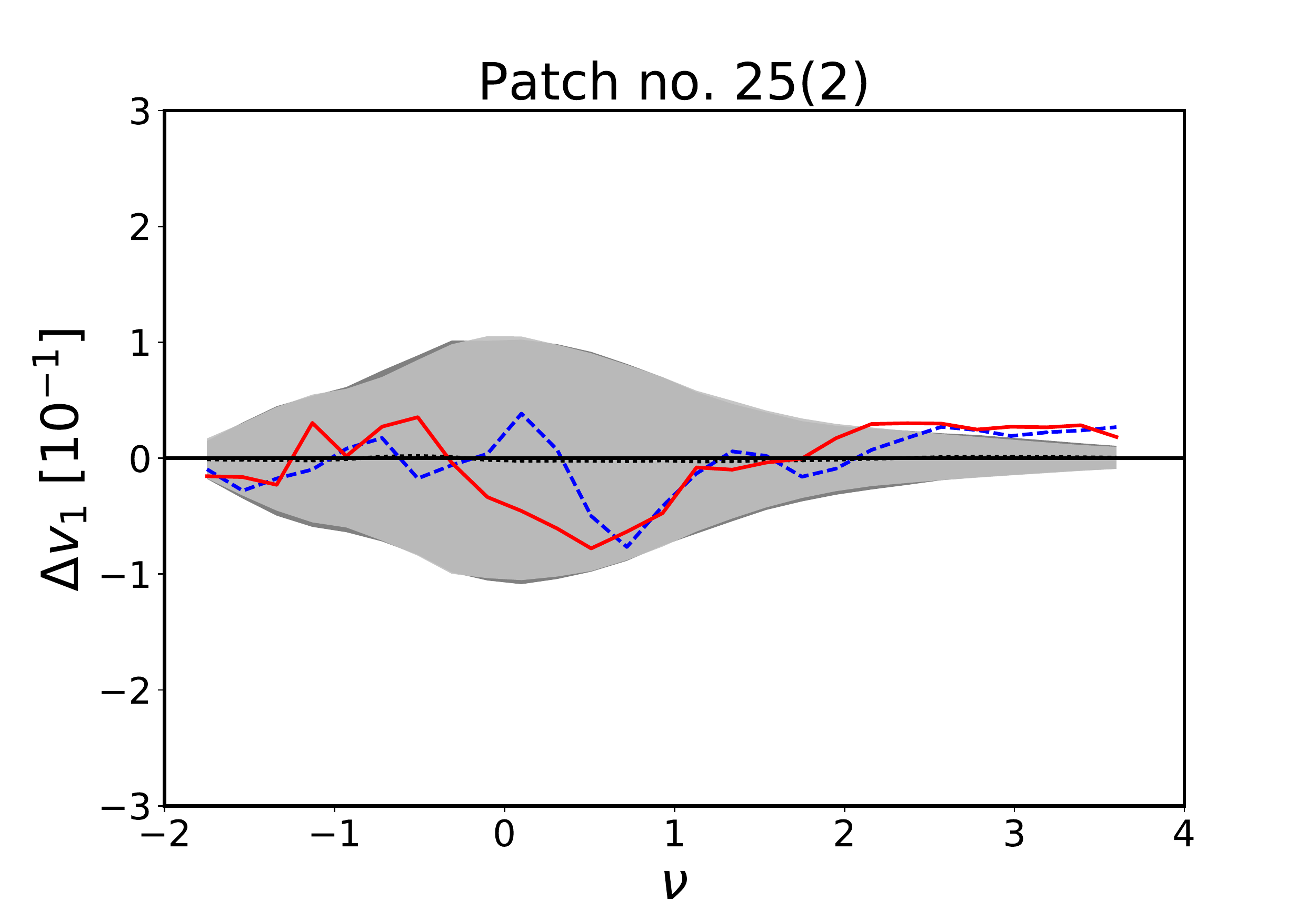}
 \includegraphics[width=0.32\textwidth]{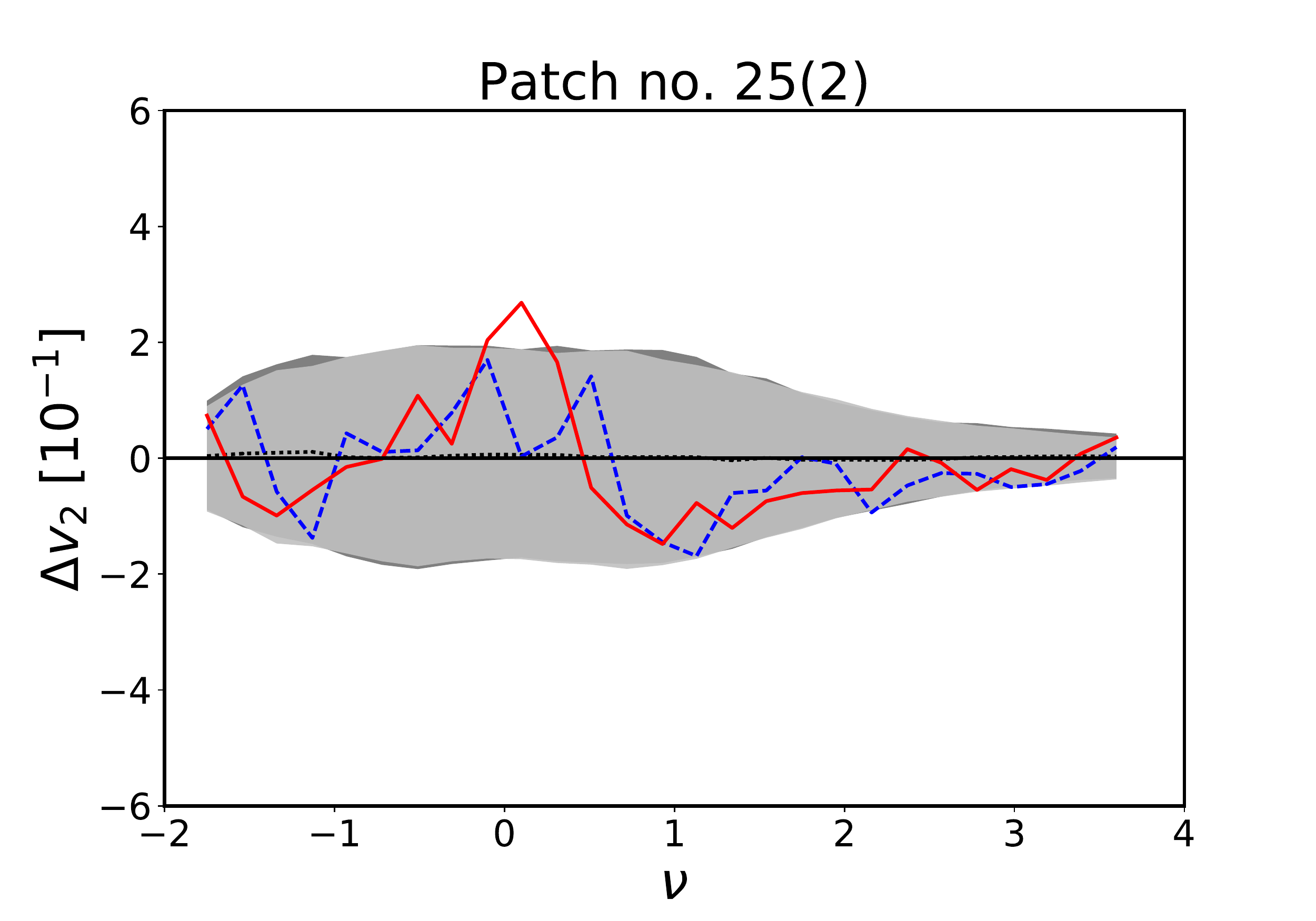}\\
 \vspace{-0.1cm}
 \includegraphics[width=0.32\textwidth]{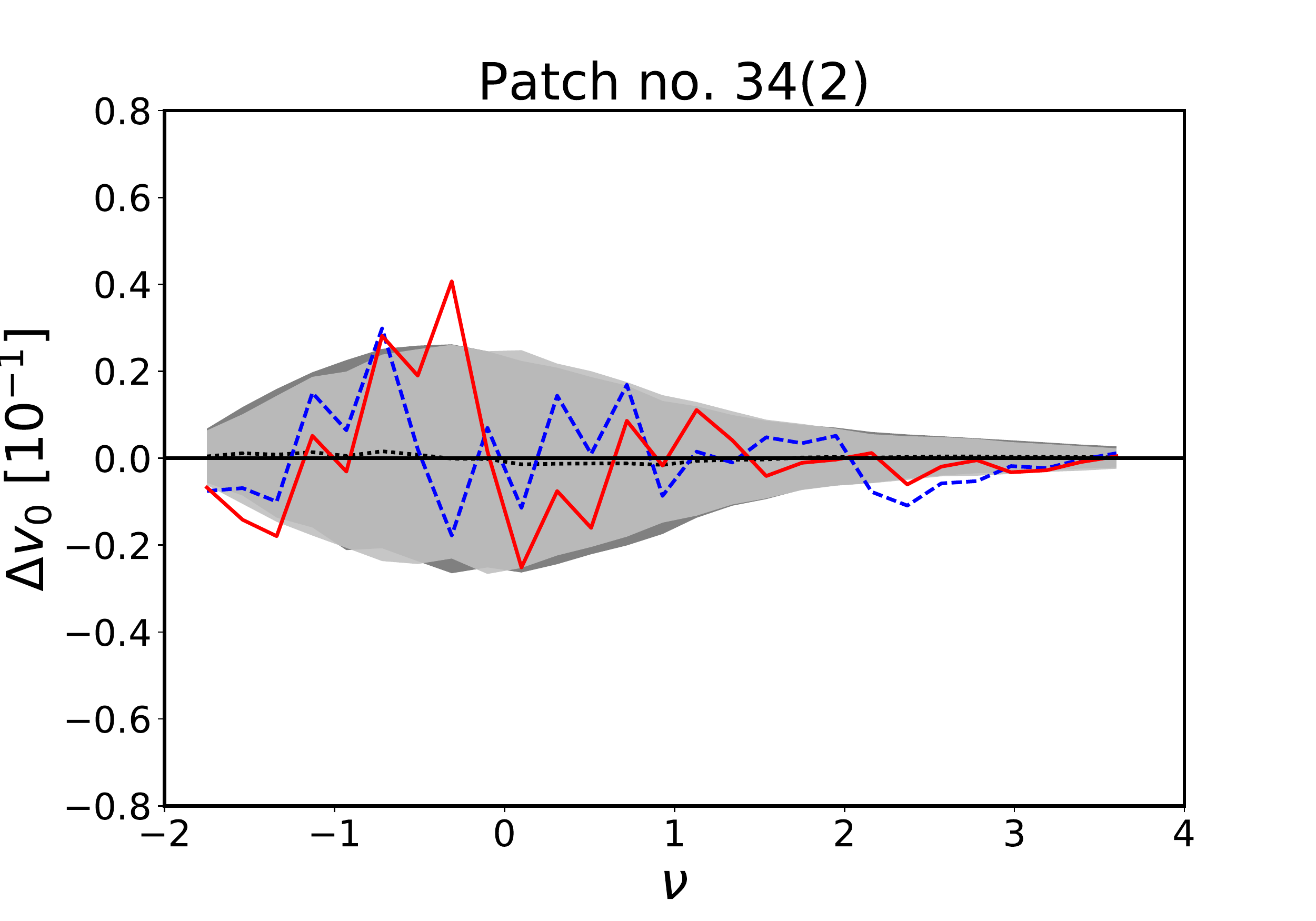}
 \includegraphics[width=0.32\textwidth]{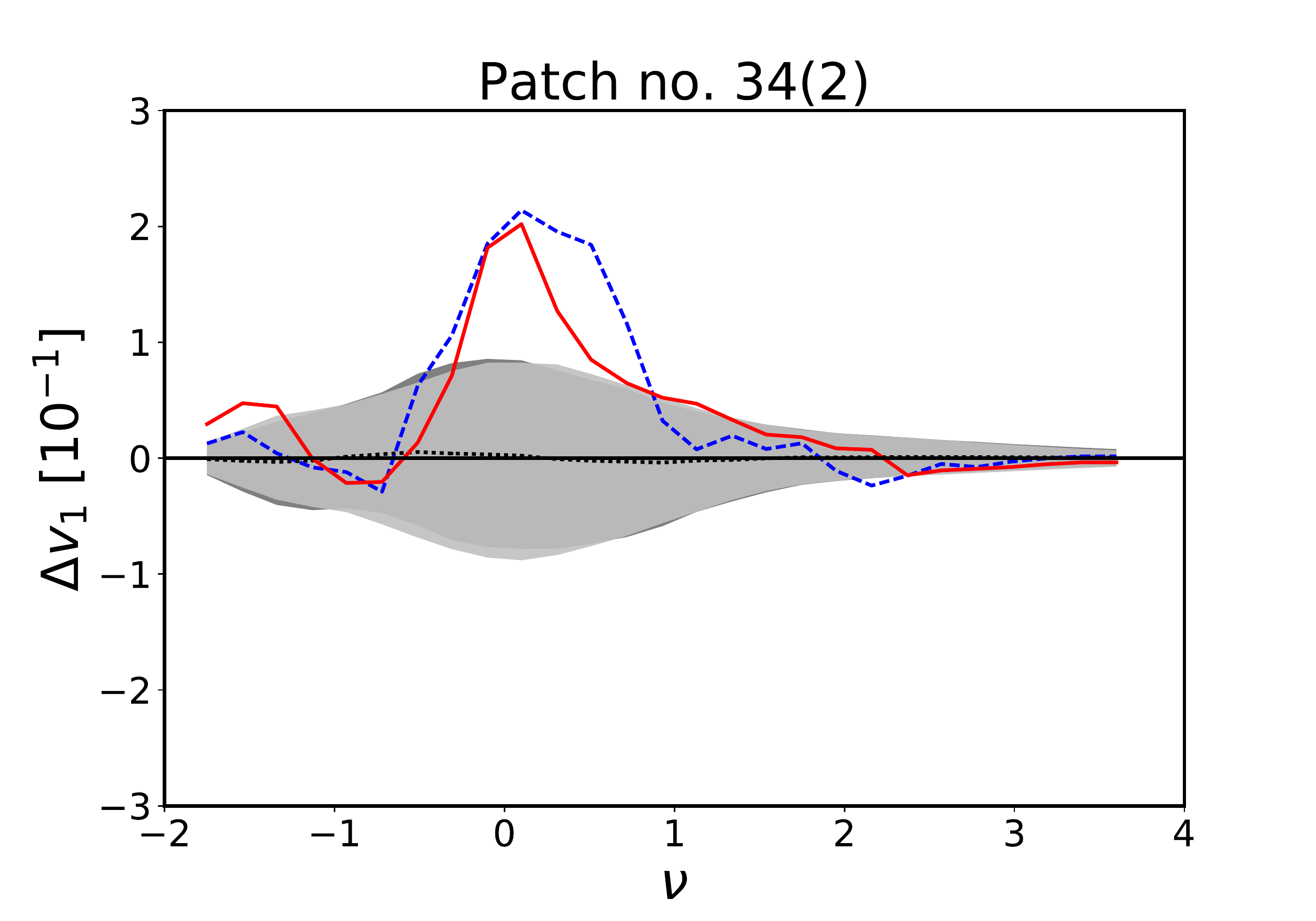}
 \includegraphics[width=0.32\textwidth]{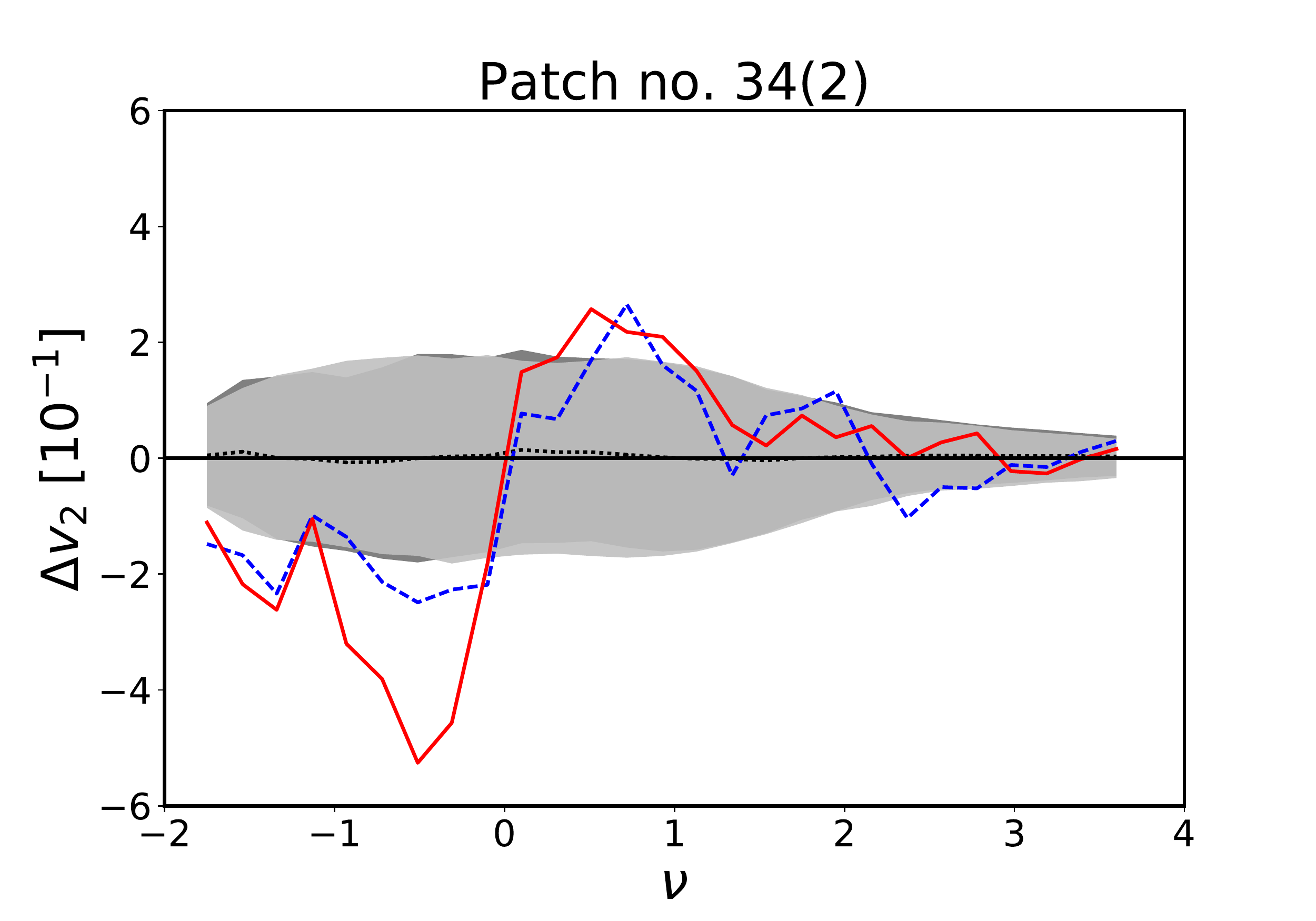}\\
 \vspace{-0.1cm}
 \includegraphics[width=0.32\textwidth]{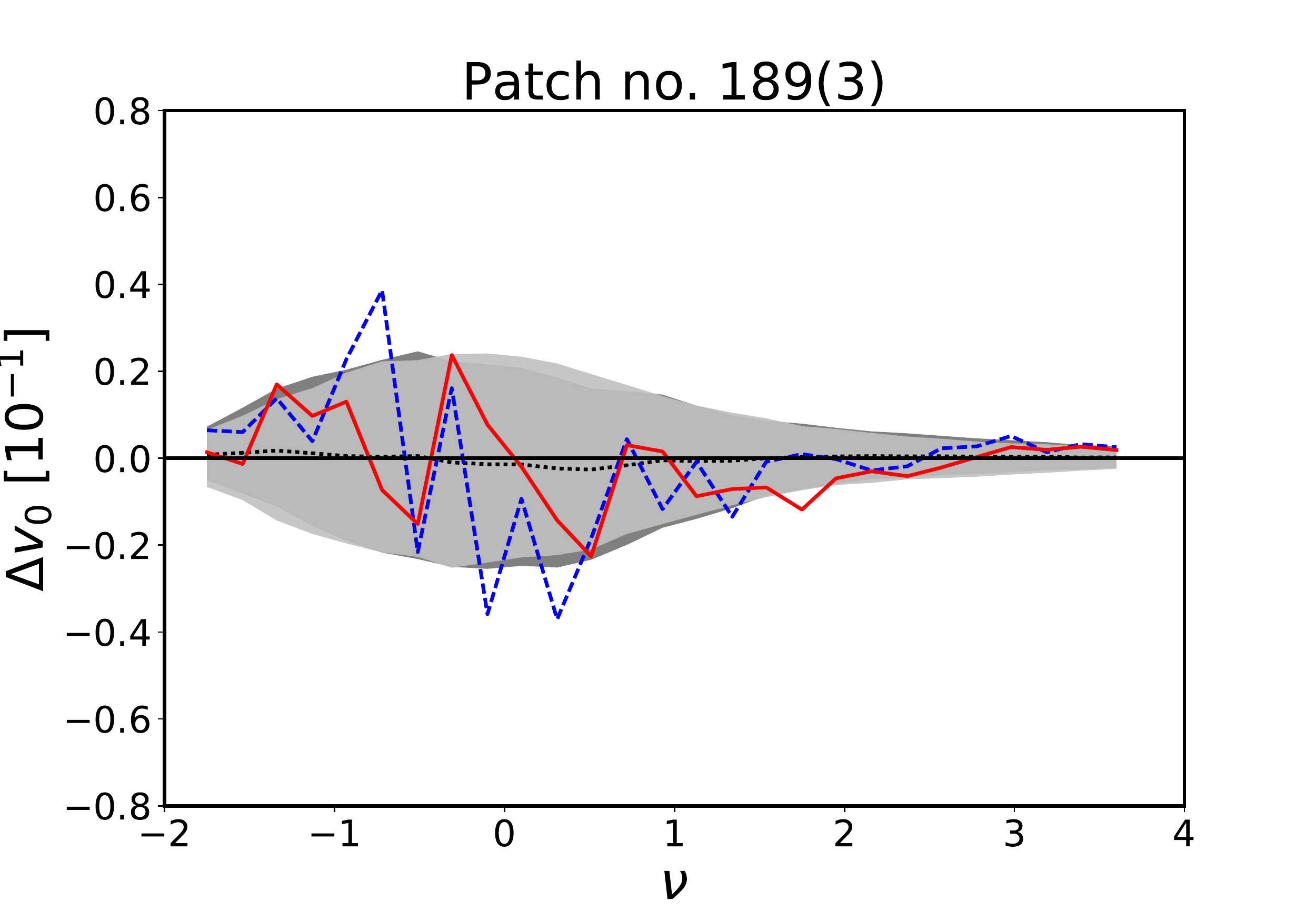}
 \includegraphics[width=0.32\textwidth]{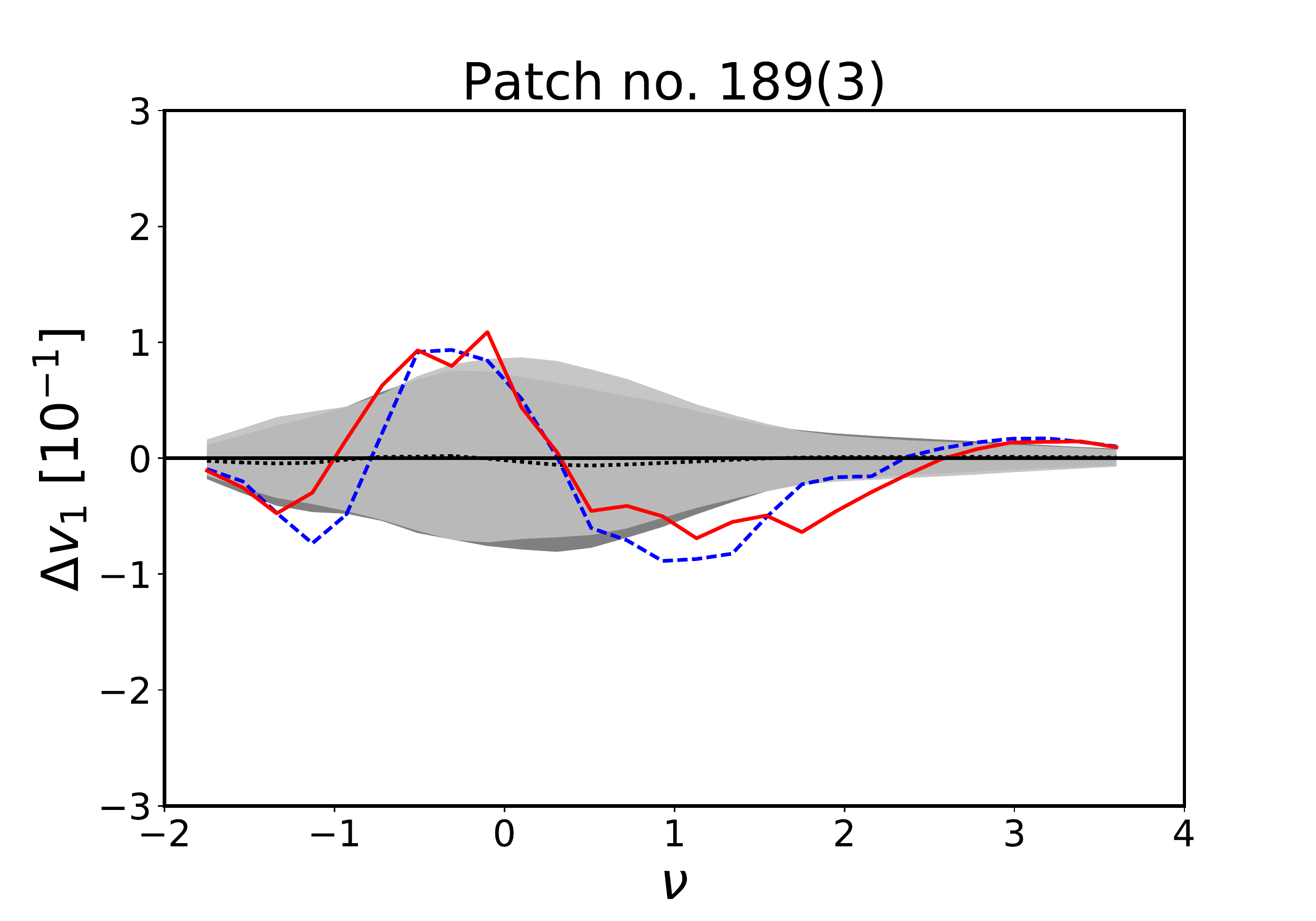}
 \includegraphics[width=0.32\textwidth]{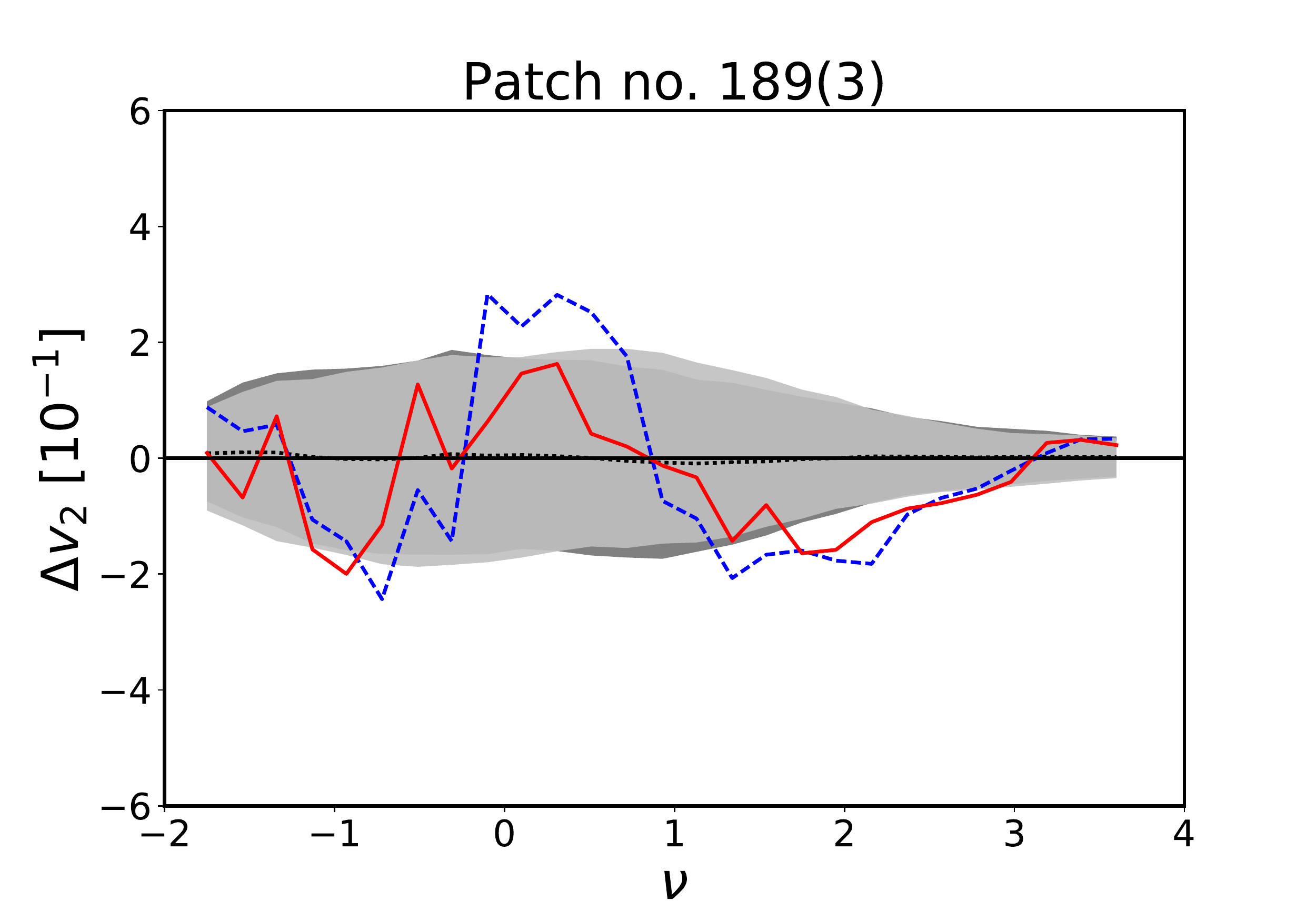}\\
 \vspace{-0.1cm}
 \includegraphics[width=0.32\textwidth]{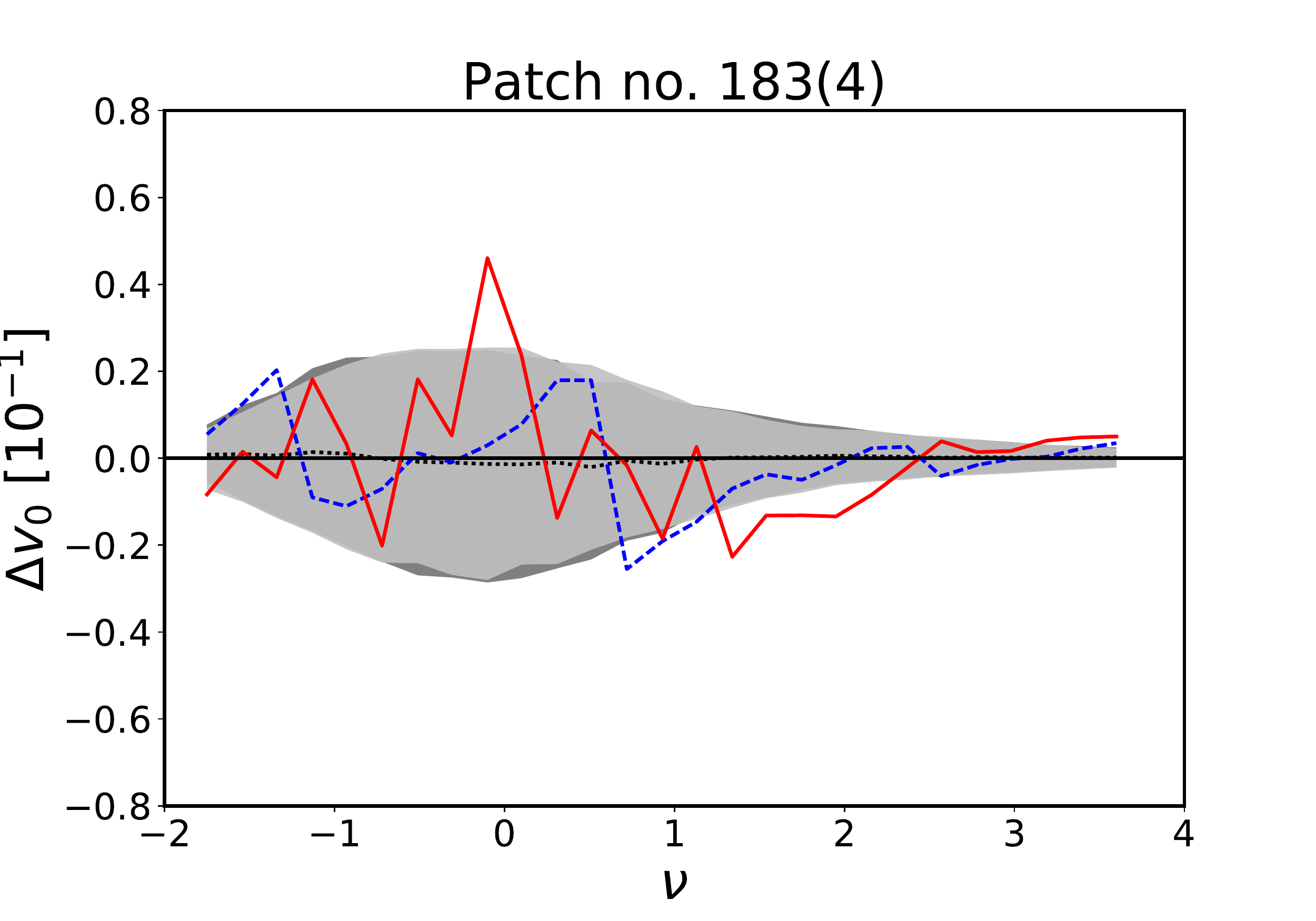}
 \includegraphics[width=0.32\textwidth]{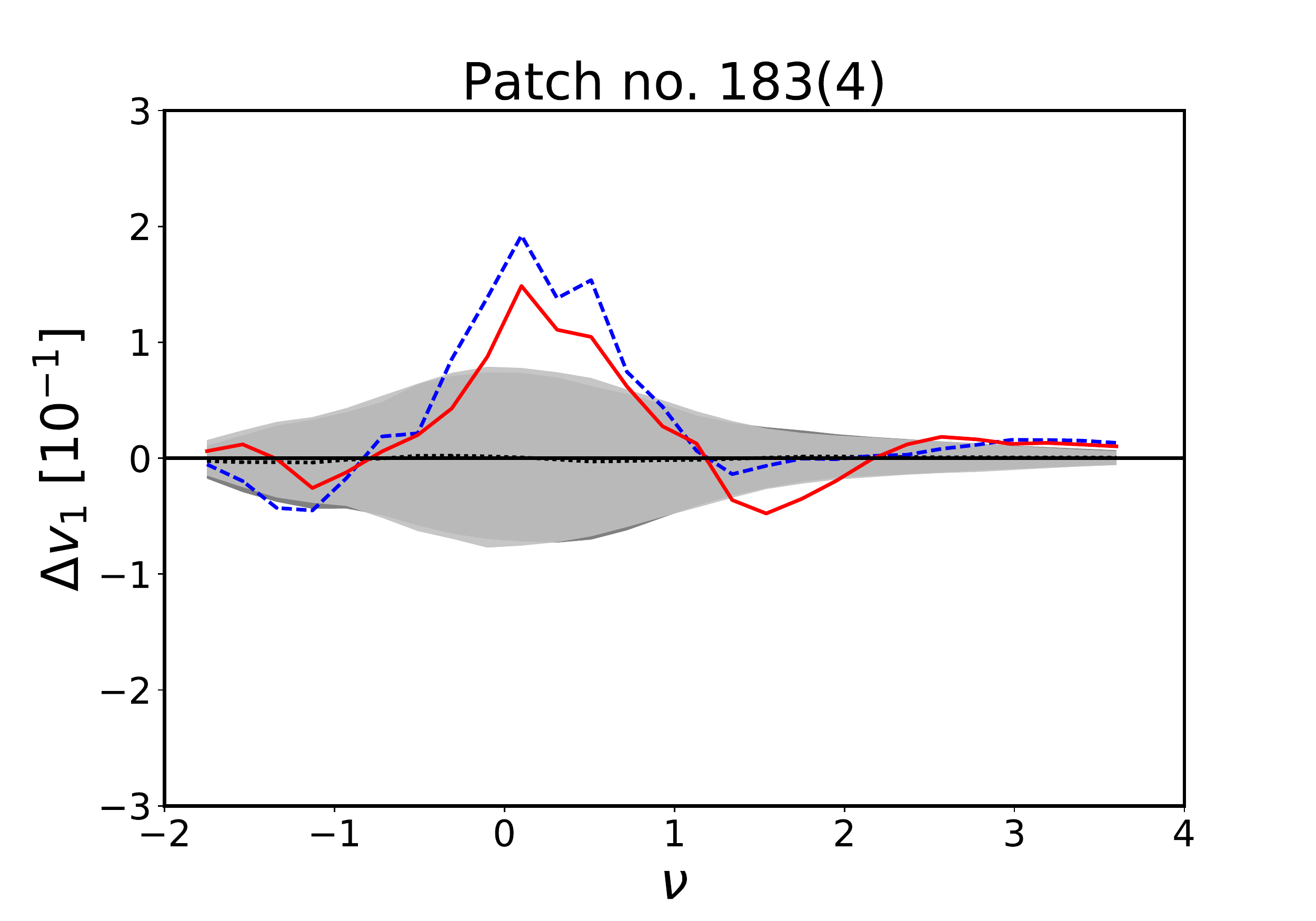}
 \includegraphics[width=0.32\textwidth]{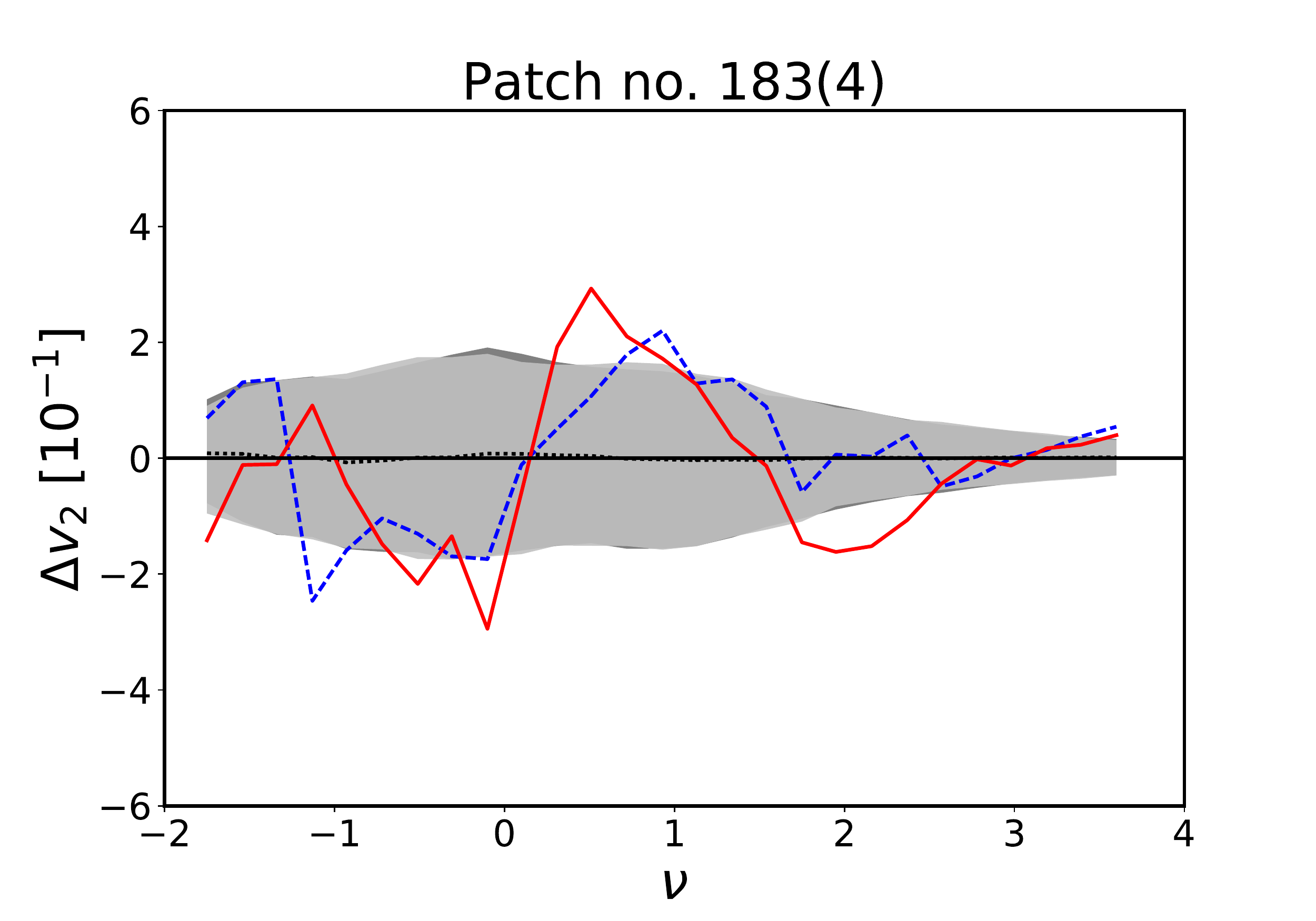}\\
\caption{Analogous to the Fig. \ref{fig:near_gal}, but for the six extreme patches far from the Galactic region.}
\label{fig:far_gal}
\end{figure*}

In order to investigate the selected patches individually and look for possible causes for their unusual features, we compare their MF vectors, $\mathrm{v}_k^{\mathrm{WSC}}$ (from each of the samples), to the corresponding mean vector calculated from the mock realisations, $\langle \mathrm{v}_k \rangle$.
This is done through the \textit{relative difference} among them, $\Delta \mathrm{v}_k$, that is, the difference between the two quantities normalized by the maximum value of the mean vector, as defined by

\begin{equation} \label{eq:relat_dif}
\Delta \mathrm{v}_k \equiv \frac{\mathrm{v}_k^{\mathrm{WSC}} - \langle \mathrm{v}_k \rangle}{\langle \mathrm{v}_k \rangle^{\rm \small MAX} },
\end{equation}
for $k$ = 0, 1, 2, i.e., Area, Perimeter, and Genus, respectively. 
The relative difference curves resulting for each MF estimated from the selected patches are displayed in Figs \ref{fig:near_gal} and \ref{fig:far_gal}. 
The two overlapped gray regions are estimated from the dispersion of the relative difference, $\Delta \mathrm{v}_k^i$, between the $k$th MF vector obtained from the $i$th GNC mock, $\mathrm{v}_k^i$, and the mean MF calculated from the 5000 non-contaminated mock realisations of the WSC-$clean$ sample, $\langle \mathrm{v}_k \rangle$, that is,

\begin{equation}  \label{eq:relat_dif_error}
\Delta \mathrm{v}_k^i \equiv \frac{\mathrm{v}_k^i - \langle \mathrm{v}_k \rangle}{\langle \mathrm{v}_k \rangle^{\rm \small MAX} } \bigg|_{\mbox{for the {\it i}th contaminated mock}}.
\end{equation}
The 2$\sigma_k$ deviation of $\Delta \mathrm{v}_k^i$ provides the shaded regions shown in Figs \ref{fig:near_gal} and \ref{fig:far_gal}, where the dark and light gray regions are derived from contaminated and non-contaminated simulations of the WSC-$clean$\footnote{Since the 2$\sigma_k$ regions from the two WSC samples are very similar, we present in Figs \ref{fig:near_gal} and \ref{fig:far_gal} only those derived from the WSC-$clean$ sample.}, respectively. 
As expected, one can observe a larger difference between the two gray regions for patches nearer the Galactic plane (Fig. \ref{fig:near_gal}), explained by the higher contamination level at low Galactic latitudes. 
The plots in Figs \ref{fig:near_gal} and \ref{fig:far_gal} also show a black dotted line representing the mean $\langle\Delta \mathrm{v}_k^i\rangle$ curve for the contaminated simulations. 
The comparison with the black solid (straight) line from the non-contaminated mocks can help to discriminate the features appearing in the relative difference curves as introduced by the galaxy distribution and coming from other effects, such as contamination.

A clear example of a contaminated region is given in the patch no. 130 in photo-$z$ bin 1, that is, 130(1). 
Their relative difference graphs not only show gray regions and dotted lines suggesting a high contamination level, but also exhibit a different behaviour between the red and blue curves obtained from WSC-$clean$ and WSC-$svm$ samples, respectively.
Note that the disagreement between the two curves appears mainly in the case of patches near the Galactic region (Fig. \ref{fig:near_gal}). 
In fact, since each sample is constructed from a different cleaning process, they are expected to differ more in such region, what seems to be confirmed by the MF results. 
Therefore, one can state that the MF are able to reveal features associated to the presence of contamination in the GNC maps, as observed in the examples of Fig. \ref{fig:near_gal} for patches no. 130(1) and 65(3).
As seen in Fig. \ref{fig:near_gal}, patches no. 137(2), and 134(3), also located near the Galactic plane, present $\Delta \mathrm{v}_k$ curves with similar patterns.

\begin{figure*}
\centering
 \includegraphics[width=0.32\textwidth]{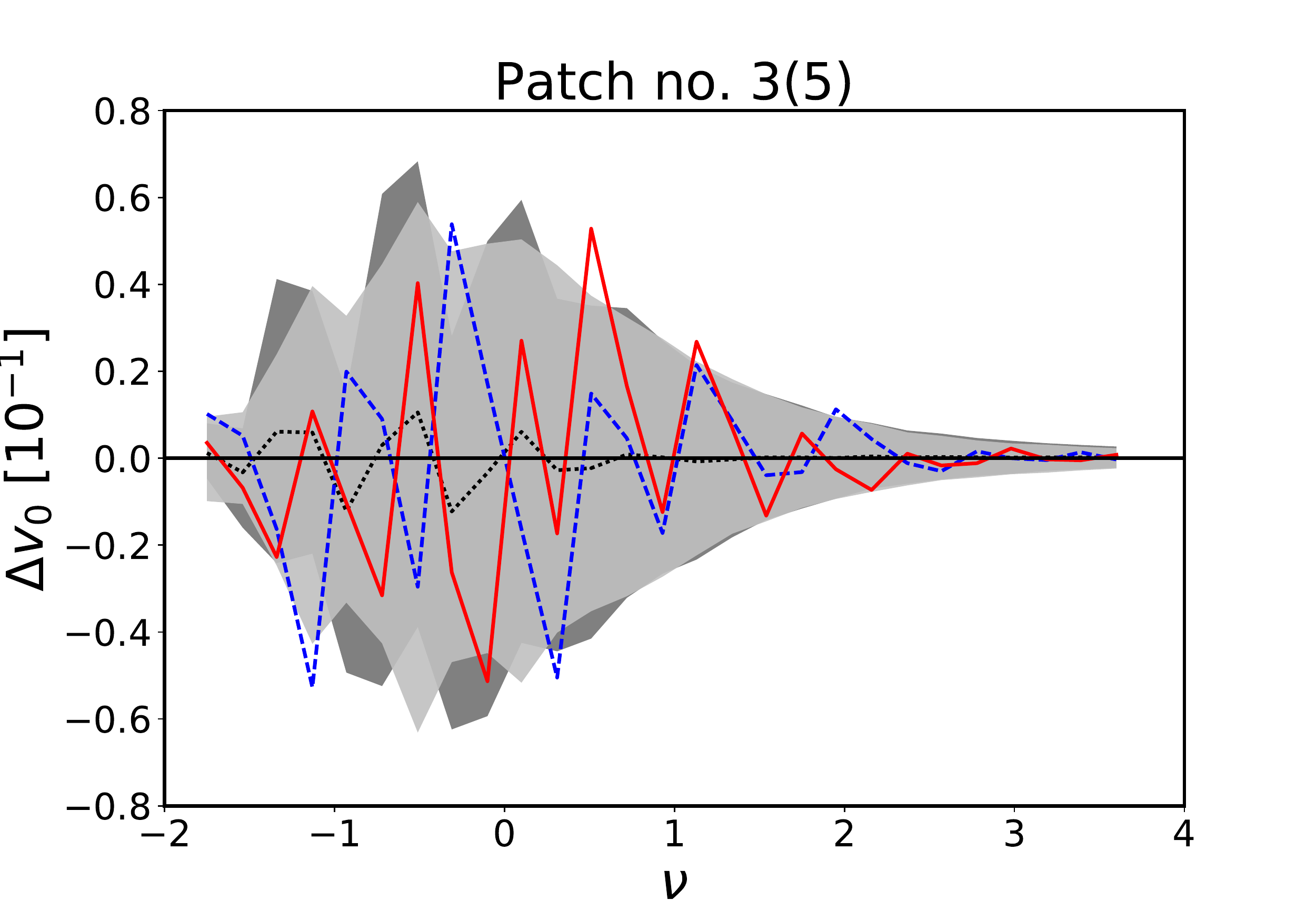}
 \includegraphics[width=0.32\textwidth]{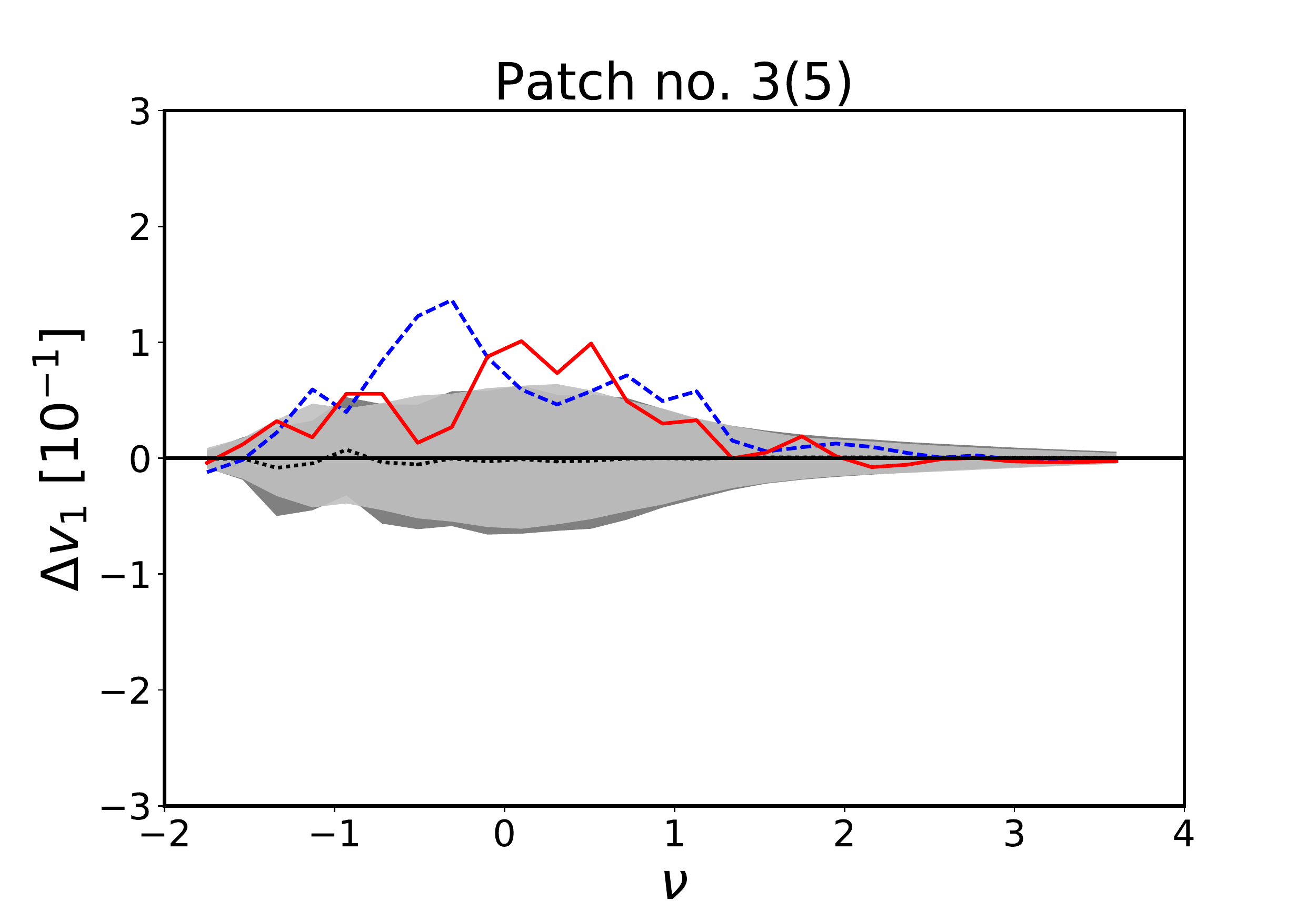}
 \includegraphics[width=0.32\textwidth]{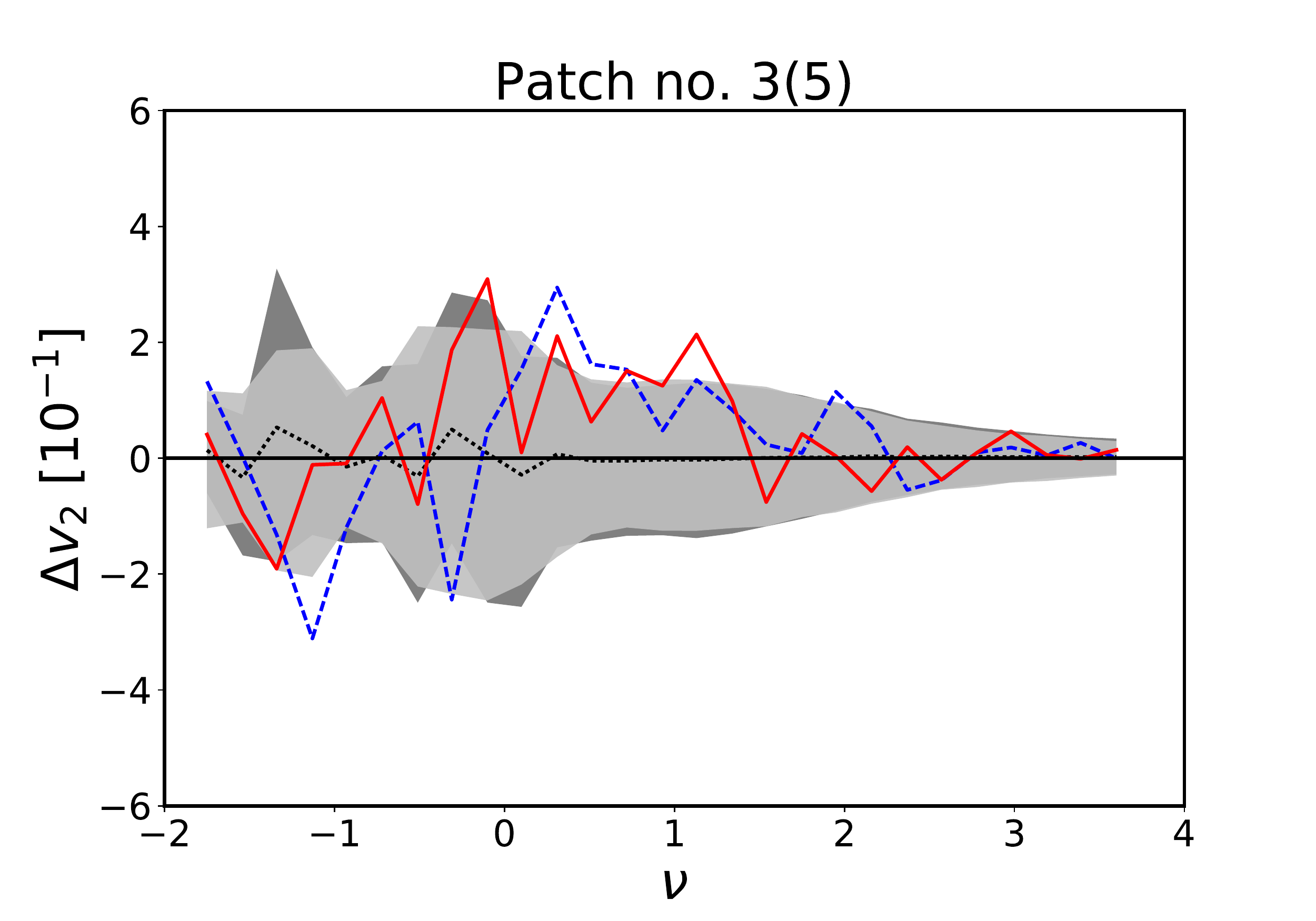}%
\caption{Analogous to the Fig. \ref{fig:near_gal}, but for an illustrative example from the patches of the last photo-$z$ bin, $0.30 < z < 0.35$.}
\label{fig:bin5}
\end{figure*}

The graphs in Fig. \ref{fig:bin5} correspond to another example of disagreements among the gray regions and also between the $\Delta \mathrm{v}_k$ curves from the two samples.
These are results from the patch no. 3(5), an illustrative example of regions far from the Galactic plane and taken from the last photo-$z$ bin.
Differently from what is observed in Fig. \ref{fig:near_gal}, in these graphs both the 2$\sigma_k$ regions and the relative difference curves present a very noisy behaviour.
Since the last photo-$z$ bin is the one with the smallest number of objects, affected by the incompleteness of the sample, these features can be associated to shot noise.
In fact, if a sample is composed by a very small number of galaxies, the derived GNC map, depending of the pixelization, can present a large number of holes (pixels with no objects), compromising the calculations of the MF, which seems to be the case here. 
This statement come specially from the fact that patch no. 3(5) presents $\Delta \mathrm{v}_k$ curves in agreement with the 2$\sigma_k$ region, which by itself seems to reveal the effect of completeness problems. 
Moreover, it is also worth to remind that the contamination of this photo-$z$ bin is expected to be larger than in the others \citep[see Section \ref{sec:section2.2} and discussion in][]{2016/bilicki}, corroborated by the disagreement observed among the red and blue curves. 
These results confirm the potential systematics of the last photo-$z$ bin. 
Interestingly, this is also the redshift range in which no patch with $p$-values $< 1.4\%$ was identified. 
Since the uncommon behaviour of the selected patches is associated to the existence of large and dense (or under-dense) structures, which appear more likely in the local Universe, one could actually expect less discrepant regions at high redshift. 

Among the four patches located near the Galactic region, and seemingly contaminated, one can still see an interesting feature in the analyses from patch no. 65(3), where the two samples agree when their relative difference curves extrapolate the gray regions around $\nu \sim 2$ (Fig. \ref{fig:near_gal}). 
Note that this is the only range of $\nu$ values where the curves extrapolate the 2$\sigma_k$ regions, which is, probably, the feature responsible for the selection of this patch as an extreme region.
Besides, the behaviour of the curves show that the three MF from the WSC samples, at this specific threshold, have amplitudes smaller than that from the mean upon the simulations.  
In general, for $\nu > 0$, a low amplitude Genus, that is, $\mathrm{v}_2^{\mathrm{WSC}} < \langle \mathrm{v}_2 \rangle$, suggests a number of over-densities, with respect to under-densities, lower than the expected by the simulations. 
It implies either a more connected region and quite high density (i.e., without many holes), or a larger number of under-densities than expected. 
However, we also see a small Perimeter, $\mathrm{v}_1^{\mathrm{WSC}} < \langle \mathrm{v}_1 \rangle$, indicating a small number of connected over-densities relative to the mocks, probably with a more regular contour, what is still confirmed by the small Area, $\mathrm{v}_0^{\mathrm{WSC}} < \langle \mathrm{v}_0 \rangle$. 
In other words, remembering that Genus can be associated to the difference between the number of over- and under-densities at a given $\nu$ threshold (equation \ref{eq:v2}), a low amplitude Genus, combined to the same behaviour from the Area and Perimeter, allows us to associate the selection of patch no. 65(3) to the presence of dense and clustered regions. 
Note that such characteristics are confirmed observing its Gnomonic projection (Fig. \ref{fig:gnomviews}), which presents two main areas of high density on its upper-left side.

For the analyses of patches far from the Galactic plane we follow a similar reasoning. 
We start from patch no. 157(1) that, as seen in Fig. \ref{fig:selectP}, is indeed very uncommon among those identified as extreme regions, 
presenting the highest statistical significance, an average $p$-value of 0.01\%. 
As observed from the first row of panels in Fig. \ref{fig:far_gal}, the {\it overall behaviour}\footnote{In opposite to the features appearing at a specific $\nu$ value, we call {\it overall behaviour} the pattern followed by the relative difference curves in the whole range of $\nu$.}
of the $\Delta \mathrm{v}_k$ curves shows that the three MF obtained from the WSC samples present an amplitude smaller than the mean from the simulations, $|\mathrm{v}_k^{\mathrm{WSC}}| < |\langle \mathrm{v}_k \rangle|$ [in absolute values -- remember Genus is negative for $\nu < 0$ (see again equation \ref{eq:relat_dif})]. 
Then, in analogy to what is observed in patch no. 65(3) at a specific $\nu$, also valid for the overall behaviour of the relative difference curves, the patch no. 157(1) is composed by connected over- and under-densities, that is, with no holes or `islands' of over-densities, respectively.

In addition to the overall behaviour of the $\Delta \mathrm{v}_k$ from patch no. 157(1), it is also important to analyse the features appearing at specific values of $\nu$, namely, $\nu \sim -1.1$ and $\nu \gtrsim 3$, where the $\Delta \mathrm{v}_k$ curves extrapolate significantly the 2$\sigma_k$ regions. 
The feature at positive $\nu$ present larger Area and Perimeter with respect to the simulations and an opposite behaviour regarding the Genus, indicating the presence of a large, connected, and very high density region. 
The corresponding Gnomonic projection confirms the presence of such structure at the bottom-right of the patch. 
Also, this projection clearly show the presence of a large and significant under-dense region at the bottom of the patch, possibly responsible for the feature appearing at the negative $\nu$ threshold. 
In fact, at $\nu \sim -1.1$, it is observed a smaller absolute value of the Genus relatively to the simulations, $|\mathrm{v}_2^{\mathrm{WSC}}| < |\langle \mathrm{v}_2 \rangle|$, indicating a lower number of under-densities, with respect to the over-densities, than expected by simulations.
This feature suggests the presence of a connected under-dense region. 
This is also corroborated by a small Perimeter, $\mathrm{v}_1^{\mathrm{WSC}} < \langle \mathrm{v}_1 \rangle$, and large Area, $\mathrm{v}_0^{\mathrm{WSC}} > \langle \mathrm{v}_0 \rangle$, appearing at $\nu \sim -1.1$. 

Very similar conclusions can be achieved analysing patch no. 180(1), not only in the overall behaviour of the $\Delta \mathrm{v}_k$ curves but also in their features appearing in specific $\nu$ thresholds. 
It differs from patch no. 157(1) only in the amplitude of the curves when extrapolating the gray regions around $\nu \sim 3.3$ and $\nu \sim -1.5$. 
In fact, the presence of a significant over-dense (under-dense) region can be seen in the left (top and right) of the Gnomonic projection of patch no. 180(1), whose difference in the significance of this area relative to the one present in the patch no. 157(1) can be seen from the corresponding color bar.

With an opposite overall behaviour relative to patch no. 157(1), patch no. 34(2) present the Perimeter and Genus with amplitude larger than that from the simulations, $|\mathrm{v}_k^{\mathrm{WSC}}| > |\langle \mathrm{v}_k \rangle|$ for $k$ = 1 and 2, but with the relative difference in Area oscillating around zero (Fig. \ref{fig:far_gal}).
Moreover, we still see the $\Delta \mathrm{v}_k$ curves extrapolating the 2$\sigma_k$ regions at the specific thresholds of $\nu \sim 2.5$ and $\nu \sim -1.5$, also with an inverted behaviour relatively to those appearing in patch no. 157(1). 
Following previous discussions, these specific and overall behaviours are associated to the presence of more than one under-dense area, possibly with irregular contours. 
This reflects what is observed in the Fig. \ref{fig:gnomviews}, that is, a large number of over-densities spread over the patch, but not of so high density.

Patch no. 189(3), on the other hand, seems to have its uncommon behaviour coming from the specific feature at $\nu \sim 2$, otherwise the $\Delta \mathrm{v}_k$ curves appear almost completely inside the shaded regions, as observed in the third row of Fig. \ref{fig:far_gal}. 
The behaviour of the three MF at this threshold, i.e., $\mathrm{v}_k^{\mathrm{WSC}} < \langle \mathrm{v}_k \rangle$  for $k$ = 0, 1, and 2, indicates the presence of connected over-dense regions (one or more), similar to the one appearing in patch no. 157(1) but of smaller sizes. 
A very similar behaviour is observed in patch no. 25(2), in the sense that their $\Delta \mathrm{v}_k$ curves appear inside the 2$\sigma_k$ regions, also having its high $\chi^2$ amplitude originated from a specific feature appearing at $\nu \sim 3$.
Presenting extrapolations of the shaded regions such that $\mathrm{v}_k^{\mathrm{WSC}} > \langle \mathrm{v}_k \rangle$, for $k$ = 0 and 1, and $\mathrm{v}_2^{\mathrm{WSC}} < \langle \mathrm{v}_2 \rangle$ at this specific $\nu$, the relative differences indicate that patch no. 25(2) contain a connected over-dense region, visible at the left side of the Gnomonic projection shown in Fig. \ref{fig:gnomviews}.

Our last patch, no. 183(4), have their $\Delta \mathrm{v}_k$ curves with overall behaviours similar to those from patch no. 34(2) -- indicating the presence of a large number of over-densities relatively to the simulations -- but with smaller absolute amplitude.
A specific feature also appears in this case at a positive threshold, namely, at $\nu \gtrsim 2$, where all the three MF of patch no. 183(4) present amplitudes smaller than expected by the simulations, $\mathrm{v}_k^{\mathrm{WSC}} < \langle \mathrm{v}_k \rangle$ for $k$ = 0, 1, and 2. 
Note that this corroborates the conclusions from their overall behaviour, i.e., the patch has a large number of over-densities relatively to the simulations. 
In fact, the Gnomonic projection of patch no. 183(4) in Fig. \ref{fig:gnomviews} shows the presence of some over-dense areas, although denser and in smaller quantity compared to patch no. 34(2).

Finally, out of curiosity, we return to the interesting case of patch no. 157(1), the highest statistical significance region among those identified by our methodology, and go beyond the analysis of its relative difference curves aiming to investigate in more detail the reasons for its very uncommon behaviour. 
As discussed, the Gnomonic projection of this patch, in agreement with our conclusions about its $\Delta \mathrm{v}_k$ curves, show clearly the presence of at least one highly dense region and one extensive under-density at the approximate positions $(l,b) = (118.4^\circ,-44.2^\circ)$ and $(l,b) = (121.2^\circ,-50.9^\circ)$, respectively.
We used the NASA/IPAC Extragalactic Database\footnote{\tt https://ned.ipac.caltech.edu/} (NED) to look for objects -- galaxies and galaxy clusters -- around these two positions, in the redshift range $0.10 < z < 0.15$. 
According to NED, there is a total of 73 objects inside a region of 60 arcmin radius centred on the over-dense area of patch no. 157(1), among them 6 galaxy clusters\footnote{This comparison is merely illustrative, we did not performed further investigations about it.}.
Inside a 60 arcmin radius region centred in the under-dense area visible in this patch, on the other hand, were found only 3 galaxies. 
Therefore, additionally to a visual inspection of Gnomonic projections, such extra information works as a consistency check of our conclusions based upon the features observed in the relative difference curves, $\Delta \mathrm{v}_k$. 
In fact, it confirms the MF to be an efficient morphological tool to investigate the clustering of galaxies and identify discrepancies with respect to the expectations of the model.
Additionally, this ability would also support their use in discriminating, for example, diverse modified gravity models, where different clustering characteristics are expected.

\section{Conclusions and final Remarks} \label{sec:section6}

The literature offers a variety of statistical tools to study the large-scale structure in the Universe. 
Among them are the Minkowski Functionals (MF), widely used to obtain morphological information about the 3D volume distributed galaxy samples. 
Here, for the first time, the MF are applied to 2D projections, that is, to galaxy number count (GNC) maps. 

We have presented through this paper the results of using the MF to perform tomographic local analyses of the  WISExSuperCOSMOS (WSC) catalogue: we divided the catalogue into 192 sky patches and into five disjoint photo-$z$ bins, converted them into GNC maps and identified regions containing uncommon distribution of objects with respect to the expected by the $\Lambda$CDM concordance model. 
This procedure allows us to identify a total of 10 patches of the GNC maps, in four different photo-$z$ bins ($0.10 < z < 0.30$), in disagreement with the mock realisations, with $p$-value $<$ 1.4\%. 
Note that the MF did not reveal any extreme region in the last photo-$z$ bin, $0.30 < z < 0.35$. 
A careful statistical analysis allowed us to evaluate the probability of finding such set of regions in the GNC maps, showing that they are not in fact discrepant but expected by the simulations, although the patch no. 157(1) appears as a highly extreme region, with $p$-value = 0.01\%.
In other words, our results indicate that the observed Universe as given by the WSC data is in agreement not only with the fiducial cosmological model, $\Lambda$CDM, but also with the contamination model, selection function, lognormal distribution of the objects, and other astrophysical features assumed to generate the mock realisations. 

Nevertheless, even though not discrepant, the selected patches are still extreme and their uncommon features motivate a thorough analysis to investigate possible reasons for their disagreement with the simulations. 
A scrutiny of the selected regions through detailed analyses of their MF curves, comparing them to the mean obtained from the $\Lambda$CDM simulations, followed by an additional investigation on the last photo-$z$ shell lead us to the main conclusions summarized bellow: 

(i) Our results show the MF to be highly efficient in obtaining a topological description of the distribution of galaxies in small regions of the sky, allowing a direct comparison with predictions of the concordance cosmological model. 
These tools have furnished crucial informations about how clustered are the galaxies, allowing us to identify the presence of unusual under- and over-densities. 

(ii) The careful analysis of the relative difference curves, $\Delta \mathrm{v}_k$, from the three MF led us to associate the observed signature to a possible cause for their uncommon behaviour. 
In particular, the divergence among the  $\Delta \mathrm{v}_k$ curves from the different samples, together with the proximity of the highlighted patches to the Galactic plane, helped us to identify those classified as extreme due to the presence of contamination, confirming the adequacy of the MF for such task. 

(iii) Our analyses of the last photo-$z$ bin of the WSC samples, differently from the first four, did not identify any extreme region. 
Analysing their $\Delta \mathrm{v}_k$ curves we do not find a statistically significant disagreement between observations and simulations (Fig. \ref{fig:bin5}), i.e. the two WSC samples either fall within the 2$\sigma_k$ regions and/or do not agree about their $p$-value distribution (the low $p$-value patches are not the same in both samples).
This behaviour might has been introduced by higher shot noise or due to a higher contamination level; it is also possible that the Universe at higher redshifts, being less evolved, is better modelled by our mocks.
Again our results reinforce the power of the MF in characterizing the galaxy distribution in GNC maps. 

Moreover, it is worth stressing that the analyses of the selected patches of the WSC GNC maps, using for comparison the mock realisations, allow us to raise possible reasons for their uncommon behaviour. 
Among these, we can also mention the possibility of inaccurate photometric calibration, instrumental problems, or even large error in the photo-$z$ estimates, leading sources to appear in a photo-$z$ bin they do not belong to. 
Nevertheless, even with all these possibilities of inaccuracy, in some cases the presence of over- and under-densities, associated to the redder and bluer pixels, respectively, appears quite clear in the relative difference curves, as shown in patch no. 157(1). 
This confirms the validity of our methodology and its capacity to discriminate features associated to the distributions of galaxies and those coming from contamination. 

Finally, we emphasize that the approach for mapping the galaxy distribution described here can also be applied to a variety of catalogues, specially due to the advantage of the MF in not depending on the size of the analysed region. 
In this sense, future experiments, providing larger and deeper catalogues, with higher completeness and better cleaning processes, are promising datasets for this kind of analysis and may also benefit from its ability to identify residual contamination that might influence cosmological analysis. 
These experiments can furnish more details and precise mapping of the galaxy distribution through the Universe, that can be efficiently studied with the methodology proposed here.

\section*{Acknowledgements}
\noindent
We thank Carlos A. P. Bengaly Jr. and Maciej Bilicki for very useful discussions.
We acknowledge the use of the code for calculating the MF, from~\cite{2013/ducout} and~\cite{2012/gay}. 
Some of the results in this paper have been derived using the {\sc healpix} package~\citep{2005/gorski}. 
CPN is supported by the FAPERJ Brazilian funding agency. 
AB acknowledges financial support from the Capes Brazilian Agency through the grant 88881.064966/2014-01. 
HSX acknowledges FAPESP Brazilian funding agency for the financial support.
GAM acknowledges Capes fellowship.




\bsp
\label{lastpage}

\end{document}